\documentclass[twocolumn,twoside]{IEEEtran}
\usepackage{mathpple}
\usepackage{times}
\usepackage{amsmath}  
\usepackage{amssymb}  
\usepackage{mathrsfs} 
\usepackage{theorem}  
\usepackage{cite}     
\usepackage{comment}  
\usepackage{upref}
\usepackage{amsfonts}
\usepackage{verbatim}
\usepackage[dvipsnames,usenames]{color}
\usepackage{enumerate}
\usepackage{cuted}
\usepackage[shortlabels]{enumitem}
\usepackage{dsfont}

\usepackage{algorithm}
\usepackage{times,color}
\usepackage{graphicx}
\usepackage{float}
\usepackage{color}
\usepackage{mathtools}
\usepackage{graphicx}
\usepackage{listings}
\usepackage{tikz}
\usepackage{multicol}
\usepackage{psfrag}

\usepackage{pgfplots}
\usepackage{hhline}
\usepackage{xcolor,colortbl}
\usepackage{algorithm}
\usepackage[noend]{algpseudocode}

\usepackage{wrapfig}
\usepackage{subfigure}
\usepackage[top=0.58in, bottom=0.58in, left=0.6in, right=0.6in]{geometry}
\usepackage{kantlipsum,cuted}
\usepackage[all]{xy}
\usepackage{algorithmicx}
\algrenewcommand\alglinenumber[1]{\scriptsize #1:}
\algrenewcommand\algorithmicindent{1em}%
\usepackage{latexsym}
\usepackage{multirow}

\usepackage{tabu}
\usepackage{lipsum}
\usepackage{blkarray, bigstrut}
\usepackage{comment}

\newcommand{\field}[1]{\mathbb{#1}}

\newcommand{\F}{\field{F}}

\newcommand{\N}{\field{N}}

\newcommand{\ut}{\cT\textmd{-}}

\newcommand{\bline}{\textbf{\textendash}}

\newcommand{\RN}[1]{%
	\textup{\uppercase\expandafter{\romannumeral#1}}%
}

\newtheorem{innercustomgeneric}{\customgenericname}
\providecommand{\customgenericname}{}
\newcommand{\newcustomtheorem}[2]{%
	\newenvironment{#1}[1]
	{%
		\renewcommand\customgenericname{#2}%
		\renewcommand\theinnercustomgeneric{##1}%
		\innercustomgeneric
	}
	{\endinnercustomgeneric}
}
\newcustomtheorem{customproposition}{Proposition}
\newcustomtheorem{customtheorem}{Theorem}
\newcustomtheorem{customlemma}{Lemma}
\newcustomtheorem{customclaim}{Claim}
\newcustomtheorem{customconjecture}{Conjecture}

\makeatletter
\def\BState{\State\hskip-\ALG@thistlm}

\makeatother

\algnewcommand{\Initialize}[1]{%
	\State \textbf{Initialize:}
	\Statex \hspace*{\algorithmicindent}\parbox[t]{.8\linewidth}{\raggedright #1}
}

\newcommand{\be}[1]{\begin{equation}\label{#1}}
\newcommand{\ee}{\end{equation}}

\newcommand{\bc}{\begin{center}}
	\newcommand{\ec}{\end{center}}

\newcommand{\ceil}[1]{\lceil{#1}\rceil}

\newcommand{\cB}{{\cal B}}
\newcommand{\cC}{{\cal C}}
\newcommand{\cD}{{\cal D}}
\newcommand{\cE}{{\cal E}}
\newcommand{\cF}{{\cal F}}

\newcommand{\cH}{{\cal H}}

\newcommand{\cO}{{\cal O}}
\newcommand{\cP}{{\cal P}}

\newcommand{\cS}{{\cal S}}
\newcommand{\cT}{{\cal T}}

\newcommand{\bfu}{{\boldsymbol u}}
\newcommand{\bfv}{{\boldsymbol v}}
\newcommand{\bfw}{{\boldsymbol w}}

\newcommand{\bfP}{{\mathbf P}}

\renewcommand{\leq}{\leqslant}

\renewcommand{\geq}{\geqslant}

\newcommand{\Cref}[1]{Co\-rol\-la\-ry\,\ref{#1}}

\theoremstyle{plain} \theorembodyfont{\normalfont\slshape}

\newtheorem{thm}{Theorem$\!$}
\newenvironment{theorem}{\begin{thm}\hspace*{-1ex}{\bf.}}{\end{thm}}

\newtheorem{prop}[thm]{Proposition$\!$}

\newtheorem{lem}[thm]{Lemma$\!$}
\newenvironment{lemma}{\begin{lem}\hspace*{-1ex}{\bf.}}{\end{lem}}

\newtheorem{cor}[thm]{Corollary$\!$}
\newenvironment{corollary}{\begin{cor}\hspace*{-1ex}{\bf.}}{\end{cor}}

\newtheorem{cons}[thm]{Construction$\!$}

\newtheorem{defi}[thm]{Definition$\!$}
\newenvironment{definition}{\begin{defi}\hspace*{-1ex}{\bf.}}{\end{defi}}

\newtheorem{cl}{Claim}
\newenvironment{claim}{\begin{cl}\hspace*{-1ex}{\bf .}}{\end{cl}}
\theorembodyfont{\normalfont}

\newtheorem{conj}{Conjecture$\!$}

\newtheorem{exam}{Example$\!$}
\newenvironment{example}{\begin{exam}\hspace*{-1ex}{\bf .}}{\end{exam}}

\newtheorem{remrk}{Remark$\!$}
\newenvironment{remark}{\begin{remrk}\hspace*{-1ex}{\bf .}}{\end{remrk}}

\newtheorem{Construction}{Construction}

\definecolor{Codecolor}{named}{White}  

\newcommand{\Copen}{\mbox{\{\kern-5.50pt\{}}
\newcommand{\Cclose}{\mbox{\}\kern-5.50pt\}}}
\newcommand{\Cslash}{\mbox{$\backslash\kern-6.02pt\backslash$}}

\begin{document}
		
		\title{Codes over Trees}
		
		\author{\large Lev~Yohananov,~\IEEEmembership{Student Member,~IEEE},  and Eitan~Yaakobi,~\IEEEmembership{Senior Member,~IEEE} 
			\thanks{L. Yohananov, and E. Yaakobi are with the Department of Computer Science, Technion --- Israel Institute of Technology, Haifa 3200003, Israel (e-mail: \texttt{\{levyohananov,yaakobi\}@cs.technion.ac.il}).}
\vspace{-3ex}		}

		\maketitle		
\begin{abstract}
	
	In graph theory, a \emph{tree} is one of the more popular families of graphs with a wide range of applications in computer science as well as many other related fields. While there are several distance measures over the set of all trees, we consider here the one which defines the so-called \emph{tree distance}, defined by the minimum number of edit operations, of removing and adding edges, in order to change one tree into another. From a coding theoretic perspective, codes over the tree distance are used for the correction of edge erasures and errors. However, studying this distance measure is important for many other applications that use trees and properties on their locality and the number of neighbor trees. Under this paradigm, the largest size of code over trees with a prescribed minimum tree distance is investigated. Upper bounds on these codes as well as code constructions are presented. A significant part of our study is dedicated to the problem of calculating the size of the ball of trees of a given radius. These balls are not regular and thus we show that while the \emph{star tree} has asymptotically the smallest size of the ball, the maximum is achieved for the  \emph{path tree}.

\end{abstract}
\vspace{-2ex}

\section{Introduction} \label{sec:intro}

In graph theory, a \emph{tree} is a  special case  of a connected graph, which comprises of $n$ labeled nodes and $n-1$ edges. Studying trees and their properties has been beneficial in numerous applications. For example, in signal processing, trees are used for the representation of waveforms~\cite{Cheng}. In programming languages, trees are used as structures to describe restrictions in the language. Trees also represent collections of hierarchical text which are used in information retrieval. In  cybersecurity  applications trees are used to represent fingerprint patterns~\cite{Moayer}. One of the biology applications includes the tree-matching algorithm to compare between trees in order to analyze multiple RNA secondary structures~\cite{Shapiro}. Trees are also used in the subgraph isomorphism problem which, among its very applications, is used for chemical substructure searching~\cite{Barnard}.

An important feature when studying trees is defining an appropriate distance function. Several distance measures over trees have been proposed in the literature. Among the many examples are the tree edit distance~\cite{Tai}, top-down distance~\cite{Selkow}, alignment distance~\cite{Jiang}, isolated-subtree distance~\cite{Tanaka}, and bottom-up distance~\cite{Valiente}. These distance measures are mostly characterized by adding, removing, and relabeling nodes and edges as well as counting differences between trees with a different number of nodes. One of the more common and widely used distance, which will be referred in this work as the \emph{tree distance}~\cite{Paulden,Gottlieb}, considers the number of edit edge operations in order to transform one tree to another. Namely, given two labeled trees over $n$ nodes, the tree distance is defined to be half of the minimum number of edges that are required to be removed and added in order to change one tree to another. This value is also equivalent to the difference between $n-1$ and the number of edges that the two trees share in common. Despite the popularity of this distance function, the knowledge of its characteristics and properties is quite limited. The goal of this paper is to close on these gaps and study trees under the tree distance from a coding theory perspective. To the best of our knowledge, this direction has not been explored rigorously so far.

Motivated by the coding theory approach, in this work we apply the tree distance, which is a metric, to study \textit{codes over trees} with a prescribed minimum tree distance. This family of codes can be used for the correction of edge erasures. 
There are several applications in which such codes can be used. For example, in data structures, a tree is a widely used abstract data type that simulates a hierarchical tree structure~\cite{Cormen}. Such tree data structures store the information in nodes and use edges as pointers between them. There are numerous examples for such tree data structures including \textit{abstract syntax trees} (AST), \textit{parsing trees} and \textit{binary search trees} (BST)~\cite{Cormen,Knuth}. AST represent the abstract syntactic structure of source code written in a programming language, while each node of the tree denotes a construct occurring in the source code. Parsing trees represent the syntactic structure of a string according to some context-free grammar. BST trees store in each node a value greater than all the values in the node's left subtree and less than those in its right subtree. These tree data structures can be implemented such that each node stores a list of pointers to other nodes in the tree. {Theoretically}, such pointers might have wrong addresses, which affects the reliability of the data structure. By adding redundancy edges and nodes, codes over trees may correct the unexpected pointer mismatches. Another family of applications include data structures such as \textit{tries} and \textit{suffix trees}~\cite{Knuth} in which the information is stored on the edges rather than the nodes. Such data structures can be implemented by a list of $n-1$ edges which is a list of node pairs together with the information on every edge. {Again, theoretically,} such an edge list may have failures that can indeed be corrected using classical error-correction codes. However, these codes will not be cardinality optimal since they do not take advantage of the structure of the tree. For the binary case, using classical error-correction codes, we show in the paper the construction of codes over trees of size $\Omega(n^{n-2d})$ where $d\leq n/2$ corresponds to the minimum tree distance of the code. Using codes over trees we show that it is possible to construct codes of cardinality $\Omega(n^2)$, while the minimum tree distance $d$ approaches $\lfloor 3n/4 \rfloor$ and $n$ is a prime number.

Another interesting problem for the tree distance is the study of the size of balls according to the tree distance. {This investigation is useful not only for applying the sphere packing bound on codes over trees, but also for other applications}. For example, in~\cite{Davies} it was claimed that recent research on nanotechnology discovered that structures of DNA molecules can be constructed into trees or lattices, and that future synthesis techniques may use physical constraints to enforce tree structures on the written base. The authors of~\cite{Davies} introduced the \textit{tree trace reconstruction problem}, in which the goal is to reconstruct a tree from several of its copies while each copy can have node deletions. In this case, the size of the tree balls may be useful. 
Another approach deals with graph matching, i.e., the problem of finding a similarity between graphs~\cite{Ganassali,Llados}.  Graph matching is an important tool used for example in computer vision and pattern recognition. 	
One of the problems under this setup is to find a \textit{model graph}, which represents the prototype symbol, as a subgraph in an \textit{input graph} that represents a diagram, which is also called \textit{subgraph isomorphism problem}~\cite{Llados}. 
If the model graph cannot be found exactly in the input graph, then the goal is to find a subgraph that is close to the model graph while the similarity is determined by edit operations on the nodes and edges. This problem is also studied for trees called \textit{subtree isomorphism problem}~\cite{Shamir}, and the size of balls of trees may be useful for this problem. 
Lastly, one of the classical problems in graph theory is finding a  minimum spanning tree (MST) for a given graph. While the MST problem is solved in polynomial time~\cite{Kruskal,Prim}, it may become NP-hard under some specific constraints. For example, in the degree-constrained MST problem ($d$-MST)~\cite{Raidl,Raidl2,Krishnamoorthy,Zhou}, it is required that the degree of every vertex in the MST is not greater than some fixed value $d$. In another example, the goal is to look for an MST in which the length of the simple path between every two vertices is bounded from above by a given value $D\geq 4$~\cite{Raidl3}. One of the common approaches for solving such problems uses \emph{evolution algorithms} (\emph{EA}). Under this setup, the goal is to find a feasible tree to the problem by iteratively searching for a candidate tree. This iterative procedure is invoked by using \textit{mutation operations} over the current tree in order to produce a new candidate tree. These mutation operations typically involve the modification of edges in the tree and as such are highly related to the tree distance. Thus, in order to analyze the complexity of such algorithms, it is necessary to study the size of the balls according to the tree distance. In fact, in~\cite{Gottlieb} the size of the radius-one ball was computed for all trees with at most 20 vertices. According to this computer search, it was observed that the smallest size of the ball is achieved when the tree is a star tree (i.e., the tree has one node connected to all other nodes), while the largest for a path tree (i.e., the tree has two leaves and the degree of all other nodes is two). In this paper, we establish this result for any number of nodes in the tree as well as for any radius. Furthermore, it is shown that the size of the radius-$t$ ball ranges between $\Omega(n^{2t})$ (for a star tree) and $\cO(n^{3t})$ (for a path tree), while the average size of  all balls  is $\Theta(n^{2.5t})$.

This paper is organized as follows.
In Section~\ref{sec:defs}, we formally define the tree distance and codes over trees as well as several more useful definitions and properties  for balls of trees, that will be defined in the sequel. An \emph{edge erasure} is the event in which one of the edges in the tree is erased and a forest is received with two connected components. This is also extended to the erasure of multiple edges. If $t$ edges are erased, then a forest with $t+1$  connected components  is received and the number of such forests is $\binom{n-1}{t}$. 
In Section~\ref{sec:main} we summarize all main results of the paper. 
In Section~\ref{ch:bound_trees}, by using several known results on the number of forests with a fixed number of  connected components we are able to derive a sphere packing bound for codes over trees. More specifically, the size of codes over trees of minimum tree distance $d$ cannot be greater than $\cO(n^{n-d-1})$.
In Section~\ref{sec:ball size}, we study balls of trees. The \emph{tree ball of trees} of a given tree $T$ consists of all trees  such that  their tree distance from $T$ is at most some fixed radius $t$. These balls are not regular. In this section, these balls are studied for radius one. Balls with a general radius are studied in Section~\ref{sec:gen}.
In Section~\ref{sec:lines_trees}, the size of star, path tree tree ball is presented, respectively. 
Lastly, in Section~\ref{tree:const}, for a fixed $d$ we show a construction of codes over trees of size $\Omega(n^{n-2d})$. It is also shown that it is possible to construct codes of cardinality $\Omega(n^2)$, while the minimum distance $d$ approaches $\lfloor 3n/4 \rfloor$ and $n$ is  a prime number. Finally, Section~\ref{sec:conc} concludes the paper.


\section{Definitions and Preliminaries}\label{sec:defs}	
Let $G=(V_n,E)$ be a graph, where $V_n=\{v_0,v_1,\ldots,v_{n-1}\}$ is a set of $n\geq 1$ labeled \emph{nodes}, also called \emph{vertices}, and $E\subseteq V_n\times V_n$ is its \emph{edge} set. In this paper, we only study undirected trees and forests. By a slight abuse of notation, every undirected edge in the graph will be denoted by $\langle v_i,v_j\rangle$ where the order in this pair does not matter, i.e., the notation $\langle v_i,v_j \rangle$ is identical to the notation $\langle v_j,v_i \rangle$.  Thus, there are $\binom{n}{2}$   possibilities for the edges  and the edge set is defined by
\begin{equation}\label{eq:En}
E_n = \{ \langle v_i,v_j \rangle ~|~ i,j \in [n] \},
\end{equation} where $[n] \triangleq \{0,1,\ldots,n-1\}$.

A finite undirected \textit{tree} over $n$ nodes is a connected undirected graph with $n-1$ edges. The \textit{degree} of a node $v_i$ is the number of edges that are incident to the node, and will be denoted by $\deg(v_i)$. 
Each node of degree $1$ is called a \textit{leaf}. The set of all trees over $n$ nodes will be denoted by $\mathbf{T}(n)$.  An undirected graph that consists of only disjoint union of trees is called a \textit{forest}.  The set of all forests over $n$ nodes with exactly $\delta$ trees will be denoted by $\mathbf{F}(n,\delta)$. 
Denote by $F(n,\delta)$ the size of $\mathbf{F}(n,\delta)$.  We sometimes use the notation $\{C_0,C_1,\dots,C_{\delta-1}\} = F \in \mathbf{F}(n,\delta)$ to explicitly denote a forest with $t$   connected components  (or subtrees) of $F$. Note that $\mathbf{F}(n,1) = \mathbf{T}(n)$.

By Cayley's formula~\cite{Aigner} it holds that $|\mathbf{T}(n)| = n^{n-2}$. The proof works by showing a bijection $\cF:\mathbf{T}(n) \rightarrow [n]^{n-2}$, where for every tree $T \in \mathbf{T}(n)$, the \textit{pr\"{u}fer sequence of $T$} is denoted by $\cF(T)=\bfw_T$. An important property is that for each $T=(V_n,E)$, the number of appearances of node $v_i\in V_n$ in  $\bfw_T$ is equal to $\deg(v_i)-1$.

\begin{definition}
	A \textit{code over trees} $\cC_\cT$, denoted by $\ut(n,M)$, is a set of $M$ trees over $n$ nodes. Each tree in the code $\cC_\cT$ is called a \textit{codeword-tree}. The \textit{redundancy} $r$ of the code $\cC_\cT$ is defined by $r= (n-2)\log(n) -\log (M)$\footnote[1]{{The base of all logarithms in the paper is assumed to be $2$.}}.
\end{definition}

{Every codeword-tree corresponds to unique information that is stored, sent, or read, i.e., the information is the structure of the codeword-tree. The storage of information depends mainly on the application that will be used. For example, in binary search trees~\cite{Cormen} the information values and pointers that represent edges are stored in nodes. In tries or suffix trees~\cite{Knuth} the symbols or strings are stored on edges.	In order to deal with \emph{erasures} and \emph{errors} of edges of trees, we initiate the study of codes over trees as will be defined next.}

{
	\begin{definition}
		An \textit{erasure of $\rho$ edges} in a tree $T\in \mathbf{T}(n)$  is the event in which $\rho$ of the edges in $T$ are erased and $T$ is separated into a forest of $\rho+1$ connected components over $n$ nodes. 	
		An \textit{error of $\psi$ edges} in a tree $T\in \mathbf{T}(n)$  is the event in which $\psi$ of the edges in $T$ are replaced with other $\psi$ edges such that we receive a new tree $T'\in \mathbf{T}(n)$. 
	\end{definition}
}

The tree distance for trees is next defined.
\begin{definition}	
	The \textit{tree distance} between two trees $T_1 = (V_n,E_1)$ and $T_2 = (V_n,E_2)$ will be denoted by $d_\cT(T_1,T_2)$ and is defined to be,
	$$ d_\cT(T_1,T_2) = n-1-|E_1\cap E_2|.$$
\end{definition}

It is clear that $d_\cT(T_1,T_2) = |E_1\setminus E_2| = |E_2\setminus E_1|$.
{
	Every tree over $n$ nodes can be represented by a binary vector of length $\binom{n}{2}$ called the \textit{characteristic vector}. Such a vector is indexed by all possible $\binom{n}{2}$ edges that the tree can have and it has ones only in the indices of the tree's edges. Using this representation, the tree distance between any two trees is one half the Hamming distance between their characteristic vectors. Thus, the tree distance is a metric as was mentioned in~\cite{Paulden} and is stated in the next lemma.}
\begin{lemma}
	The tree distance is a metric.
\end{lemma}

The \emph{tree distance} of a code over trees $\cC_{\cT}$ is denoted by $d_\cT(\cC_\cT)$, which is the minimum tree distance between any two distinct trees in $\cC_{\cT}$, that is,
$$d_\cT(\cC_\cT) = \min_{T_1\neq T_2~T_1,T_2\in\cC_\cT} \{d_\cT(T_1,T_2)\}.$$

\begin{definition}\label{def:5}
	A \textit{code over trees $\cC_\cT$ of tree distance $d$},  denoted by $\ut(n,M,d)$, has $M$ trees over $n$ nodes and its tree distance is $d_\cT(\cC_\cT) = d$.
\end{definition}

Since the tree distance is a metric the following theorem holds straightforwardly. 
\begin{theorem}\label{th:dist}
	A  $\ut(n,M)$ code over trees $\cC_\cT$ is of tree distance {at least} $d$ if and only if it can correct any $d-1$ edge erasures and if and only if it can correct any $\lfloor(d-1)/2\rfloor$ edge errors.
\end{theorem}

{Next}, we define the largest size of a code over trees with a prescribed tree distance.
\begin{definition}\label{def:7}
	The \textit{largest size of a code over trees} with tree distance $d$ is denoted by $A(n,d)$. The \textit{minimum redundancy of a code over trees} will be defined by $r(n,d) = (n-2)\log(n) - \log(A(n,d))$.
\end{definition}

A tree  will be called a \textit{star tree} (or a \textit{star} in short) if it has a node $v_i,i \in [n]$ such that $\deg(v_i) = n-1$, and all the other nodes $ v_j, j \in [n]$, $j\neq i$ satisfy $\deg(v_j)=1$. A \textit{path graph} or a {\textit{path tree}} over $n$ nodes is a graph whose nodes can be listed in the order $v_{i_0}, v_{i_1}, \dots, v_{i_{n-1}}$, where $i_0,i_1,\dots,i_{n-1} \in [n]$, such that its edges are $\langle v_{i_j},v_{i_{j+1}}\rangle$ for all $j\in[n-1]$.

\begin{figure}[h!]\label{fig:star-line}
	\hfill
	\subfigure[The star tree.]{\includegraphics[width=30mm]{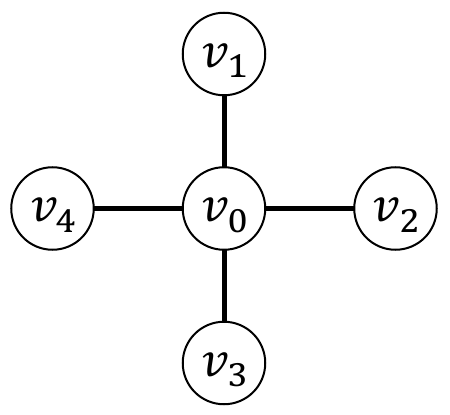}}
	\hfill
	\subfigure[The path tree.]{\includegraphics[width=43.2mm]{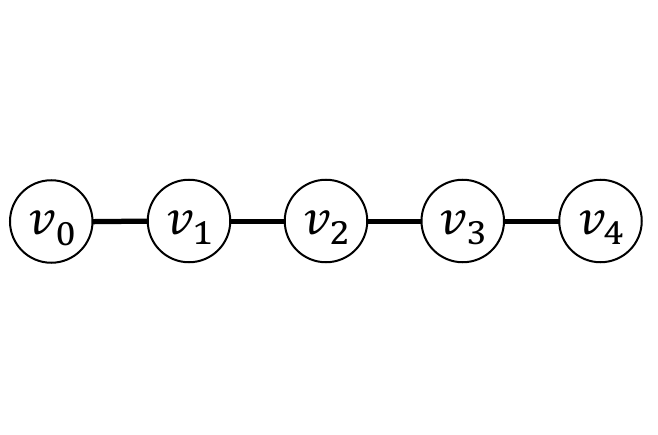}}
	\hfill
	\caption{{For $n=5$ a star and a path trees are presented.}}	
\end{figure}

\begin{definition}\label{def25}
	The \textit{tree ball of a tree of radius $t$ in $\mathbf{T}(n)$ centered at $T\in \mathbf{T}(n)$ } is defined to be
	$$\cB_T(n,t) = \{T' \in\mathbf{T}(n) ~|~ d_{\cT}(T',T) \leq t \}.$$
	The size of the tree ball of trees of $T$, $\cB_T(n,t)$, is denoted by $V_T(n,t)$.
\end{definition}
Note that $V_T(n,t)$ depends on the choice of its center $T$. For example, we will show that if $T$ is a star then $V_T(n,1)= (n-1)(n-2)+1$ and if $T$ is a path tree, then $V_T(n,1) = (n-1)(n-2)(n+3)/6+1$.
If $T$ is a star, path tree the size of $V_T(n,t)$ is denoted by $V^{\star}(n,t), V^{\bline}(n,t)$, respectively. 
We define the \emph{average ball size} {of radius $t$} to be the average value of all tree balls of trees {of radius $t$}, that is,
\begin{align*}
V(n,t) = \frac{\sum_{T\in {\mathbf{T}(n)} } V_T(n,t) } {n^{n-2}}.
\end{align*}

\begin{definition}\label{def:S}
	The \textit{sphere of radius $t\geq 0$ centered at $T\in \mathbf{T}(n)$ } is defined to be
	$$\cS_T(n,t) = \cB_T(n,t)\setminus \cB_T(n,t-1),$$
	where $\cS_T(n,0) = \cB_T(n,0) = \{T\}$, by definition.
	The size of the sphere of radius $t$ is
	equal to the number of all trees in $\cS_T(n,t)$ and is denoted by $S_T(n,t)$. If $T$ is a star, path tree then we denote the sphere $S_T(n,t)$ by $S^{\star}(n,t), S^{\bline}(n,t)$, respectively. 
\end{definition}

{
	For each $T =(V_n, E)\in \mathbf{T}(n)$ and for each $ E' \subseteq E, |E'| =t $, denote the forest $F_{T,E'} = (V_n,E\setminus E')$. Note that $F_{T,E'} \in \mathbf{F}(n,t+1)$. 
	\begin{definition}\label{def9}
		The \textit{forest ball of a tree $T=(V_n,E)$ of radius $t$ in $\mathbf{F}(n,t+1)$} is defined to be
		$$\cP_T(n,t) = \{F_{T,E'}\in \mathbf{F}(n,t+1) ~~|~~ E' \subseteq E,|E'| = t \}.$$
\end{definition}}


{
	Given a tree $T = (V_n,E)$ and an edge-set $E'\in E, |E'|=t$, let $F_{T,E'} = (V_n,E\setminus E') \in \cP_T(n,t)$ be the forest which is also denoted by $F_{T,E'} =  \{ C_0,C_1,\dots,C_{t} \}$, such that $|C_0| \leq |C_1| \leq \cdots \leq |C_{t}|$. The \textit{profile vector of $T$ and $E'$} is denoted by ${\bfP}_T(E') = (|C_0|,|C_1|,\dots,|C_{t}|)$ and the multi-set $P_T(n,t)$ is given by
	\begin{align}\label{P_T}
	P_T(n,t) = \{ {\bfP}_T(E') ~|~ E' \subseteq E, |E'|= t\}.
	\end{align}
	It is can be verified that $|P_T(n,t) | = |\cP_T(n,t)|  = \binom{n-1}{t}$. 
}

\begin{definition}\label{def26}
	The \textit{tree ball of a forest (or the forest's ball in short) of radius $t$ centered at $F\in \mathbf{F}(n,t+1)$ } is defined to be
	$$\cB_F(n,t) = \{T \in\mathbf{T}(n) ~|~ F\in \cP_T(n,t)  \}.$$
	The size of the forest's ball of radius $t$ is equal to the number of all trees in $\cB_F(n,t)$ and is denoted by $V_F(n,t)$.
\end{definition}
Notice that for every two distinct trees $T_1,T_2 \in \cB_F(n,t)$ it holds that $d_{\cT}(T_1,T_2) \leq t$.
Note also that we have three different ball definitions, the forest ball of trees of Definition~\ref{def9} denoted by $\cP_T(n,t)$,  the tree ball of trees of Definition~\ref{def25}, denoted by $\cB_T(n,t)$,  the forest's ball of Definition~\ref{def26}, denoted by $\cB_F(n,t)$.

Furthermore, for the convenience of the reader, relevant notation and terminology referred to throughout the paper is summarized in Table~\ref{tab:cql}.
\begin{table}[!h]
	\caption{Table of Definitions and Notations}\label{tab:cql}
	\begin{center}		\vspace{-4mm}
		\begin{tabular}{ccc}
			\hline
			{Notation} & Meaning & Remarks   \\
			\hline
			\hspace{-2ex}$n$ & \hspace{-2ex}{The number of nodes} & \hspace{-2ex}Sec.~\ref{sec:defs}  \\
			\hspace{-2ex}$\mathbf{T}(n)$ & \hspace{-2ex}The set of all labeled trees over $n$ nodes & \hspace{-2ex}Sec.~\ref{sec:defs}  \\
			\hspace{-2ex}$\mathbf{F}(n,\delta)$ & \hspace{-2ex}The set of all forests with $\delta$ connected components &\hspace{-2ex} Sec.~\ref{sec:defs}	 \\		
			\hspace{-2ex}$F(n,\delta)$ & \hspace{-2ex}The size of $\mathbf{F}(n,\delta)$ & \hspace{-2ex}Sec.~\ref{sec:defs}  \\
			\hspace{-2ex}$d$ & \hspace{-2ex}{The tree distance} &\hspace{-2ex} Def.~\ref{sec:defs}  \\		
			\hspace{-2ex}$\ut(n,M,d)$ & \hspace{-2ex}{A code over trees of size $M$} &\hspace{-2ex} Def.~\ref{def:5}  \\			
			\hspace{-2ex}$A(n,d)$ &\hspace{-2ex} {The largest size of a $\ut(n,M,d)$ code} &\hspace{-2ex} Def.~\ref{def:7}  \\
			\hspace{-2ex}$r(n,d)$ &\hspace{-2ex} {The minimum redundancy of a $\ut(n,M,d)$ code} &\hspace{-2ex} Def.~\ref{def:7}  \\
			\hspace{-2ex}$t$ & \hspace{-2ex}{The radius of a ball} & \hspace{-2ex} Def.~\ref{def25}  \\
			\hspace{-2ex}$\cB_T(n,t)$ & \hspace{-2ex}The tree ball of a tree {of radius $t$} centered at $T$ & \hspace{-2ex}Def.~\ref{def25}  \\
			\hspace{-2ex}$V_T(n,t)$ & \hspace{-2ex}The size of $\cB_T(n,t)$ & \hspace{-2ex}Def.~\ref{def25}  \\
			\hspace{-2ex}$V(n,t)$ & \hspace{-2ex} {The average ball size of radius $t$} & \hspace{-2ex}Def.~\ref{def25}  \\
			\hspace{-2ex}$V^{\star}(n,t))$ & \hspace{-2ex} {The value of $V_T(n,t)$ if $T$ is a star} & \hspace{-2ex} Sec.~\ref{sec:defs}  \\
			\hspace{-2ex}$V^{\bline}(n,t)$ & \hspace{-2ex} {The value of $V_T(n,t)$ if $T$ is a path tree} & \hspace{-2ex} Sec.~\ref{sec:defs}  \\	
			\hspace{-2ex}$\cS_T(n,t)$ & \hspace{-2ex}The sphere of a tree  {of radius $t$}  centered at $T$ & \hspace{-2ex}Def.~\ref{def:S}  \\
			\hspace{-2ex}$S_T(n,t)$ & \hspace{-2ex}The size of $\cS_T(n,t)$ &\hspace{-2ex} Def.~\ref{def:S}  \\
			\hspace{-2ex}$S^{\star}(n,t))$ & \hspace{-2ex} {The value of $S_T(n,t)$ if $T$ is a star} & \hspace{-2ex} Sec.~\ref{sec:defs}  \\
			\hspace{-2ex}$S^{\bline}(n,t)$ & \hspace{-2ex} {The value of $S_T(n,t)$ if $T$ is a path tree} & \hspace{-2ex} Sec.~\ref{sec:defs}  \\
			\hspace{-2ex}$\cP_T(n,t)$ & \hspace{-2ex}The forest ball of a tree  {of radius $t$}  centered at $T$ & \hspace{-2ex} Def.~\ref{def9}  \\
			\hspace{-2ex}$P_T(n,t)$ & \hspace{-2ex}The set of profiles of $\cP_T(n,t)$ & \hspace{-2ex} Def.~\ref{def9}  \\
			\hspace{-2ex}$\cB_F(n,t)$ & \hspace{-2ex}The tree ball of a forest  {of radius $t$}  centered at $F$ & \hspace{-2ex}Def.~\ref{def26}  \\
			\hspace{-2ex}$V_F(n,t)$ &\hspace{-2ex} The size of $\cB_F(n,t)$ & \hspace{-2ex}Def.~\ref{def26}  \\
			
			\hline
		\end{tabular}
	\end{center}
\end{table}

{
	\section{Main Results}\label{sec:main}
}

{
	This section summarizes the main results in the paper. Theorem~\ref{theo:1} states three main upper bounds which will be presented in Section~\ref{ch:bound_trees}. The first bound is a sphere packing bound that will be proved in Theorem~\ref{th:trees_bounds0}. The second, third bound is an improved upper bound in case that $d=n-2,d=n-3$ that will be derived in Theorem~\ref{th:trees1_1},~\ref{th:trees2}, respectively.
	\begin{theorem}\label{theo:1}
		\begin{enumerate}
			\item 	For all $n\geq 1$ and fixed $d$,
			\begin{align*}
			A(n,d)\leq F(n,d)/\binom{n-1}{d-1} = \cO (n^{n-1-d}).
			\end{align*}
			\item 	For all positive integers $n$, $A(n,n-2) \leq n$.
			\item 	For all $n \geq 9$, $A(n,n-3) \leq n^2$.
		\end{enumerate}	
	\end{theorem}
}
{While in Theorem~\ref{theo:1} we obtained upper bounds on $A(n,d)$ using forest balls of trees, in Theorem~\ref{theo:2} we show another approach to obtain both lower and upper bounds on codes over trees using tree balls of trees. For that}, in Section~\ref{sec:ball size} tree balls of trees of radius one are studied and the main results on these balls are summarized in the next theorem.
\begin{theorem}\label{theo:2}
	\begin{enumerate}
		\item 	For any $T \in \mathbf{T}(n)$,
		\begin{align*}
		V^{\star}(n,1) \leq 	V_T(n,1) \leq 	V^{\bline}(n,1).
		\end{align*}
		\item For all $n\geq 1$, $V(n,1) \approx 0.5\sqrt{\frac{\pi} {2} }n^{2.5}=\Theta(n^{2.5})$.
		\item For all $n\geq 1$, $V^{\star}(n,1) = \Theta(n^2)$,  $V^{\bline}(n,1) = \Theta(n^3)$.
	\end{enumerate}
\end{theorem}
{
	The first, second result of Theorem~\ref{theo:2} is proved in Theorem~\ref{theo1},~\ref{cor32}, respectively, while the third is deduced using Lemma~\ref{lemma:21}. 
	The reader can verify that the hardest part of this theorem is to prove, using inductive and recursive arguments, that $ V_T(n,1) \leq 	V^{\bline}(n,1)$. In fact this is shown for arbitrary $t$ in Section~\ref{sec:lines_trees}.
	By using the fact that $V^{\star}(n,1) = \Theta(n^2)$, the upper bound $A(n,3) = \cO(n^{n-4})$ is concluded which is the same upper bound result as the sphere packing bound. Applying the \textit{generalized Gilbert-Varshamov} bound\footnote[1]{{ Let $X$ be a finite set with some distance function $d:X\times X \rightarrow \N$. Assume that the volume of every ball is
			$B_r(x)  =\{y \in X |d(x,y) \leq r\}$.   It was proved in~\cite{Tolhuizen} that if $\overline{\Delta}_r =   \Big( \sum_{x\in X} |B_r(x)|\Big)/|X|$, then the generalized Gilbert-Varshamov bound asserts that there exists a code with minimum distance $r + 1$ and of size at least $|X|/ \overline{\Delta}_r$. }}~\cite{Tolhuizen}, while using the fact that $V(n,1) =\Theta(n^{2.5})$, it is then deduced that  $A(n,2) = \Omega(n^{n-4.5})$. This bound is improved in Section~\ref{tree:const}.
}

{
	In Section~\ref{sec:gen} similar results, summarized in the next theorem, of tree balls of trees with arbitrary radius are shown. 
	\begin{theorem}\label{theo:3}
		For all $T \in \mathbf{T}(n)$ and fixed $t$, it holds that
		\begin{enumerate}
			\item			$V_T(n,t) = \Omega(n^{2t}),~   V_T(n,t) = \cO(n^{3t}).$
			\item 			$V(n,t) = \Theta(n^{2.5t}).	$
			\item   $V^{\star}(n,t) = \Theta(n^{2t})$ and $V^{\bline}(n,t) = \Theta(n^{3t})$.
		\end{enumerate}
	\end{theorem}  
	The first result is shown in Theorem~\ref{theo:40}, the second is deduced in Corollary~\ref{cor:40}, and the third one is shown in Section~\ref{sec:lines_trees} as a result of Theorem~\ref{theo:29} and Theorem~\ref{theo:35}. These results are derived from recursive formulas that calculate the size of the tree balls of trees of radius $t$.}

{
	Again, using the fact that  $V^{\star}(n,t) = \Theta(n^{2t})$, it is deduced that for all $d=2t+1$, $A(n,d) = \cO(n^{n-1-d})$ which matches the upper bound results by the sphere packing bound.
	Applying the generalized Gilbert-Varshamov lower bound and using the fact that $V(n,t) = \Theta(n^{2.5t})$, it is also derived that for $d=t+1$, $A(n,d) = \Omega(n^{n-2-2.5(d-1)})$. This bound is also improved in Section~\ref{tree:const}.
}

{
	In Section~\ref{sec:lines_trees}, we study the sizes of tree balls of trees of stars and path trees for arbitrary radius. 
	{Our main contribution in this section is formulated in recursive formulas for the sizes of tree balls of trees for arbitrary trees. We then show upper and lower bounds on these formulas using the sizes of tree balls of trees of the star and path trees.}
	We present these results in the following theorem.
	\begin{theorem}
		For all $n$ and fixed $t$ let 
		$$P =  n^{t-1}  \binom{n-1}{t}(n-t),~~Q =  n^{t-1} \binom{n+t}{2t+1}.$$ 
		The following properties holds:
		\begin{enumerate}
			\item 		
			\begin{align*}
			& \sum^{t}_{i=0} 	\binom{n-2-t+i}{i} V^{\star}(n,t-i)  = P.
			\end{align*}
			\item 		
			\begin{align*}
			& \sum^{t}_{i=0} 	\binom{n-2-t+i}{i} V^{\bline}(n,t-i) =  Q.
			\end{align*}
			\item 	For all $T \in \mathbf{T}(n)$ 
			\begin{align*}
			P \leq \sum^{t}_{i=0} 	\binom{n-2-t+i}{i} V_T(n,t-i)   \leq Q.
			\end{align*}
		\end{enumerate}
	\end{theorem}
}{
	The first result is deduced from Theorem~\ref{theo35} and is also shown in Equation~\eqref{eq:32}. The second result is due to Theorem~\ref{theo:35} and the last result can be found in the proof of Theorem~\ref{theo45}. 
	The reader will find out that the challenging part of this theorem is to prove that $$ \sum_{i=0}^t\binom{n-2-t+i}{i} V_T(n,t-i)   \leq Q,$$ 
	which is also used in order to prove that $ V_T(n,1) \leq 	V^{\bline}(n,1)$.	
	This section concludes with conjecturing that for fixed $t$ and $n$ large enough, $$V^{\star}(n,t) \leq  V_T(n,t) \leq V^{\bline}(n,t).$$
}

{
	Lastly, in Section~\ref{tree:const} we provide several constructions that improve upon the generalized Gilbert-Varshamov lower bounds. The results of these constructions are summarized in the following  theorem.
	\begin{theorem}
		It holds that
		\begin{enumerate}
			\item There exists an $\ut(n,\lfloor n/2 \rfloor,n-1)$ code.
			\item There exists an $\ut(n,n,n-2)$ code.
			\item   For any positive integer $d \leq n/2$, there exists an $\ut(n,M,d)$ code such that $ M =\Omega( n^{n-2d})$.
			\item 	For fixed $m$ and prime $n$, there exists an $\ut(n,\frac{n-1}{2}\cdot \lfloor \frac{n-1}{m} \rfloor ,\lfloor  \frac{3n}{4} \rfloor - \lceil\frac{3n}{2m}\rceil  - 2 )$ code.
		\end{enumerate}
	\end{theorem}
	The result in a) is proved in Theorem~\ref{th:trees0} using Construction~\ref{const:tree0}, the result in b) is due to Theorem~\ref{th:trees1} and Construction~\ref{const:tree1}, the result in c) holds according to Corollary~\ref{cor:2} and Construction~\ref{const:tree3_1}, and the result in d) follows from Construction~\ref{const:tree2} and is proved in Theorem~\ref{th:tree4}.  
	The result in d) assures that it is possible to construct codes of cardinality $\Omega(n^2)$, while the minimum distance $d$ approaches $\lfloor 3n/4 \rfloor$ and $n$ is a prime number. 
	Comparing to Theorem~\ref{th:trees2} in which it was  shown that $A(n,n-3) = \cO(n^2)$, the result of Theorem~\ref{th:tree4} shows that $A(n,d) = \Omega(n^2)$, when $d$ approaches $\lfloor 3n/4 \rfloor$ and $n$ is prime. Thus, finding the range of values of $d$ for which $A(n,d) = \Theta(n^2)$ is left for future work. 
}

\section{{Upper Bounds on Codes over Trees}}\label{ch:bound_trees}
In this section we show upper bounds for codes over trees. 
Remember that $F(n,\delta)$ is the size of $\mathbf{F}(n,\delta)$, i.e., the number of forests with $n$ nodes and $\delta$   {connected components}. The value of $F(n,\delta)$ was shown in~\cite{Moon70countinglabelled}, to be
\begin{align*}
& F(n,\delta) =   \binom{n}{\delta}n^{n-\delta-1}\sum^{\delta}_{i=0}\left((-\frac{1}{2})^i\binom{\delta}{i}\frac{(\delta+i)(n-\delta)!}{n^i(n-\delta-i)!}\right)
\end{align*}
or another representation of it in~\cite{Bollobas},
$$F(n,\delta) = n^{n-\delta}\left(\sum_{i=0}^{\delta} (-\frac{1}{2})^i \binom{\delta}{i} \binom {n-1}{\delta-1+i}\frac{ (\delta+i)!}{n^i\delta!}\right).$$
The next corollary summarizes some of these known results.
\begin{corollary}\label{cor:1} 
	The following properties hold for all $n$. 
	\begin{enumerate}
		\item $ F(n,1) =n^{n-2}$,
		\item $ F(n,2) =\frac{1}{2}n^{n-4}(n-1)(n+6)$,
		\item $ F(n,3) =\frac{1}{8}n^{n-6}(n-1)(n-2)(n^2+13n+60)$,
		\item $F(n,n-4) =\frac{1}{16}\binom{n}{4}(n^2+3n+10)(n-4)(n+3)$,
		\item $ F(n,n-3) =\frac{1}{2}\binom{n}{4}(n^2+3n+4)$\label{bound_n-4},
		\item $ F(n,n-2) =3\binom{n+1}{4}$\label{bound_n-3},
		\item $ F(n,n-1) =\binom{n}{2}$\label{bound_n-2},
		\item $ F(n,n) = 1.$ 
	\end{enumerate}
\end{corollary}


\subsection{Sphere-Packing Bound}

The following theorem proves the sphere packing bound for codes over trees.
\begin{theorem}\label{th:trees_bounds0}
	For all $n\geq 1$ and $1 \leq d \leq n$, it holds that $A(n,d)\leq F(n,d) /\binom{n-1}{d-1}$.
\end{theorem}

\begin{IEEEproof}
	Let $\cC_\cT$ be a $\ut(n,M,d)$ code such that $n\geq 1$ and $1 \leq d \leq n$. 
	Using Theorem~\ref{th:dist}, it is deduced that {given a codeword-tree $T_1$, each $d-1$ of its edge erasures can be corrected. Thus, every forest $F$ in the forest ball of trees $\cP_{T_1}(n,d-1)$ cannot appear in any other forest ball of trees $\cP_{T_2}(n,d-1)$, for all $T_2 \in \cC_\cT\setminus\{T_1\}$. }
	Thus, for every two distinct codeword-trees $T_1,T_2 \in \cC_\cT$ it holds that 
	$$\cP_{T_1}(n,d-1)\cap \cP_{T_2}(n,d-1) = \emptyset.$$
	As already mentioned, for all $T = (V_n,E)$ it holds that $ |\cP_T(n,d-1)| = \binom{n-1}{d-1}$. Therefore,
	$$ M\cdot \binom{n-1}{d-1} = M\cdot |\cP_T(n,d-1)| \leq F(n,d),$$
	which leads to the fact that  
	$$A(n,d)\leq \frac{F(n,d)}{\binom{n-1}{d-1}}.$$
\end{IEEEproof}

It was also proved in~\cite{Moon70countinglabelled} that for any fixed $\delta$,
\begin{align*}
\lim_{n\rightarrow \infty} \frac{F(n,\delta)}{n^{n-2}} = \frac{1}{2^{\delta- 1}(\delta-1)!},
\end{align*}
which immediately implies the following corollary.
\begin{corollary}\label{cor1}
	For all $n\geq 1$ and fixed $d$, it holds that	
	\begin{align*}
	A(n,d)\leq F(n,d)/\binom{n-1}{d-1} = \cO (n^{n-1-d}),
	\end{align*}
	and thus $r(n,d) = (d-1)\log(n) + \cO(1)$.
\end{corollary}

Notice that by Corollary~\ref{cor:1}\eqref{bound_n-2} it holds that
\begin{align}\label{boundC2_0}
A(n,n-1) \leq \binom{n}{2}/(n-1)= n/2.
\end{align}
In Section~\ref{tree:const} we will show that 
$$A(n,n-1)= \lfloor n/2 \rfloor,$$
by showing a construction of a $\ut(n,\lfloor n/2 \rfloor ,n-1)$ code over trees for all $n\geq 1$.
Similarly, by Corollary~\ref{cor:1}\eqref{bound_n-3},
\begin{align}\label{boundC2_1}
A(n,n-2) \leq 3\binom{n+1}{4}/\binom{n-1}{n-3} = \frac{1}{2}\binom{n+1}{2},
\end{align}
however, we will next show how to improve this bound such that $A(n,n-2)  \leq n$. In Section~\ref{tree:const}, a construction of $\ut(n,n,n-2)$ codes over trees will be shown, leading to $A(n,n-2) = n$.
Finally, by Corollary~\ref{cor:1}\eqref{bound_n-4},
\begin{align}\label{boundC2_2}
\nonumber A(n,n-3) \leq & \frac{1}{2}\binom{n}{4}(n^2+3n+4)/\binom{n-1}{n-4} \\
=& \frac{1}{8}n(n^2+3n+4),
\end{align}
where a better upper bound will be shown in the sequel, which improves this bound to be $A(n,n-3)  \leq  1.5n^2$. Finding a construction for this case is left for future work.

Before we show the improved upper bound for $A(n,n-3)$, a few more definitions are presented. 
{ The girth of a graph is the length of a shortest cycle contained in the graph. If the graph does not contain any cycles (i.e. it is an acyclic graph), its girth is defined to be infinity.
}
For a positive integer $n$, let $E_n$ be the set of all $\binom{n}{2}$ edges as defined in~\eqref{eq:En}.
A graph $G = (U \cup V,\cE)$ is a \textit{bipartite graph} with {two sets of nodes $U$ and $V$ such that} $U \cap V = \emptyset$ and every edge connects a vertex from $U$ to a vertex from $V$, i.e., $\cE \subseteq U\times V$. 
{Reiman's} inequality in~\cite{Neuwirth} and~\cite{Reiman} states that if $|V| \leq  |U|$, then every bipartite graph $G = (U \cup V,\cE)$ {with} girth at least $6$ satisfies
\begin{align}\label{rei_ineq}
|\cE|^2 - |U|\cdot|\cE| - |V|\cdot|U|\cdot(|V|-1) \leq 0.
\end{align}

\subsection{ An Improved Upper Bound for $A(n,n-2)$}
According to Theorem~\ref{th:trees_bounds0},  $A(n,n-2) \leq  \frac{1}{2}\binom{n+1}{2}$ and in the next theorem this bound will be improved to be $A(n,n-2) \leq n$. 
\begin{theorem}\label{th:trees1_1}
	For all positive integers $n$, $A(n,n-2) \leq n$.
\end{theorem}

\begin{IEEEproof}
	Let $\cC_{\cT}$ be a  $\ut(n,M,n-2)$ code. Let $G = (U\cup V, \cE)$ be a bipartite graph such that $ V= \cC_{\cT} , U = E_n $ (defined in (\ref{eq:En})) and $(T,e) \in \cE$ if and only if the tree $T \in \cC_{\cT}$ has the edge $e \in E_n$.
	Clearly, $|V| = M, |U|=\binom{n}{2}$ and $|\cE| = M(n-1)$. 
	
	Since $\cC_{\cT}$ is a  $\ut(n,M,n-2)_{\cD}$ code it holds that for all $T_1 = (V_n,E_1),T_2 = (V_n, E_2) \in \cC_{\cT}$, $|E_1\cap E_2| \leq 1$. That is, there are no  two codeword-trees in $\cC_{\cT}$ that share the same two edges. Hence, there does not exist a cycle of length four in $G$.
	If the girth of $G$ is at least $6$ {(including the case in which $G$ is acyclic by definition of the girth)}, by~\eqref{boundC2_1}, for all $ n \geq 3$, it holds that $ |V| = M \leq \frac{1}{2}\binom{n+1}{2} \leq \binom{n}{2} = |U|$, so the inequality stated in~\eqref{rei_ineq} will be used next.
	Since $|V| = M,|U|=\binom{n}{2}$ and  $|\cE| = M(n-1)$,
	\begin{align*}
	M^2(n-1)^2 - \binom{n}{2}M(n-1) - M\binom{n}{2}(M-1) \leq 0,
	\end{align*}
	or equivalently 
	\begin{align*}
	M(n-1) - \frac{n}{2}(M-1)  \leq   \binom{n}{2},
	\end{align*}
	which is equivalent to
	\begin{align*}
	M(\frac{n}{2}-1)   \leq   \binom{n}{2} - \frac{n}{2},
	\end{align*}
	and since 
	\begin{align*}
	\frac{\binom{n}{2} - \frac{n}{2}}{(\frac{n}{2}-1)} =   \frac{\frac{n}{2}(n-1) - \frac{n}{2}}{(\frac{n}{2}-1)} = n\frac{(\frac{n}{2}-1)}{(\frac{n}{2}-1)} =   n,
	\end{align*}
	we deduce that $M \leq n$.
\end{IEEEproof}	
As mentioned  above, in Section~\ref{tree:const} we will show that $A(n,n-2) = n$.

\subsection{An Improved Upper Bound for $A(n,n-3)$}
We showed in~\eqref{boundC2_2} that $A(n,n-3)\leq \frac{1}{8}n(n^2+3n+4)= \cO(n^3)$. In this section this bound will be improved by proving that $A(n,n-3) \leq n^2$.

Denote by $\cH_n$ the set of forest-sets
\begin{align*}
\cH_n = \left\{ \cF \subseteq \mathbf{F}(n,2)  \middle|
\begin{array}{cc}
\forall F_1 = (V_n, E_1), F_2 = (V_n,E_2) \in \cF,\\ 
|E_1 \cap E_2| \leq 1 
\end{array}
\right\}.
\end{align*}

\begin{example}\label{ex:forest}
	For $n=4$ we partially show an example of the forest-sets in $\cH_4$.
	\begin{figure}[h!]\label{fig:forests_example}
		\centering
		\subfigure{\includegraphics[width=90mm]{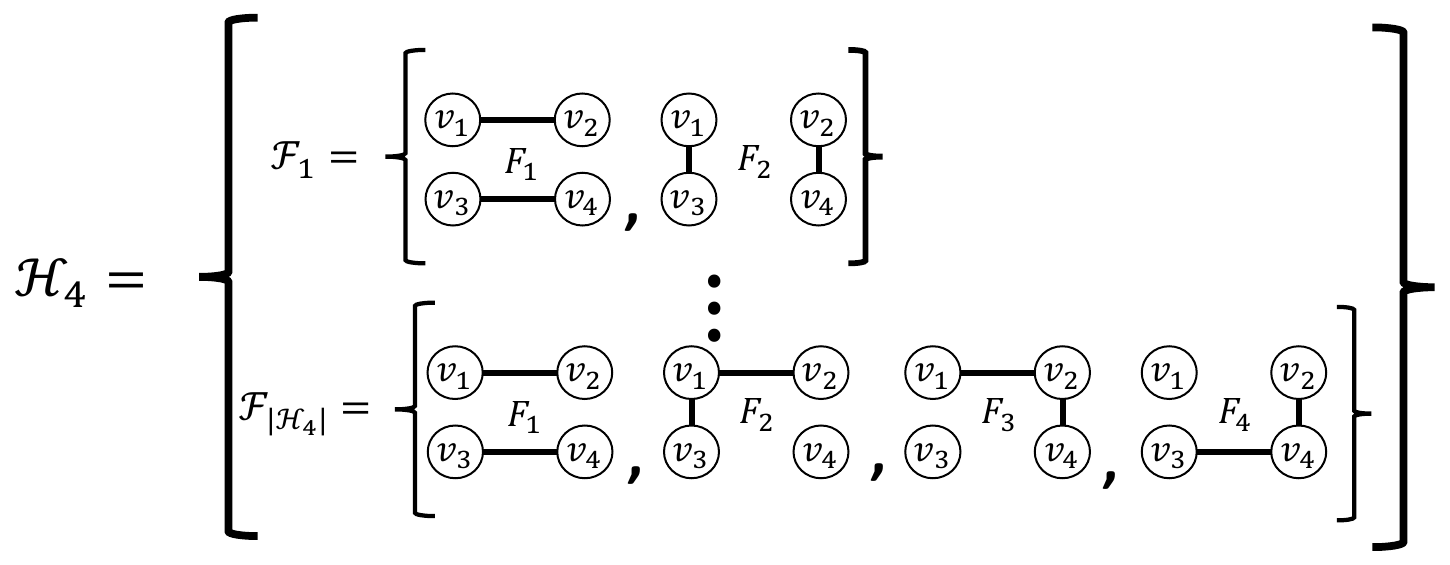}}
		\caption{{An example of the $\cH_4$ set.  Given a forest-set $\cF \in \cH_4$, every two forests $ F_1 = (V_n, E_1) \in \cF,F_2 = (V_n,E_2) \in \cF$ hold $|E_1 \cap E_2| \leq 1$.}}
	\end{figure}
	
\end{example}

We start with showing the following lemma. 
\begin{lemma}\label{lem:forest1}
	For $n \geq 9$ and for all $\cF \in\cH_n$ it holds that $|\cF| \leq 2n$.
\end{lemma}
\begin{IEEEproof}
	Let  $\cF$ be a forest-set in $\cH_n$, and let $G = (U\cup V, \cE)$ be a  bipartite graph such that
	$ V= \cF , U = E_n $ and $(F,e) \in \cE$ if and only if the forest $F \in \cF$ has the edge $e \in E_n$.
	Clearly $|V|=|\cF|,|U|=\binom{n}{2},$ and  $|\cE| = |\cF|(n-2)$. Note that $G$ does not have girth $4$ since for all $F_1 = (V_n,E_1),F_2 = (V_n, E_2) \in \cF$ it holds that $|E_1\cap E_2| \leq 1$. 
	
	Assume that the girth of $G$ is at least $6$.  We consider the following two cases regarding the sizes of the $V$ and $U$. In {the} first case, where $|V|\leq |U|$ we receive the bound stated in the lemma and we will show that the latter case cannot hold. 	
	
	\textbf{Case 1:} Assume that $|V| = |\cF| \leq \binom{n}{2} = |U|$. By~\eqref{rei_ineq}
	\begin{align*}
	|\cF|^2(n-2)^2 - \binom{n}{2}|\cF|(n-2) - |\cF|\binom{n}{2}(|\cF|-1) \leq 0,
	\end{align*}
	or equivalently 
	\begin{align*}
	|\cF|(n-2)^2  - \binom{n}{2}(|\cF|-1) \leq  \binom{n}{2}(n-2),
	\end{align*}
	which is equivalent to
	\begin{align*}
	|\cF|\Big((n-2)^2  - \binom{n}{2}\Big) \leq  \binom{n}{2}(n-2)-\binom{n}{2}.
	\end{align*}
	Next it is deduced that
	\begin{align*}
	|\cF|\Big((n-2)^2  - \binom{n}{2}\Big) \leq  \binom{n}{2}(n-3),
	\end{align*}
	which is equivalent to 
	\begin{align*}
	|\cF| \leq \frac{n^3 -4n^2 +3n}{n^2 - 7n +8},
	\end{align*}
	and therefore $|\cF| \leq 2n$ for all $ n \geq 9$. 
	
	\textbf{Case 2:}	 Assume that $ |V| = |\cF| > \binom{n}{2} = |U|$.
	Again, since the girth is {at least six} we have that 
	\begin{align*}
	|\cF|^2(n-2)^2 - |\cF|^2(n-2) - \binom{n}{2}|\cF|(\binom{n}{2}-1) \leq 0,
	\end{align*}
	or equivalently 
	\begin{align*}
	|\cF|(n-2)^2 - |\cF|(n-2)  \leq \binom{n}{2}(\binom{n}{2}-1),
	\end{align*}
	which is equivalent to
	\begin{align*}
	|\cF|(n-3)\leq \frac{\binom{n}{2}(\binom{n}{2}-1)}{(n-2)}.
	\end{align*}
	Hence for all $n\geq 9$
	\begin{align*}
	|\cF|\leq \frac{n(n^2-1)}{4(n-3)} \leq \binom{n}{2},
	\end{align*}
	which results with a contradiction.
\end{IEEEproof}



Let $\cC_{\cT}$ be a $\ut(n,M,n-3)$ code. For all $e \in E_n$, denote by $c(\cC_{\cT},e)$ the number of codeword-trees of $\cC_{\cT}$ having the edge $e$. 

\begin{lemma}\label{lem:bound_n-4}
	Let  $\cC_{\cT}$ be a $\ut(n,M,n-3)$ code, where $n \geq 9$. Then, for all $e \in E_n$ it holds that $c(\cC_{\cT},e) \leq 2n$.
\end{lemma}
\begin{IEEEproof}
	For $e\in E_n$, denote $k=c(\cC_{\cT},e)$ and let $T_0= (V_n,E_0),T_1= (V_n,E_1),\dots, T_{k-1} = (V_n,E_{k-1}) \in \cC_{\cT}$ be the $k$ codeword-trees such that
	\begin{align}\label{eq:e}
	e \in \bigcap_{i \in [k]}E_i. 
	\end{align}
	Denote by $\cF \subseteq \mathbf{F}(n,2)$ the set of $k$ different forests received by removing the edge $e$ from  $T_0,T_1,\dots,T_{k-1}$.
	Notice that since  $\cC_{\cT}$ is a $\ut(n,M,n-3)$ code it holds that $|E_i \cap E_j| \leq 2, i,j\in [k]$ and by~\eqref{eq:e} we deduce that for all distinct $F_i = (V_n, \cE_i),F_j = (V_n,\cE_j) \in \cF, |\cE_i \cap \cE_j| \leq 1$. By Lemma~\ref{lem:forest1}, for all $n \geq 9, k = |\cF| \leq 2n$ which leads to the fact that $c(\cC_{\cT},e) \leq 2n$.
\end{IEEEproof} 

Lastly, the main result for this section is shown.
\begin{theorem}\label{th:trees2}
	For all $n \geq 9$, $A(n,n-3) \leq n^2$.
\end{theorem}

\begin{IEEEproof}
	Let $n \geq 9$ and let $\mathcal{C}_{\cT}$ be a $\ut(n,M,n-3)$ code over trees. Since for all $e\in E_n$, $c(\cC_{\cT},e)$ is the number of codeword-trees of $\cC_{\cT}$ having the edge $e$, we deduce that $\sum_{e \in E_n}c(\mathcal{C}_{\cT},e) = M(n-1)$. By Lemma~\ref{lem:bound_n-4}, for all $e\in E_n$, $c(\cC_{\cT},e) \leq 2n$. Therefore, 
	\begin{align*}
	M(n-1) = \sum_{e \in E_n}c(\mathcal{C}_{\cT},e) \leq \binom{n}{2}\cdot 2n = n^2(n-1),
	\end{align*} 
	and therefore, $M \leq n^2$.
	
\end{IEEEproof}
Lastly, we verified that for $4\leq n\leq 8$, it holds that $A(n,n-3) \leq 1.5n^2$.

\section{Balls of Trees of Radius One}\label{sec:ball size}

{In previous section} we introduced and studied the forest ball of a tree in order to derive a sphere packing bound on codes over trees with a prescribed minimum tree distance. In this section we study the size behavior of tree balls of trees. These results will also be used to apply the generalized Gilbert Varshamov bound~\cite{Tolhuizen} on codes over trees. We start from some definitions.

Our main goal in this section is to study the size of the radius-one tree ball of trees for all trees. This result is proved in the next lemma.
\begin{lemma}\label{lemma:21}
	For any $T \in \mathbf{T}(n)$ it holds that
	\begin{align}
	V_T(n,1) = \sum_{ (i,n-i) \in P_T(n,1)} \Big(i(n-i)-1 \Big) + 1.
	\end{align}
\end{lemma}

\begin{IEEEproof}
	Let $T = (V_n,E) \in \mathbf{T}(n)$. For any tree $T' = (V_n,E') \in \cB_T(n,1)\setminus \{T\}$, if $e \in E\setminus E'$ and $e' \in E'\setminus E$, then $T'$ is generated uniquely by removing an edge $e$ from $E$, yielding two  {connected components} (subtrees) $\{C_0,C_1\} \in \cP_T(n,1), |C_0|\leq |C_1|$, and adding the edge $e'\neq e$ between $C_0$ and $C_1$. Thus,
	\begin{align*}
	|\cB_T(n,1)\setminus \{T\}| = \sum_{ (|C_0|,|C_1|) \in P_T(n,1)} \Big(|C_0||C_1|-1 \Big).
	\end{align*}
	By denoting $|C_0|=i$ and $|C_1| = n-i$,
	\begin{align*}
	V_T(n,1) = \sum_{ (i,n-i)\in P_T(n,1)} \Big(i(n-i)-1 \Big) + 1.
	\end{align*}
\end{IEEEproof}	
Note that if $T$ is a star, then 
$$P_T(n,1) = \Big\{ \underbrace{ (1,n-1), \dots, (1,n-1)  }_{ \text{$n-1$ times } } \Big\}.$$
Therefore,
\begin{align*}
V^{\star}(n,1) &= \sum_{ (1,n-1) \in P_T(n,1)} \Big(1\cdot(n-1)-1 \Big) + 1 \\
&= (n-1)(n-2)+1.
\end{align*}

If $T$ is a path tree, for odd $n$, 
\begin{align*}
P_T(n,1) =\Big\{(i,n-i),(i,n-i)~|~ 1\leq i \leq  \frac {n-1}{2}   \Big\},
\end{align*}
and for even $n$,
\begin{align*}
& P_T(n,1) =\Big\{(i,n-i),(i,n-i)~|~ 1\leq i \leq  \frac {n-2}{2}   \Big\} \\
& ~~~~~~~~~~~~~~~~~~~~~~~\cup \{(n/2,n/2)\}.
\end{align*}
In both cases,
\begin{align*}
V^{\bline}(n,1) &= \sum^{n-1}_{ i=1} \Big(i\cdot(n-i)-1 \Big) + 1 \\
&= \sum^{n-1}_{ i=1} \Big(i\cdot(n-i )\Big) - (n-1) + 1 \\
&\stackrel{(a)}{=} \binom{n+1}{3} - (n-1) + 1\\
&= (n+1)n(n-1)/6-6(n-1)/6 +1 \\
&= (n-1)(n^2+n-6)/6 +1 \\
&= (n-1)(n-2)(n+3)/6 +1,
\end{align*}
where $(a)$ and its general case is shown in the proof of Theorem~\ref{theo:35}.

Our next goal is to show that for any $T \in \mathbf{T}(n)$ it holds that
\begin{align*}
V^{\star}(n,1) \leq 	V_T(n,1) \leq 	V^{\bline}(n,1).
\end{align*}
The following claim is easily proved.
\begin{claim}\label{claim10}
	Given positive integers $i,n$ such that	$i  \in [n]$, it holds that $n-1 \leq  i(n-i)$.
\end{claim}
Next we state that for all $T\in \mathbf{T}(n)$,
\begin{align}\label{eq13}
\sum_{ (i,n-i) \in P_T(n,1)}i(n-i) \leq \binom{n+1}{3},
\end{align}
while the proof will be shown in the general case in Lemma~\ref{lemm:39} where more than one edge is erased.
\begin{theorem}\label{theo1}
	For any $T \in \mathbf{T}(n)$ it holds that
	\begin{align*}
	V^{\star}(n,1) \leq 	V_T(n,1) \leq 	V^{\bline}(n,1).
	\end{align*}
\end{theorem}
\begin{IEEEproof}
	First we prove the lower bound. 
	For all $T\in \mathbf{T}(n)$
	\begin{align*}
	V_T(n,1) &= \sum_{ (i,n-i) \in P_T(n,1)} \Big(i\cdot(n-i)-1 \Big) + 1  \\
	&\geq  \sum_{ (i,n-i)  \in P_T(n,1)} \Big(1\cdot(n-1)-1 \Big) + 1 \\
	& = (n-1)(n-2)+1 =V^{\star}(n,1) ,
	\end{align*}
	where the inequality holds due to Claim~\ref{claim10}. 	Next, due to~\eqref{eq13},
	\begin{align*}
	V_T(n,1)  
	&=\sum_{ (i,n-i)  \in P_T(n,1) } \Big(i\cdot(n-i)-1 \Big) + 1 \\
	& = \sum_{ (i,n-i) \in P_T(n,1)} \Big(i\cdot(n-i )\Big) - (n-1) + 1 \\       
	&\leq  \binom{n+1}{3}- (n-1) + 1  \\
	& = (n-1)(n-2)(n+3)/6 +1=  V^{\bline}(n,1),
	\end{align*}	
	which leads to the fact that $V_{T}(n,1) \leq V^{\bline}(n,1)$.

\end{IEEEproof}

Our next goal is to show an approximation for the average ball of radius one, that is, the value $V(n,1)$. The first step in this calculation is established in the next lemma, {where its proof can be found in Appendix~\ref{app:0}.}
\begin{lemma}\label{lem29}
	For a positive integer $n$ it holds that 
	\begin{align*}
	\sum_{T\in \mathbf{T}(n) } V_T(n,1) = \sum_{F\in {\mathbf{F}(n,2)}  } (V_F(n,1))^2- 	(n-2)n^{n-2} .
	\end{align*}	
\end{lemma}
{
	In proof of Lemma~\ref{lem29} we use the equality
	\begin{align}\label{eq:20}
	\sum_{T\in \mathbf{T}(n) } \sum_{F\in \cP_T(n,1) } 1& = \sum_{F\in {\mathbf{F}(n,2)} } \sum_{T\in \cB_F(n,1) } 1.
	\end{align}
	which holds by changing the order of summation of all distinct couples of trees and forests.} One can check that~\eqref{eq:20} is true also for $t>1$, and we will use it in {Lemma}~\ref{theo36} which is in the next section. Notice also that from this equality it is deduced that 
\begin{align*}
\sum_{F\in {\mathbf{F}(n,t+1)}  } V_F(n,t)= \binom{n-1}{t}n^{n-2}.
\end{align*}

Now, we are ready to show the following theorem.
\begin{theorem}\label{th:ave_one}
	For all $n$,
	\begin{align*}
	\sum_{T\in \mathbf{T}(n) } V_T(n,1) =\frac{1}{2} n!\sum^{n-2}_{k=0} \frac{n^k} {k!}- 	(n-2)n^{n-2}.
	\end{align*}
\end{theorem}
\begin{IEEEproof}
	It was shown in~\cite{Moon70countinglabelled} that
	$$F(n,2) = \frac{1}{2}\sum^{n-1}_{i=1}\binom{n}{i} i^{i-2} (n-i)^{n-i-2},$$
	where $i$ and $n-i$ represent the sizes of two {connected components} of each forest in $\mathbf{F}(n,2)$.  
	Furthermore, since for all $\{C_0,C_1\} = F\in {\mathbf{F}(n,2)}$, if $|C_0| = i$ then $V_F(n,1) = i(n-i)$, it is deduced that,
	\begin{align*}
	\sum_{F\in {\mathbf{F}(n,2)}  } (V_F(n,1))^2 &= \frac{1}{2}\sum^{n-1}_{i=1}\binom{n}{i} i^{i-2} (n-i)^{n-i-2} [i(n-i)]^2 \\
	& =\frac{1}{2}\sum^{n-1}_{i=1}\binom{n}{i} i^{i} (n-i)^{n-i} \stackrel{(a)}{=}  \frac{1}{2} n!\sum^{n-2}_{k=0} \frac{n^k} {k!},
	\end{align*}
	where $(a)$ holds according to Theorem 5.1 in~\cite{Amdeberhan}.
	Using Lemma~\ref{lem29} it is deduced  that 
	\begin{align*}
	\sum_{T\in \mathbf{T}(n) } V_T(n,1) =\frac{1}{2} n!\sum^{n-2}_{k=0} \frac{n^k} {k!}- 	(n-2)n^{n-2} .
	\end{align*}	
\end{IEEEproof}

For two functions $f(n)$ and $g(n)$ we say that $f(n) \approx g(n)$ if $	\lim_{n\rightarrow \infty}\frac{f(n)}{g(n)} = 1.$ As a direct result of Theorem~\ref{th:ave_one} the next corollary follows.
\begin{corollary}\label{cor32}
	It holds that,
	$$V(n,1) \approx 0.5\sqrt{\frac{\pi} {2} }n^{2.5}.$$
\end{corollary}
\begin{IEEEproof}
	It was shown in~\cite{Flajolet}
	that $$n!\sum^{n-2}_{k=0} \frac{n^k} {k!} \approx \sqrt{\frac{\pi} {2} }n^{n+0.5},$$ and therefore,
	\begin{align*}
	V(n,1) = \frac{\sum_{T\in {\mathbf{T}(n)} } V_T(n,1) } {n^{n-2}} & \approx \frac{1}{2}\sqrt{\frac{\pi} {2} }n^{n+0.5 -(n-2)} & \\
	& = \frac{1}{2}\sqrt{\frac{\pi} {2} }n^{2.5}. & 
	\end{align*}
\end{IEEEproof}

To summarize the results of this section, we proved that for every $T \in \mathbf{T}(n)$ it holds that $V_T(n,1) = \Omega(n^2)$,  $V_T(n,1) = \cO(n^3)$ and the average ball size satisfies $V(n,1) = \Theta(n^{2.5})$. In order to apply the sphere packing bound for the tree balls of trees of radius one, we can only use the lower bound $V_T(n,1) = \Omega(n^2)$ and get that 
$$A(n,3) \leq \frac {n^{n-2}}{\alpha n^2} = \frac{1}{\alpha}n^{n-4},$$
for some constant $\alpha$. This bound is equivalent in its order to the one achieved in Corollary~\ref{cor1}. 

While we could not use the average ball size in applying the sphere packing bound, this can be done for the generalized Gilbert-Varshamov lower bound~\cite{Tolhuizen}. 
{Intuitively, it is done by dividing the number of all trees over $n$ nodes by the average ball size.}
Namely, according to~\cite{Tolhuizen}, the following lower bound on $A(n,2)$ holds
$$A(n,2) = \Omega(n^{n-2-2.5}) = \Omega(n^{n-4.5}).$$ 
{
	The reader can find Construction~\ref{const:tree3_1} in Section~\ref{tree:const} for codes over trees with tree distance $d$ and cardinality $\Omega(n^{n-2d})$. In case that $d=2$ the cardinality is {$$\Omega(n^{n-2d}) = \Omega(n^{n-4}),$$}
	which improves upon the generalized Gilbert-Varshamov lower bound of this case. In the next section, we show similar results of the ball $\cB_T(n,t)$ for general radius $t$.
}
\section{Balls of Trees of Arbitrary Radius}\label{sec:gen}
The main goal of this section is to calculate for each $T \in \mathbf{T}(n)$ the size of its ball $\cB_T(n,t)$ and sphere $\cS_T(n,t)$  for general radius $t$. For that, in Subsection~\ref{subsec:1}, it is first shown how to calculate the forest's ball. Using this result, in Subsection~\ref{subsec:2}, a recursive formula for the tree ball of trees is given and finally in Subsection~\ref{sub-sec:3} we study the average ball size of trees. 

\subsection{The Size of the Forest's Ball}\label{subsec:1} 
In this subsection it is shown how to explicitly find the size of the forest's ball $\cB_F(n,t)$. By using this result, we will be able to proceed to the next step, which is calculating the size of the tree ball of trees $\cB_T(n,t)$. Throughout this section we use the notation $\deg_\cT(v_i)$ for the degree of the node $v_i$ in a tree $\cT$ in order to emphasize over which tree the degree is referred to.
We start with several definitions and claims.

\begin{figure}[h!]
	\centering
	\hfill
	\subfigure[The forest $F$.]{\includegraphics[width=60mm]{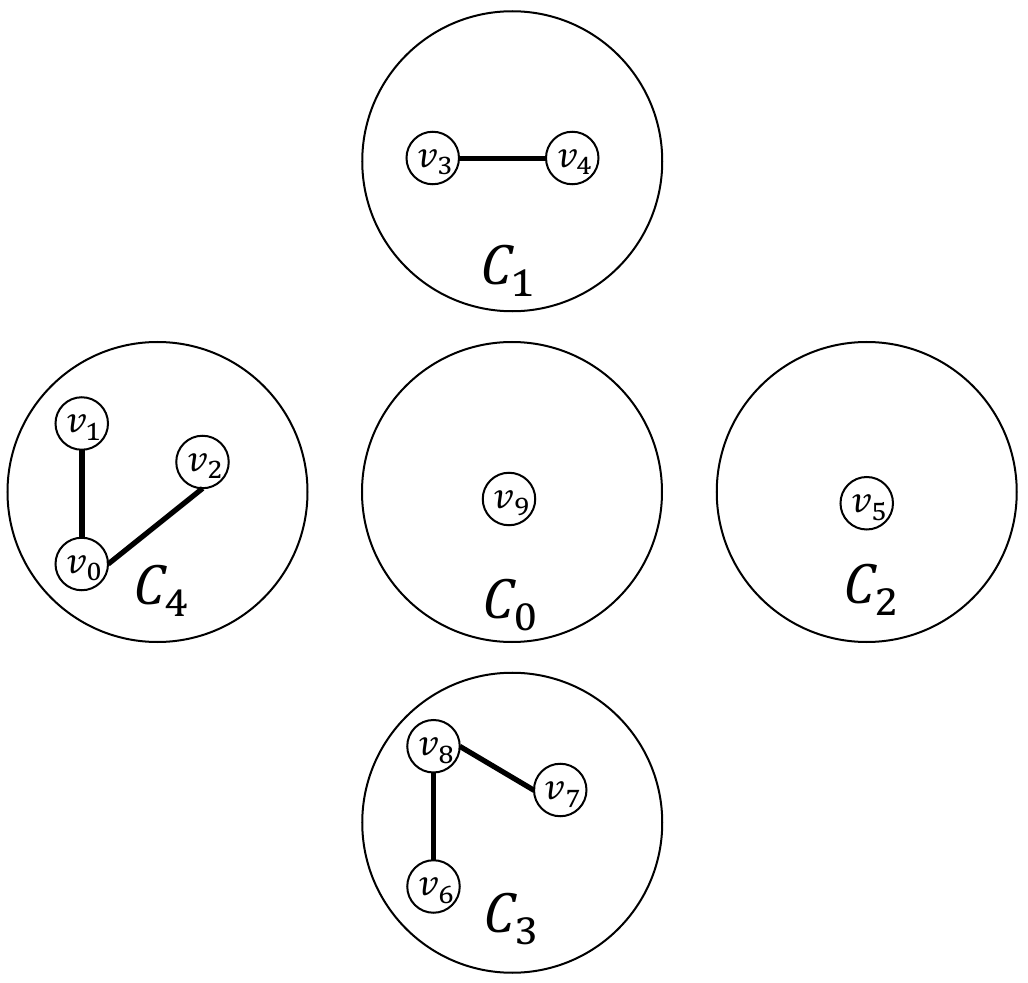}}
	\hfill
	\subfigure[The tree $\cT$.]{\includegraphics[width=25mm]{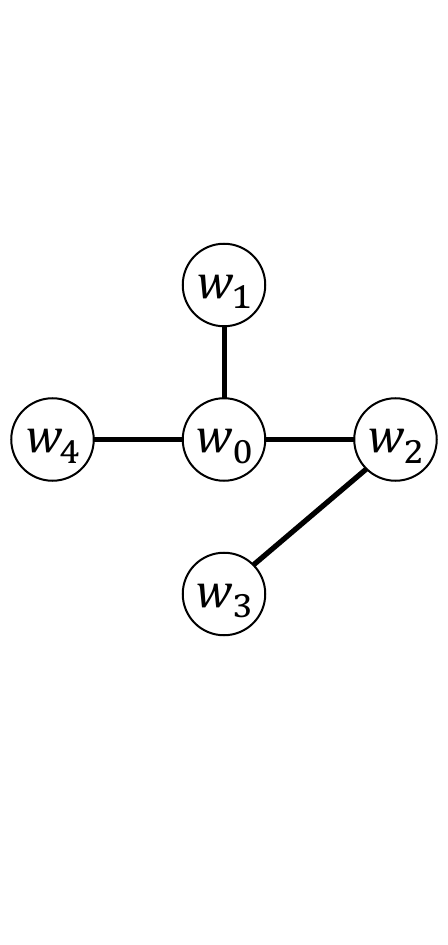}}
	\hfill
	\caption{{For $n=10$ and $t=4$, a forest $F = \{C_0,C_1,C_2,C_3,C_4\} \in \mathbf{F}(10,5)$  over the set of nodes $\{v_i~|~ i \in [10] \}$, and a tree $\cT \in \mathbf{T}(4)$ over the set of nodes $\{w_i~|~ i \in [5]\}$, are presented. Notice that $|C_0|=1,|C_1|= 2,|C_2|=1,|C_3|= 3,|C_4|=3$, and thus, $
			\bfP_1(F,\cT) = |C_0|\cdot|C_1|\cdot|C_0|\cdot|C_2|\cdot|C_0|\cdot|C_4|\cdot|C_2|\cdot|C_3| = 18$.}}	\label{fig:comp}				
\end{figure}

Let $\cT = (V_t,E)\in \mathbf{T}(t)$ be a tree, where  $V_t=\{v_0,v_1,\ldots,v_{t-1}\}$, and let $F = \{C_0,C_1,\dots,C_{t-1}\}\in \mathbf{F}(n,t)$ be a forest. Let $\bfP_1(F,\cT):\mathbf{F}(n,t)\times \mathbf{T}(t)\rightarrow \N$ be the following mapping. For all $F$ and $\cT$, 
\begin{align*}
\bfP_1(F,\cT) =\prod_{ 	\langle v_i,v_j \rangle \in E }  |C_i||C_j|.
\end{align*}
The mapping $\bfP_1$ counts the number of options to complete a forest $F$ with $t$ connected components into a complete tree, according to a specific tree structure $\cT$ with $t$ nodes, corresponding to the $t$ connected components of $F$. Since every $|C_i|$ appears in this multiplication exactly  $\deg_\cT(v_i)$ times ($v_i$ is a node in $\cT$), it is deduced that,
\begin{align}\label{eq:11}
\bfP_1(F,\cT) =\prod_{ 	\langle v_i,v_j \rangle \in E }  |C_i||C_j|= \prod_{ C_i \in F } |C_i|^{\deg_\cT(v_i)}.
\end{align}
Fig.~\ref{fig:comp} demonstrates the mapping $\bfP_1$.

Let $F = \{C_0,C_1,\dots,C_{t-1}\}\in \mathbf{F}(n,t)$ be a forest and let $E_F$ be its edge set. For all $T = (V_n,E_T) \in V_F(n,t)$  we denote its \textit{component edge set} $E_{F,T}$ by 
$$E_{F,T} = E_T\setminus E_F.$$
The component edge set is the set of edges that were added to the forest $F$ in order to receive the tree $T$. We are ready to show the following claim.

\begin{claim}\label{claim:2}
	For all $F\in \mathbf{F}(n,t+1)$ it holds that
	\begin{align*}
	V_F(n,t) = \sum_{ \cT  \in \mathbf{T}(t+1)}\bfP_1(F,\cT).
	\end{align*}
\end{claim}
\begin{IEEEproof}
	Let $F = \{C_0,C_1,\dots,C_t\}$ be a forest.
	Let $H$ be a mapping  $H: V_F(n,t) \rightarrow \mathbf{T}(t+1)	$ that will be defined as follows.  For each $T \in V_F(n,t)$ with a component edge set $E_{F,T}$, it holds that $ H(T) = \cT $ if for all $ e \in E_{F,T}$ such that $e$ connects between $C_k$ and $C_\ell$, the edge $\langle v_k,v_\ell \rangle $ exists in $\cT$.
	Clearly, every $T \in V_F(n,t)$ is mapped and $H$ is well defined. Moreover, for any $\cT = (V_{t+1},E)\in \mathbf{T}(t+1)$, $H$ maps exactly
	\begin{align*}
	\prod_{ 	\langle v_i,v_j \rangle \in E }  |C_i||C_j|.
	\end{align*}	
	trees from $V_F(n,t)$ into $\cT$, which is exactly the value of $\bfP_1(F,\cT)$. Thus,
	\begin{align*}
	V_F(n,t) = \sum_{ \cT  \in \mathbf{T}(t+1)}\bfP_1(F,\cT).
	\end{align*}	
	
\end{IEEEproof}

Next, another mapping $\bfP_2(F,\cT):\mathbf{F}(n,t)\times \mathbf{T}(t)\rightarrow \N$ is defined. 
For every forest $F = \{C_0,C_1,\dots,C_{t-1}\}\in \mathbf{F}(n,t)$ and a tree $\cT\in \mathbf{T}(t)$ with a pr\"{u}fer sequence $$ \bfw_{\cT} =(i_0,i_1,\dots,i_{t-3}) \in [t]^{t-2},$$ 
we let 
\begin{align*}
\bfP_2(F,{\cT}) =|C_{i_0}|\cdot|C_{i_1}|\cdots|C_{i_{t-3}}|.
\end{align*}
Using the fact that each number $i$ of node $v_i$ appears in the pr\"{u}fer sequence $\bfw_{\cT}$ of $\cT$ exactly $\deg_\cT(v_i)-1$ times, we deduce that 
\begin{align}\label{eq:12}
\bfP_2(F,{\cT}) = \prod_{ C_i \in F } |C_i|^{\deg_\cT(v_i)-1}.
\end{align}

Let $g_F(x)$ be the generating function of $F$, defined by
\begin{align*}
g_F(x) = \sum^{t-1}_{i=0} x^{|C_i|}.
\end{align*}
This generating function will be used in the proof of the following claim.
\begin{claim}\label{claim:3}
	Let $F$ be a forest in $ \mathbf{F}(n,t+1)$. Then,
	\begin{align*}
	\sum_{ \cT \in \mathbf{T}(t+1) } \bfP_2(F,\cT)= \Big (\sum_{  C_i \in F} |C_i|  \Big)^{t-1} = n^{t-1}.
	\end{align*}
\end{claim}
\begin{IEEEproof}
	Let $F\in \mathbf{F}(n,t+1)$ be a forest and let $g_F(x)$ be its generating function.
	Let $G(x) = (g_F(x))^{t-1}$ and we deduce that
	\begin{align*}
	G(x) & = (g_F(x))^{t-1} = \Big( \sum^{t}_{i=0} x^{|C_i|}\Big)^{t-1} & \\
	&= \sum_{(i_0,i_1,\dots,i_{t-2} )\in [t+1]^{t-1}} x^{|C_{i_0}|\cdot|C_{i_1}|\cdots|C_{i_{t-2}}|}.& 
	\end{align*}
	Since each monomial of $G(x)$ is of the from $x^{|C_{i_0}|\cdot|C_{i_1}|\cdots|C_{i_{t-2}}|}$ for $(i_0,i_1,\dots,i_{t-2}) \in [t+1]^{t-1}$, it holds that the sum of all the powers of $x$ in $G(x)$ is 
	$$\sum_{(i_0,i_1,\dots,i_{t-2} )\in [t+1]^{t-1}} {|C_{i_0}|\cdot|C_{i_1}|\cdots|C_{i_{t-2}}|},$$
	which is equal to the sum 
	$$ \Big (\sum_{  C_i \in F} |C_i|  \Big)^{t-1}.$$ 
	Furthermore, each vector $ (i_0,i_1,\dots,i_{t-2}) \in [t+1]^{t-1} $ is a pr\"{u}fer sequence $ \bfw_{\cT}$ of some $\cT \in \mathbf{T}(t+1)$.
	Thus we deduce that 
	\begin{align*}
	G(x) =  \sum_{\cT \in \mathbf{T}(t+1)} x^{\bfP_2(F,{\cT})},
	\end{align*}
	and  the powers sum of $x$ is exactly 
	$$\sum_{ \cT \in \mathbf{T}(t+1) } \bfP_2(F,\cT).$$
	Therefore,
	$$	\sum_{ \cT \in \mathbf{T}(t+1) } \bfP_2(F,\cT)= \Big (\sum_{  C_i \in F} |C_i|  \Big)^{t-1}.$$
	Lastly, since  $\sum_{  C_i \in F} |C_i| = n$, it holds that 	
	$$\Big (\sum_{  C_i \in F} |C_i|  \Big)^{t-1} = n^{t-1},$$ 
	which concludes the proof.
\end{IEEEproof}

According to the last two claims, the next corollary is derived and provides an explicit expression to calculate the forest's ball size.
\begin{corollary}\label{cor:3}
	For any $\{C_0,C_1,\dots,C_t\} = F \in \mathbf{F}(n,t+1)$ it holds that
	\begin{align*}
	V_F(n,t) = n^{t-1} \prod_{ C_i \in F } |C_i|.
	\end{align*}	
\end{corollary}
\begin{IEEEproof}
	The proof will hold by the following sequence of equations, that will be explained below,
	\begin{align*}
	V_F(n,t) & \stackrel{(a)}{=} \sum_{ \cT  \in \mathbf{T}(t+1)}\bfP_1(F,\cT) \\
	& \stackrel{(b)}{=}\ \sum_{ \cT  \in \mathbf{T}(t+1)} \prod_{ C_i \in F } |C_i|^{\deg_\cT(v_i)} \\
	& \stackrel{(c)}{=}  \prod_{ C_i \in F } |C_i| \sum_{ \cT  \in \mathbf{T}(t+1)} \prod_{ C_i \in F } |C_i|^{\deg_\cT(v_i)-1} \\
	& \stackrel{(d)}{=} \prod_{ C_i \in F } |C_i| \sum_{ \cT  \in \mathbf{T}(t+1)} \bfP_2(F,\cT)  \\
	& \stackrel{(e)}{=}  \prod_{ C_i \in F } |C_i| \Big (\sum_{  C_i \in F} |C_i|  \Big)^{t-1} \\	
	& \stackrel{(f)}{=}n^{t-1} \prod_{ C_i \in F } |C_i| .
	\end{align*}
	Equality $(a)$ holds by Claim~\ref{claim:2}. 
	Equality $(b)$ holds due to~\eqref{eq:11}. 
	Equality $(c)$ is a result of taking the common factor $\prod_{ C_i \in F } |C_i|$ from the summation. Note also that for all $i \in [t+1]$, $\deg_\cT(v_i) > 0$.  
	Equality $(d)$ holds due to~\eqref{eq:12}. 
	Equality $(e)$ holds by Claim~\ref{claim:3}.  
	Equality $(f)$ holds since $(|C_0|+|C_1|+\dots+|C_t|) = n$. 	
\end{IEEEproof}

\subsection{The Size of the Tree Ball of Trees}\label{subsec:2}
In this subsection we present a recursive formula for the tree ball of trees $\cB_T(n,t)$ and its sphere $\cS_T(n,t)$, as well as asymptotic bounds on their sizes. First, according to Corollary~\ref{cor:3}, we immediately get the following corollary.
\begin{corollary}\label{lem10}
	For all $T \in \mathbf{T}(n)$ it holds that
	\begin{align*}
	\sum_{F\in \cP_T(n,t) } V_F(n,t) = n^{t-1} \sum_{(i_0,i_1,\dots,i_{t}) \in P_T(n,t)} i_0i_1\cdots i_t.
	\end{align*}
\end{corollary}

Next, a recursive connection between the sizes of forest's balls and spheres (of trees) is shown. 
\begin{lemma}\label{lemma:34}
	For all $T \in \mathbf{T}(n)$ it holds that
	\begin{align}\label{eq20}
	\sum_{F\in \cP_T(n,t) } V_F(n,t)  = \sum^{t}_{i=0} 	\binom{n-1-t+i}{i} S_T(n,t-i).
	\end{align}
\end{lemma}
\begin{IEEEproof}
	Let $T = (V,E)\in \mathbf{T}(n)$. First notice that for all $0 \leq i \leq t$, 
	$$\bigcup^{t}_{i=0} S_T(n,i) = \bigcup_{F\in \cP_T(n,t) }  V_F(n,t).$$
	Therefore, our main goal in this proof is finding, for a given tree $T_i=(V,E_i) \in S_T(n,i)$, the number of forests in $\cP_T(n,t)$ in which the tree belongs to their ball of trees. This number equals to the size of the intersection $\cP_T(n,t) \cap \cP_{T_i}(n,t)$ since all of these forests belong also to $\cP_{T_i}(n,t)$. Thus, every forest $F\in\cP_T(n,t) \cap \cP_{T_i}(n,t)$  is received in two steps. First, remove from $T_i$ the $t-i$ edges in $E_i \setminus E$. Then, $i$ more edges from $E\cap E_i$ are chosen, where  $|E\cap E_i| = n-1-(t-i)$. 
	Note that indeed every forest in $\cP_T(n,t) \cap \cP_{T_i}(n,t)$ is generated by this procedure. Thus,
	\begin{align*}
	|\cP_T(n,t) \cap \cP_{T_i}(n,t)| = \binom{n-1-(t-i)}{i}.
	\end{align*}
	Therefore, in~\eqref{eq20} each tree $T_i \in  S_T(n,t-i) $ belongs to the forest's balls of $\binom{n-1-(t-i)}{i}$ different forests in $ \cP_T(n,t)$. Since it is true for all $0 \leq i \leq t$ we conclude the lemma's statement. 
\end{IEEEproof}

Combining Corollary~\ref{lem10} and Lemma~\ref{lemma:34}, a recursive formula for the size of a sphere is presented.
\begin{corollary}\label{cor:10}
	For any $T \in \mathbf{T}(n)$ it holds that
	\begin{align*}
	& \sum^{t}_{i=0} 	\binom{n-1-t+i}{i} S_T(n,t-i) = n^{t-1} \hspace{-3ex}\sum_{(i_0,i_1,\dots,i_{t}) \in P_T(n,t)} i_0i_1\cdots i_t.
	\end{align*}	
\end{corollary}

Using Corollary~\ref{cor:10}, a recursive formula for the tree ball of trees is immediately deduced, {see Appendix~\ref{app:1.1}.}
\begin{theorem}\label{theo35}
	For any $T \in \mathbf{T}(n)$ it holds that
	\begin{align*}
	& \sum^{t}_{i=0} 	\binom{n-2-t+i}{i} V_T(n,t-i)  =n^{t-1} \hspace{-3ex}\sum_{(i_0,i_1,\dots,i_{t}) \in P_T(n,t)} i_0i_1\cdots i_t.
	\end{align*}
\end{theorem}

{The proof of the following lemma can be found in Appendix~\ref{app:1.2}, and it is the last step before presenting the main result of this section.}
\begin{lemma}\label{lemma:30}
	For any positive integer $\alpha$, if
	\begin{align*}
	& \sum^{t}_{i=0} 	\binom{n-2-t+i}{i} V_T(n,t-i)  = \Omega(n^{\alpha t}),
	\end{align*}	
	and $V_T(n,0) = 1$, then  $V_T(n,t) =  \Omega(n^{\alpha t})$.
\end{lemma}

Finally, the main result of this section is shown.
\begin{theorem}\label{theo:40}
	For all $T \in \mathbf{T}(n)$ and fixed $t$, it holds that
	\begin{align*}
	V_T(n,t) = \Omega(n^{2t}),~ & V_T(n,t) = \cO(n^{3t}).
	\end{align*}
\end{theorem}
\begin{IEEEproof}
	First we will prove that $	V_T(n,t) = \Omega(n^{2t})$. 
	Given positive integers $i_0,i_1,\dots,i_{t-1},i_t,n$ such that
	$i_0+i_1+\dots+i_{t-1}+i_t=n,$  
	it holds that
	\begin{equation}\label{eq:ineq}
	n-t \stackrel{(a)}{\leq}  i_0i_1\cdots i_t \stackrel{(b)}{\leq} \Big(\frac{n}{t+1} \Big)^{t+1},
	\end{equation}
	where $(a)$ is well known and $(b)$ holds by using the arithmetic-geometric mean inequality. 
	Thus, for all $T\in \mathbf{T}(n)$
	\begin{align*}
	\sum^{t}_{i=0} 	\binom{n-2-t+i}{i} &V_T(n,t-i) \\
	&  \stackrel{(a)}{=}n^{t-1} \sum_{(i_0,i_1,\dots,i_{t}) \in P_T(n,t)} i_0i_1\cdots i_t \\
	& \stackrel{(b)}{\geq } n^{t-1}  \binom{n-1}{t}(n-t) = \Omega(n^{2t}),
	\end{align*}
	and 
	\begin{align*}
	\sum^{t}_{i=0} 	\binom{n-2-t+i}{i} & V_T(n,t-i)  \\
	&  \stackrel{(a)}{=}n^{t-1} \sum_{(i_0,i_1,\dots,i_{t}) \in P_T(n,t)} i_0i_1\cdots i_t \\
	& \stackrel{(b)}{\leq } n^{t-1} \binom{n-1}{t} \Big(\frac{n}{t+1}\Big)^{t+1} =\cO(n^{3t}),	
	\end{align*}
	where in both cases 	$(a)$ holds by Theorem~\ref{theo35} and inequality $(b)$ holds according to~(\ref{eq:ineq}).
	Therefore, it immediately deduced that $V_T(n,t) =  \cO(n^{3t})$. The result $V_T(n,t) =  \Omega(n^{2t})$ is deduced according to Lemma~\ref{lemma:30}.
\end{IEEEproof}

\subsection{The Average Ball Size}\label{sub-sec:3}
In this section we study the asymptotic behavior of the average ball size (of trees). First, using Theorem~\ref{theo35} and Lemma~\ref{lemma:34} we deduce that for all $T \in \mathbf{T}(n)$
\begin{align}\label{eq21}
\sum_{F\in \cP_T(n,t) } V_F(n,t)  = \sum^{t}_{i=0} 	\binom{n-2-t+i}{i} V_T(n,t-i).
\end{align}
The following recursive relation on the average ball size is presented. 
\begin{lemma}\label{theo36}
	For all $n$ and $t$, it holds that 
	\begin{align*}
	& \sum^{t}_{i=0} 	\binom{n-2-t+i}{i} V(n,t-i)  \\
	&=  n^{2t-n} \frac{1}{(t+1)!}\sum_{ \substack{ 1\leq  i_0, i_1 , \dots ,i_{t} \leq n \\ i_0+i_1+ \dots +i_{t} = n }}\binom{n}{i_0,i_1,\dots,i_t} i_0^{i_0}i_1^{i_1}\cdots i_t^{i_t}.
	\end{align*}
\end{lemma}
\begin{IEEEproof}
	The following holds,
	\begin{align*}
	\sum^{t}_{i=0} & 	\binom{n-2-t+i}{i} \sum_{T\in \mathbf{T}(n)} V_T(n,t-i)  \\
	&   \stackrel{(a)}{=} \sum_{T\in \mathbf{T}(n)}\sum^{t}_{i=0} 	\binom{n-2-t+i}{i} V_T(n,t-i) \\
	&  \stackrel{(b)}{=}  \sum_{T\in \mathbf{T}(n)}\sum_{F\in \cP_T(n,t) } V_F(n,t)  \stackrel{(c)}{=}  \sum_{F\in \mathbf{F}(n,t+1)}\sum_{T\in \cB_F(n,t) } V_F(n,t) \\
	&  =  \sum_{F\in \mathbf{F}(n,t+1)} (V_F(n,t))^2  \stackrel{(d)}{=}\sum_{F\in \mathbf{F}(n,t+1)} (n^{t-1} \prod_{ C_i \in F } |C_i|)^2  \\
	& \stackrel{(e)}{=} \frac{n^{2t-2}}{(t+1)!}\sum_{ \substack{ 0< i_0 , \dots ,i_{t} < n \\ i_0+ \dots +i_{t} = n }}\binom{n}{i_0,\dots,i_t} i_0^{i_0-2}\cdots i_t^{i_t-2} (i_0\cdots i_t) ^2 \\
	& =  n^{2t-2} \frac{1}{(t+1)!}\sum_{ \substack{ 0 <  i_0, i_1 , \dots ,i_{t} <n \\ i_0+i_1+ \dots +i_{t} = n }}\binom{n}{i_0,i_1,\dots,i_t} i_0^{i_0}i_1^{i_1}\cdots i_t^{i_t}.
	\end{align*}
	Equality $(a)$ holds by changing the summation order.
	Equality $(b)$ holds due to~\eqref{eq21} and Theorem~\ref{theo35}.
	Equality $(c)$ holds by changing the summation order of trees and forests as it was done in~\eqref{eq:20}.
	Equality $(d)$ holds by Corollary~\ref{cor:3}.
	We deduce equality $(e)$ as follows. It was shown in~\cite{Moon70countinglabelled} that
	\begin{align*}
	F(n,t+1) =  \frac{1}{(t+1)!}\sum_{ \substack{ 0< i_0 , \dots ,i_{t} < n \\ i_0+ \dots +i_{t} = n }}\binom{n}{i_0,\dots,i_t} i_0^{i_0-2}\dots i_t^{i_t-2}.
	\end{align*}
	For each $F \in \mathbf{F}(n,t+1)$ we denote $|C_j| = i_j$, $j\in [t+1]$. Thus,
	$$  (\prod_{ C_i \in F } |C_i|)^2 = (i_0\cdots i_t) ^2,$$
	which verifies the equality in step $(e)$.
	After dividing the last expression in the series of equations by $n^{n-2}$, the proof is concluded. 
\end{IEEEproof}	

Next, we seek to show the main result of  this section, that is, the asymptotic size of the average ball. For that, we first show the following claim, {where} its proof is shown in Appendix~\ref{app:2}.
\begin{claim}\label{claim:4}
	For a positive integer $n$ and a fixed $t$ it holds that
	\begin{align*}
	& \sum^{n-1}_{i=1}\binom{n}{i} i^{i} (n-i)^{n-i} \Theta (i^{t/2})  =  \Theta(n^{t/2})\sum^{n-1}_{i=1}\binom{n}{i} i^{i} (n-i)^{n-i}.
	\end{align*}
\end{claim}
The following lemma is now presented.
\begin{lemma}\label{lem:avg}
	\begin{align*}
	\sum_{ \substack{ 0 <  i_0, i_1 , \dots ,i_{t} <n \\ i_0+i_1+ \dots +i_{t} = n }}\binom{n}{i_0,i_1,\dots,i_t} i_0^{i_0}i_1^{i_1}\dots i_t^{i_t}= \Theta(n^{n+t/2}).
	\end{align*}
\end{lemma}
\begin{IEEEproof}
	Consider the sequence of integeres $ 1^1,2^2,3^3,\dots$, that is $a_n = n^n$, for $n\geq 1$. Let $G(x)$ be its generating function, i.e. 
	$$G(x) = \sum^{\infty}_{n=1} a_n \frac{x^n}{n!} = \sum^{\infty}_{n=1} n^n \frac{x^n}{n!}.$$
	Denote by $F_t(x)$ the function $F_t(x) = (G(x))^{t+1}$. Thus,
	\begin{align*}
	& F_t(x) = \Big(\sum^{\infty}_{i_0=1} i_0^{i_0} \frac{x^{i_0}}{i_0!} \Big) \dots \Big(\sum^{\infty}_{i_t=1} i_t^{i_t} \frac{x^{i_t}}{i_t!} \Big) \\
	&  = \sum^{\infty}_{n=1} \Big(\sum_{ \substack{0 <  i_0, i_1 , \dots ,i_{t} <n\\ i_0+i_1+ \dots +i_{t} = n }}\binom{n}{i_0,i_1,\dots,i_t} i_0^{i_0}i_1^{i_1}\dots i_t^{i_t} \Big)\frac{x^n}{n!},
	\end{align*} 
	and the coefficient of $x^n/n!$ in $F_t(x)$ is exactly 
	$$ \sum_{ \substack{0 <  i_0, i_1 , \dots ,i_{t} <n \\ i_0+i_1+ \dots +i_{t} = n }}\binom{n}{i_0,i_1,\dots,i_t} i_0^{i_0}i_1^{i_1}\dots i_t^{i_t}.$$
	Next, it is shown by induction on $t$ that the order of the coefficient of $x^n/n!$ in $F_t(x)$ is $\Theta(n^{n+t/2})$.\\
	\textbf{Base:}	 Clearly the coefficient of $x^n/n!$ in $F_0(x)$ is $n^n$, since $F_0(x) = G(x)$.\\
	\textbf{{Inductive} Step:} Assume that the coefficient of $x^n/n!$ in $F_t(x)$ is $\Theta(n^{n+t/2})$. Thus, the coefficient of $x^n/n!$ in $F_{t+1}(x)$ is exactly
	\begin{align*}
	F_{t+1}&(x)  = F_t(x)F_0(x) \\
	&  = \Big( \sum^{\infty}_{i=1} \Big(\sum_{ \substack{ 0 <  i_0, i_1 , \dots ,i_{t} < i \\ i_0+i_1+ \dots +i_{t} = i }}\binom{i}{i_0,i_1,\dots,i_t} i_0^{i_0}i_1^{i_1}\dots i_t^{i_t} \Big)\frac{x^i}{i!} \Big) \\
	&\cdot\Big( \sum^{\infty}_{j=1} j^{j} \frac{x^j}{j!}\Big) \stackrel{(a)}{=} \Big( \sum^{\infty}_{i=1} \Theta (i^{i+t/2}) \frac{x^i}{i!}\Big) \cdot\Big(\sum^{\infty}_{j=1} j^{j} \frac{x^j}{j!}\Big) \\
	&\stackrel{(b)}{=} \sum^{n-1}_{i=1}\binom{n}{i} i^{i} (n-i)^{n-i} \Theta (i^{t/2})  \frac{x^n}{n!}  \\
	& \stackrel{(c)}{=}  \Theta(n^{t/2})\sum^{n-1}_{i=1}\binom{n}{i} i^{i} (n-i)^{n-i}  \frac{x^n}{n!} \\
	& \stackrel{(d)}{=}\Theta(n^{t/2})\sum^{n-1}_{i=1} \Theta(n^{n+0.5})  \frac{x^n}{n!}= \sum^{n-1}_{i=1}\Theta(n^{n+(t+1)/2}) \frac{x^n}{n!},\\
	\end{align*}
	where equality $(a)$ holds by the induction assumption,
	and equality $(b)$ holds by denoting $i+j = n$.  
	Equality $(c)$ holds by Claim~\ref{claim:4} and equality $(d)$ holds due to Corollary~\ref{cor32}, where we showed that the coefficient of $x^n/n!$ in $F_1(x)$ is $\Theta(n^{n+0.5})$.
	
\end{IEEEproof}

We are now ready to find the asymptotic size of the average ball.
\begin{corollary}\label{cor:40}
	It holds that
	\begin{align*}
	V(n,t) = \Theta(n^{2.5t}).
	\end{align*}
\end{corollary}
\begin{IEEEproof}
	It holds that
	\begin{align*}
	\sum^{t}_{i=0} & \binom{n-2-t+i}{i} V(n,t-i)  \\
	&\stackrel{(a)}{=}  n^{2t-n} \frac{1}{(t+1)!}\sum_{ \substack{ 1\leq  i_0, i_1 , \dots ,i_{t} \leq n \\ i_0+i_1+ \dots +i_{t} = n }}\binom{n}{i_0,i_1,\dots,i_t} i_0^{i_0}i_1^{i_1}\cdots i_t^{i_t} \\
	& \stackrel{(b)}{=} \Theta(n^{2t-n})\Theta(n^{n+t/2}) = \Theta(n^{2.5t}),
	\end{align*}
	where $(a)$ holds by {Lemma}~\ref{theo36} and $(b)$ holds using Lemma~\ref{lem:avg}. Therefore it is deduced that $ V(n,t) = \cO(n^{2.5t})$.  The result $ V(n,t) = \Omega(n^{2.5t})$ is proved according to Lemma~\ref{lemma:30}. 
\end{IEEEproof}

In summary, we proved that for every $T \in \mathbf{T}(n)$ and fixed $t$ it holds that $V_T(n,t) = \Omega(n^{2t})$,  $V_T(n,t) = \cO(n^{3t})$ and the average ball size satisfies $V(n,t) = \Theta(n^{2.5t})$. The sphere packing bound for smallest tree ball of trees size of radius $t$ for $\ut(n,M,d=2t+1)$ codes over trees in this case shows that
$$A(n,d) \leq \frac {n^{n-2}}{\alpha n^{2t}} = \frac{1}{\alpha}n^{n-2-2t}= \frac{1}{\alpha}n^{n-1-d},$$
for some constant $\alpha$. Thus, we derive a similar result as in Corollary~\ref{cor1}.

By using the generalized Gilbert-Varshamov lower bound for the average ball size~\cite{Tolhuizen} for $\ut(n,M,d=t+1)$ codes over trees, we get,
$$A(n,d) = \Omega(n^{n-2-2.5(d-1)}) = \Omega(n^{n+0.5-2.5d}).$$ 
However, in Section~\ref{tree:const}, based upon Construction~\ref{const:tree3_1}, we will get that
$$A(n,d) =  \Omega(n^{n-2d}).$$

In the next section similar results are shown for stars and path trees. While the exact size of the tree balls of trees is found for stars, for path trees we only find its asymptotic behavior and finding its exact expression is left for future work. It is also shown that for a fixed $t$ the star tree has asymptoticly the smallest size of the tree of ball of trees, while the path tree achieves asymptoticly the largest size.

\section{The Tree Balls of Trees for Stars and Path Trees}\label{sec:lines_trees}
Several more interesting results on the size of the tree balls of trees and more specifically for stars and path trees are shown in this section. First we show an exact formula for $V^{\star}(n,t)$ and conclude that $V^{\star}(n,t) = \Theta(n^{2t})$. Then we simplify the recursive formula in Theorem~\ref{theo35} for path trees and we will show that $V^{\bline}(n,t) = \Theta(n^{3t})$. Finally, we will show the following explicit upper bound on the recursive formula in Theorem~\ref{theo35}, that will not depend on the structure of the tree,
\begin{align*}
& \sum^{t}_{i=0} 	\binom{n-2-t+i}{i} V_T(n,t-i)  \leq n^{t-1} \binom{n+t}{2t+1}.
\end{align*} 
This result will be shown in Theorem~\ref{theo45}. 

First, we derive some interesting properties from the recursive formula in Theorem~\ref{theo35}, which proved that
\begin{align*}
& \sum^{t}_{i=0} 	\binom{n-2-t+i}{i} V_T(n,t-i)  =n^{t-1} \hspace{-3ex}\sum_{(i_0,i_1,\dots,i_{t}) \in P_T(n,t)} i_0i_1\cdots i_t.
\end{align*}
Notice also that for all $T\in \mathbf{T}(n)$ and $t=n-1$, 
\begin{align*}
n^{n-2} &=  n^{(n-1)-1} \sum_{(1,1,\dots,1) \in P_T(n,n-1)} 1 \\
& = \sum^{n-1}_{i=0} 	\binom{n-2-(n-1)+i}{i} V_T(n,n-1-i)   \\
& = \sum^{n-1}_{i=0} 	\binom{i-1}{i} V_T(n,n-1-i)  =  V_T(n,n-1),
& 
\end{align*}
where $\binom{-1}{0}$ is defined to be $1$, and indeed $V_T(n,n-1) = n^{n-2}$. 
Similarly, if $t=n-2$ then
\begin{align*}
& 2(n-1)n^{n-3} =  n^{(n-2)-1} \sum_{(i_0,i_1,\dots,i_{t}) \in P_T(n,n-2)} 2 \\
& = \sum^{n-2}_{i=0} 	\binom{i}{i} V_T(n,n-2-i)    = \sum^{n-2}_{i=0} V_T(n,n-2-i),
\end{align*}
and thus,
\begin{align} 
& \sum^{n-2}_{i=0}  V_{T}(n,i)  = 2(n-1)n^{n-3}.
\end{align}

As for stars, applying Theorem~\ref{theo35}, we simply draw the following formula
\begin{align} \label{eq:32}
& \sum^{t}_{i=0} 	\binom{n-2-t+i}{i} V^{\star}(n,t-i)  = n^{t-1}  \binom{n-1}{t}(n-t).
\end{align}

Using this result and the proof of Theorem~\ref{theo35}, the following interesting result holds. 
\begin{corollary}\label{theo43}
	For any $T \in \mathbf{T}(n)$ it holds that
	\begin{align*}
	\sum^{t}_{i=0} \binom{n-2-t+i}{i} \Big(	 V_T(n,t-i) - V^{\star}(n,t-i)  \Big) \geq 0.
	\end{align*}
\end{corollary}

Next an exact formula of the size of the tree ball of trees for stars is presented. The proof of this theorem is shown in Appendix~\ref{app:1}.
\begin{theorem}\label{theo:29}
	The size of the sphere for a star satisfies
	\begin{align*}
	S^{\star}(n,t)= \binom{n-1}{t}(n-1)^{t-1}(n-t-1),
	\end{align*}
	and the size of the tree ball of trees for a star satisfies
	\begin{align*}
	V^{\star}(n,t)= \sum^t_{j=0}\binom{n-1}{j}(n-1)^{j-1}(n-j-1).
	\end{align*}
\end{theorem}

Note that while in Theorem~\ref{theo:40} it was shown that for all $T\in \mathbf{T}(n)$ it holds that $V_T(n,t) = \Omega(n^{2t})$, for stars it is  deduced that $ S^{\star}(n,t) = \Theta(n^{2t})$ and $V^{\star}(n,t) = \Theta(n^{2t})$, which verifies that stars have asymptotically the smallest size of the tree ball of trees.

We turn to study the size of the tree ball of trees for path trees. We first simplify the formula of Theorem~\ref{theo35} in the path tree case.
\begin{theorem}\label{theo:35}
	The size of the tree ball of trees for a path tree satisfies 
	\begin{align*}
	& \sum^{t}_{i=0} 	\binom{n-2-t+i}{i} V^{\bline}(n,t-i) =  n^{t-1} \binom{n+t}{2t+1}.
	\end{align*}
\end{theorem}
\begin{IEEEproof}
	Denote by $A$ the set
	\begin{align*}
	A =  \left\{ (j_0,j_1,\dots,j_t)  \middle|
	\begin{array}{cc}
	1 \leq j_0 \leq n-t  \\ 
	1 \leq j_1 \leq n-(t-1)-j_0  \\ 
	\vdots \\
	1 \leq j_{t-1} \leq n-1-\sum^{t-2}_{s=0}j_s \\
	j_{t} = n-\sum^{t-1}_{s=0}j_s
	\end{array} 
	\right\}.
	\end{align*}
	Let $T\in \mathbf{T}(n)$ be a path tree. The following equations hold. 
	\begin{align*}
	\frac{1}{n^{t-1}} \sum^{t}_{i=0} & \binom{n-2-t+i}{i} V^{\bline}(n,t-i) \\
	& \stackrel{(a)}{=}\sum_{(i_0,i_1,\dots,i_{t}) \in P_T(n,t)} i_0i_1\cdots i_t \\
	& \stackrel{(b)}{=} \sum_{(j_0,j_1,\dots,j_t) \in A} j_0j_1\cdots j_t \stackrel{(c)}{=}   \binom{n+t}{2t+1}.
	\end{align*}
	Equality $(a)$ holds due to Theorem~\ref{theo35}.
	As for equality $(b)$, note that after an erasure of $t$ edges of $T$, we get $t+1$ connected components of $T$ where each of them is a path tree. The value of $j_i$ represents a path {subtree} as follows. The first path subtree will be of size $j_0$ which can be at least of size $1$ and at most of size $n-t$. Similarly, the size $j_1$ of the second path subtree ranges by between $ 1$ and $n-(t-1)-j_0$, i.e.  $1\leq  j_1 \leq n-(t-1)-j_0$. Continuing with this analysis, the size $j_t$ of the last path subtree satisfies $ j_{t} = n-\sum^{t-1}_{s=0}j_s$. Hence, the set of all vectors $(j_0,j_1,\dots,j_{t})$ is exactly the set $A$, which verifies equality $(b)$.
	Equality $(c)$ holds using combinatorial proof. Consider the problem of counting the number of options to choose $2t+1$ numbers from the set of numbers $[n+t]$. The right hand side is trivial. As for the left hand side, denote by $(x_1,x_2,\dots,x_{2t+1})$ a vector such that $x_1<x_2<\dots<x_{2t+1}$ representing an option of chosen $2t+1$ numbers. We choose these $2t+1$ in two steps. First we choose the values of $x_2,x_4,\ldots,x_{2t}$. We translate choosing these numbers to choosing the values of $j_0,j_1,\ldots, j_t$ such that 
	\begin{align*}
	x_2 = j_0, x_4 = j_1+j_0, \ldots, x_{2t} = \sum^{t-1}_{s=0}j_s =  n-j_t.
	\end{align*}
	In the next step we choose the values of $x_1,x_3,\ldots,x_{2t+1}$. Since $x_1 <x_2$, there are $j_0$ options to pick $x_1$. Similarly since $x_2<x_3<x_4$, there are $x_4-x_2 = j_1$ options to pick $x_3$. Lastly, since $x_{2t} <x_{2t+1}<n$, there are $j_t$ options to pick $x_{2t+1}$.
	Thus, every option of choosing $j_0,j_1,\ldots, j_t$ counts $ j_0j_1\dots j_{t-1}j_t $ solutions, and since all options of this problem are counted, the proof is concluded.
\end{IEEEproof}

Similarly to the case of stars, we showed in Theorem~\ref{theo:40} that for all $T\in \mathbf{T}(n)$ it holds that $V_T(n,t) = \cO(n^{3t})$, and it is also true for path trees as we can see in Theorem~\ref{theo:35}. According to Lemma~\ref{lemma:30} we also deduce that  $ V^{\bline}(n,t) = \Theta(n^{3t})$, that is, a path tree has asymptoticly the largest size of the tree ball of trees.

Although a path tree has asymptoticly the largest size of the tree ball of trees, it is not necessarily true that for every $n$ and $t$ its size is strictly the largest. We will show such an example at the end of this section.

Our last goal of this section is a stronger upper bound on the size of the tree ball of trees. According to Theorem~\ref{theo:40}, it was shown that for every tree $T \in \mathbf{T}(n)$ it holds that 
\begin{align*}
\sum_{(i_0,i_1,\dots,i_{t}) \in P_T(n,t)} i_0i_1\cdots i_t{\leq } \binom{n-1}{t} \Big(\frac{n}{t+1}\Big)^{t+1} , 
\end{align*}
while our goal is to improve this upper bound to be
\begin{align}\label{eq:23}
\sum_{ (i_0,i_1,\dots,i_t)  \in P_T(n,t) } i_{0} i_{1}\cdots i_t \leq  \binom{n+t}{2t+1}.
\end{align}
While this result does not improve asymptotic upper bound of $\cO(n^{3t})$, we believe that this upper bound is interesting and furthermore it verifies the statement of Theorem~\ref{theo45}.

First the definition of~\eqref{P_T} is slightly modified. Let $\bfv_\ell = (v_{i_0},v_{i_1},\dots, v_{i_{\ell-1}})$ be a vector of $\ell$ not necessary distinct nodes of $T\in \mathbf{T}(n)$ where  $1 \leq  \ell \leq t+1$. Let $F_{T,E'}(\bfv_\ell) = (V_n,E\setminus E') \in \cP_T(n,t)$ be a forest which is also denoted by $F_{T,E'}(\bfv_\ell) =  \{ C_0,C_1,\dots,C_{t} \}$, such that $ v_{i_j} \in C_j, j\in [\ell]$ and $ |C_{\ell}| \leq |C_{\ell+1}|\leq  \cdots \leq |C_{t}| $. In case there is more than one way to order to connected components $C_0,C_1,\dots,C_{t}$, we choose one of them arbitrarily. For  $1 \leq  \ell \leq t+1$ denote the multi-set $P_T(n,t;\bfv_\ell)$
\begin{align*}
& P_T(n,t;\bfv_\ell) = \\
& \Big\{ (|C_0|, \dots,|C_{t}|) |  \{C_0, \dots,C_{t} \} = F_{T,E'}(\bfv_\ell) \in \cP_T(n,t)   \Big\},
\end{align*}
and for $\ell = 0$, $$P_T(n,t;\bfv_\ell) = P_T(n,t).$$
Intuitively, this multi-set consist of profiles of forests such that all the nodes of the vector $\bfv_\ell$ are in different connected components of these forests. 
From this definition, in case that not all of the nodes in $\bfv_\ell$ are distinct, then $P_T(n,t;\bfv_\ell) = \emptyset$.
Another property is that for all $\ell \in [t+2]$ it holds that $$|P_T(n,t;\bfv_\ell)| \leq |P_T(n,t)| = \binom{n-1}{t}. $$
Next, for all $\ell \in [t+2]$, denote by $f_T(n,t;\bfv_\ell)$ the function
\begin{align*}
& f_T(n,t;\bfv_\ell)=  \sum_{ (c_0,c_1,\dots,c_t)  \in P_T(n,t;\bfv_\ell) } c_{\ell}c_{\ell+1}\cdots c_t,
\end{align*}
where in case that  $\ell = t+1$, the function $f_T$ is defined as
\begin{align*}
f_T(n,t,\bfv_{t+1}) & = \sum_{ (c_0,c_1,\dots,c_t)  \in P_T(n,t;\bfv_{t+1}) } 1 \\
& =  |P_T(n,t;\bfv_{t+1}) | \leq  \binom{n-1}{t}.
\end{align*}
Again, if not all of the nodes in $\bfv_\ell$ are distinct, by definition $P_T(n,t;\bfv_\ell) = \emptyset$ and 
$$f_T(n,t;\bfv_\ell) = 0 .$$
Since in case that $t=n-1$, each connected component is of size $1$, the following property is immediately deduced,
\begin{align}\label{eq:30}
& f_T(n,n-1;\bfv_\ell) = \hspace{-1ex} \sum_{ (c_0,c_1,\dots,c_{n-1})  \in P_T(n,n-1;\bfv_\ell) } \hspace{-1ex}c_{\ell}c_{\ell+1}\cdots c_{n-1} =  1.
\end{align}
The main goal in this part is to show that 
\begin{align}
f_T(n,t;\bfv_\ell) \leq \binom{n+t-\ell}{2t+1-\ell},
\end{align}
where in case that $\ell=0$, the equality~\eqref{eq:23} is immediately deduced.

Let $T \in \mathbf{T}(n)$ for $n\geq 2$. For two integers $t$ and $\ell$   such that $0 \leq \ell<   t+1 < n$, let $\bfv_\ell = (v_{i_0},\dots,v_{i_{\ell-2}},v_{i_{\ell-1}})$ be a vector of $\ell$ nodes in $T$.
For $j \in [\ell+1]$ denote $\bfv_j =  (v_{i_0},\dots,v_{i_{j-1}})$. For any node $v_x$ in $T$ denote $\bfv_{j+1}(v_x) =  (v_{i_0},\dots,v_{i_{j-1}},v_x)$.
By a slight abuse of notation, given a vector $(c_0, \dots,c_{t})\in P_T(n,t;\bfv_\ell)$, the connected component $C_j$ is referred to the value $c_j$, or in other words $(c_0, \dots,c_{t}) = (|C_0|,\dots,|C_t|)$.
For any node $v_x$ in $T$ denote by $A_T(n,t;\bfv_\ell,v_x)$ the set 
\begin{align*}
A_T(n,t;\bfv_\ell,v_x) = & \Big\{ (|C_0|,\dots,|C_t|) \in P_T(n,t;\bfv_\ell) | v_x \in \bigcup_{i\in [\ell]} C_i \Big\}.
\end{align*}
Let $v_x$ be a leaf connected to a node denoted by $v_y$ in $T \in \mathbf{T}(n)$, and let $T_1 \in \mathbf{T}(n-1)$ be a tree generated by removing $v_x$ from $T$. 
The definitions introduced above are used in the next claims and lemmas. 
\begin{claim}\label{claim:5}
	The following properties hold
	\begin{enumerate}
		\item It holds that	
		\begin{align*}
		& \sum_{ (c_0,\dots,c_{t})  \in P_T(n,t;\bfv_\ell) } c_{\ell}\cdots c_{t} \\
		& = \sum_{ (c_0,\dots,c_{t})  \in 	A_T(n,t;\bfv_\ell,v_y) } \hspace{-3ex}c_{\ell}\cdots c_{t}  + \hspace{-3ex}\sum_{ (c_0,\dots,c_{t})  \in P_T(n,t;\bfv_{\ell+1}(v_y)) } \hspace{-3ex}c_{\ell}\cdots c_{t}.
		\end{align*}
		\item  If $v_x$ is not in $\bfv_\ell$ then,	
		\begin{align*}
		&    \sum_{ (c_0,\dots,c_t)  \in 	A_{T_1}(n-1,t;\bfv_\ell,v_y)  }\hspace{-3ex} c_{\ell}\cdots c_t = \hspace{-3ex}  \sum_{ (c'_0,\dots,c'_t)  \in 	A_T(n,t;\bfv_\ell,v_x)  }\hspace{-3ex} c'_{\ell}\cdots c'_{t}.
		\end{align*}
		\item If $v_x$ is not in $\bfv_\ell$ then,
		\begin{align*}
		&  \sum_{ (c_0,\dots,c_{t})  \in P_T(n,t;\bfv_{\ell+1}(v_x)) } \hspace{-3ex}c_{\ell}\cdots c_{t} \\	
		& = \sum_{ (c'_0,\dots,c'_{t})  \in P_{T_1}(n-1,t;\bfv_{\ell+1}(v_y)) } (c'_{\ell}+1)c'_{\ell+1}\cdots c'_{t}  \\
		& +  \sum_{ (c_0,\dots,c_{\ell-1},1,c_{\ell+1},\dots, c_t) \in P_T(n,t;\bfv_{\ell+1}(v_x)) } 1\cdot c_{\ell+1}\cdots c_{t}.
		\end{align*}
		\item If $v_x=v_{i_{\ell-1}}$ then 
		\begin{align*}
		& \sum_{ (c_0,\dots,c_t)  \in P_T(n,t;\bfv_\ell(v_x))} c_\ell\cdots c_t \\
		& =  \sum_{ (c'_0,\dots,c'_t)  \in P_{T_1}(n-1,t;\bfv_\ell(v_y)) } c'_\ell  \cdots c'_t   \\
		& +  \sum_{ (c_0,\dots,c_{\ell-2},1,c_{\ell},\dots, c_t) \in P_T(n,t;\bfv_{\ell}(v_x)) } c_{\ell}\cdots c_{t}.
		\end{align*}
	\end{enumerate}
\end{claim}

The proof of Claim~\ref{claim:5} can be found in Appendix~\ref{app:4}. Next we show a recursive formula with respect to $f_T$.

\begin{figure*}[t!]
	\centering
	\label{rec:1}\subfigure[The tree $T\in \mathbf{T}(10)$.]{\includegraphics[width=40mm]{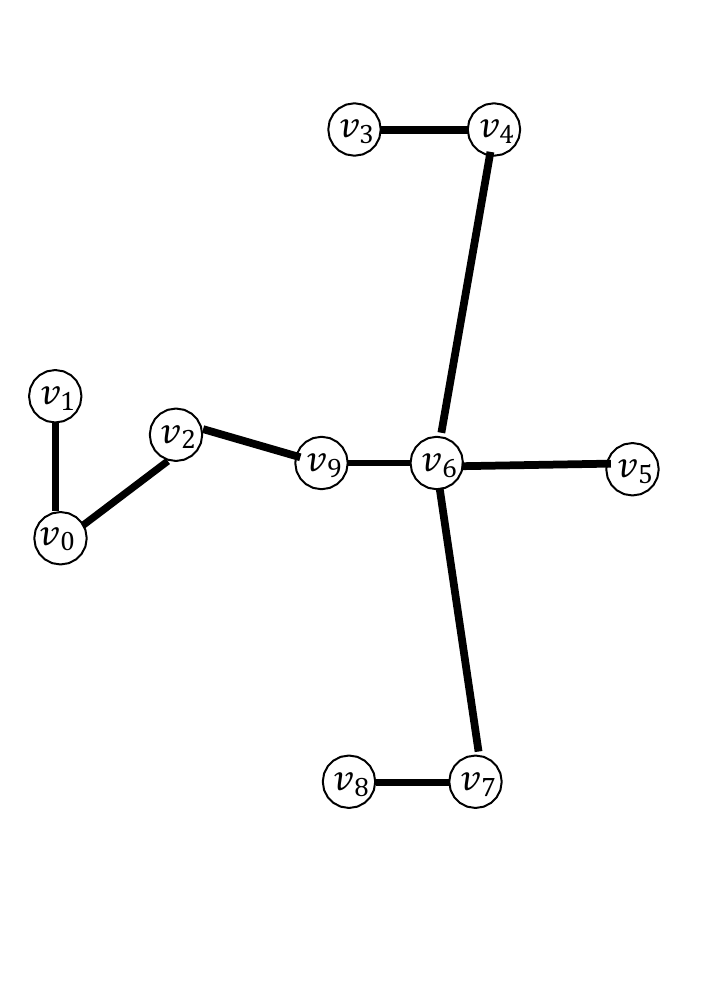}}
	\hfill
	\label{rec:2}\subfigure[An erasure of $4$ edges of $T$ including the edge $\langle v_5,v_6 \rangle$, and $|C_0|\cdot|C_1|\cdot|C_3|\cdot|C_4|$ is counted in $ f_{T_{1}}(9,3;(v_7))$.]{\includegraphics[width=57mm]{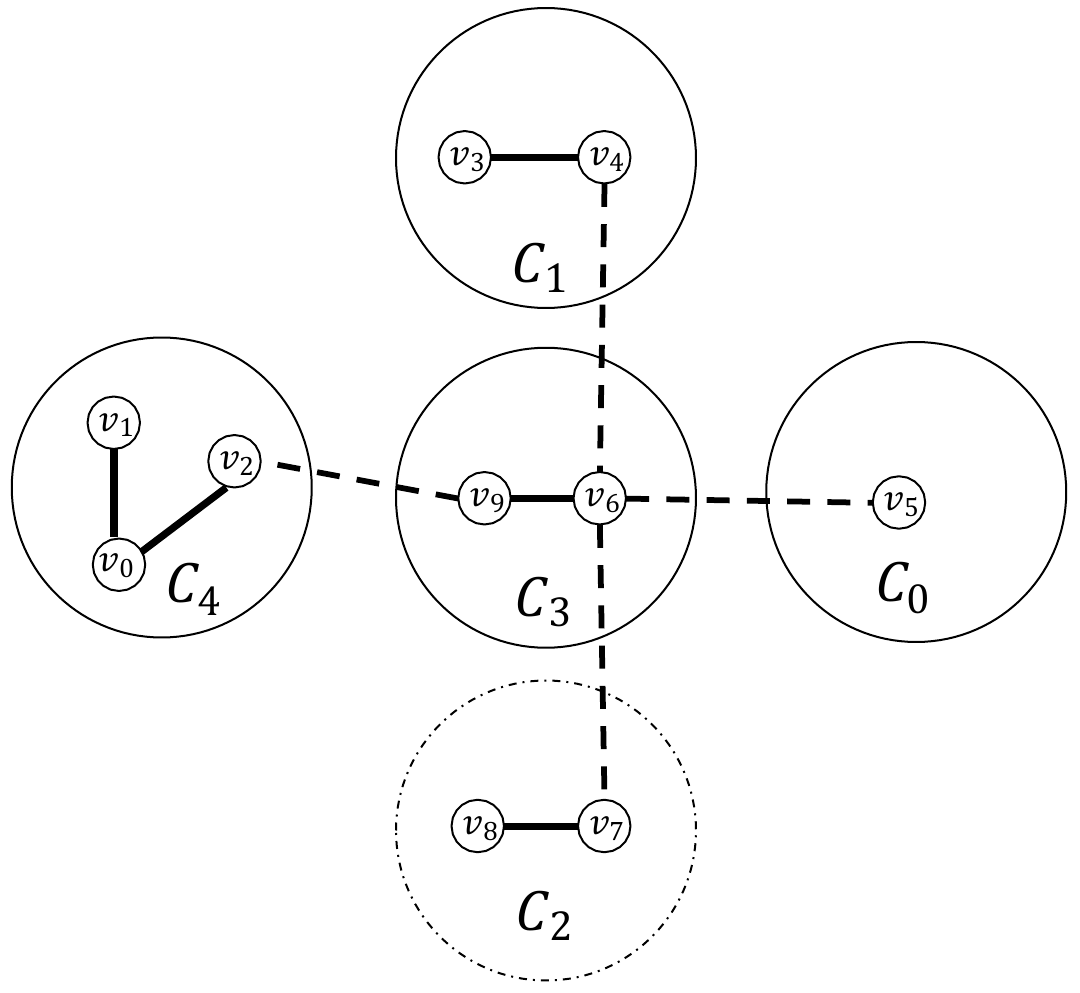}}
	\hfill
	\label{rec:3}\subfigure[An erasure of $4$ edges of $T$ without the edge $\langle v_5,v_6 \rangle$, and $|C_0|\cdot|C_1|\cdot\Big(|C_3|-1\Big)\cdot|C_4|$ is counted in $ f_{T_{1}}(9,4;(v_7))$.]{\includegraphics[width=40mm]{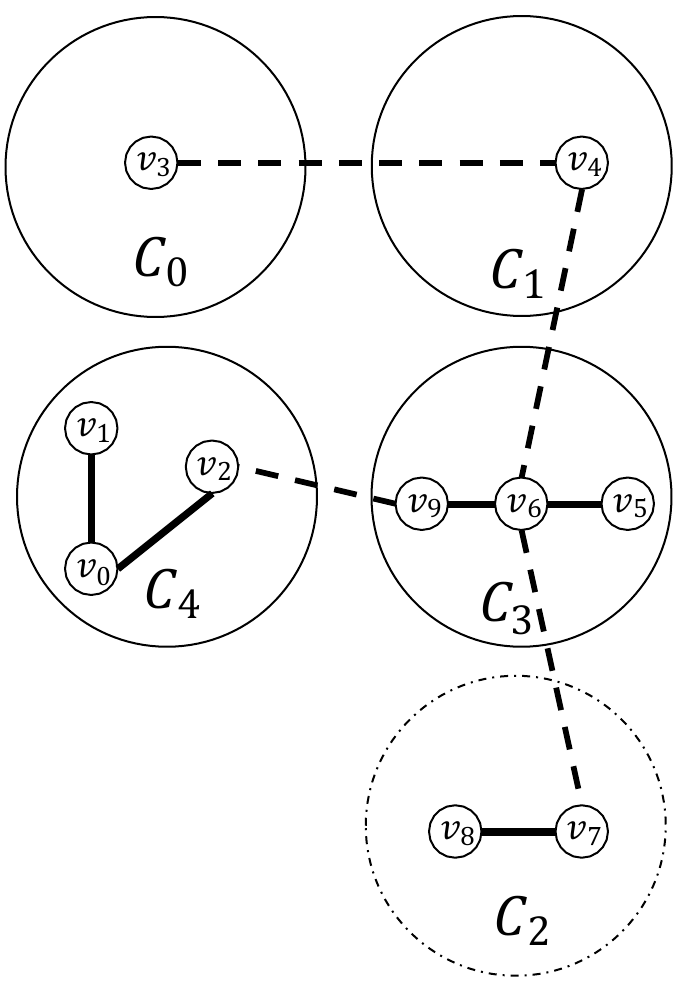}}
	\hfill
	\label{rec:4}\subfigure[An erasure of $4$ edges of $T$ without the edge $\langle v_5,v_6 \rangle$, and $|C_0|\cdot|C_1|\cdot|C_4|$ is counted in $ f_{T_{ 1}}(9,4;(v_7,v_6) )$.]{\includegraphics[width=40mm]{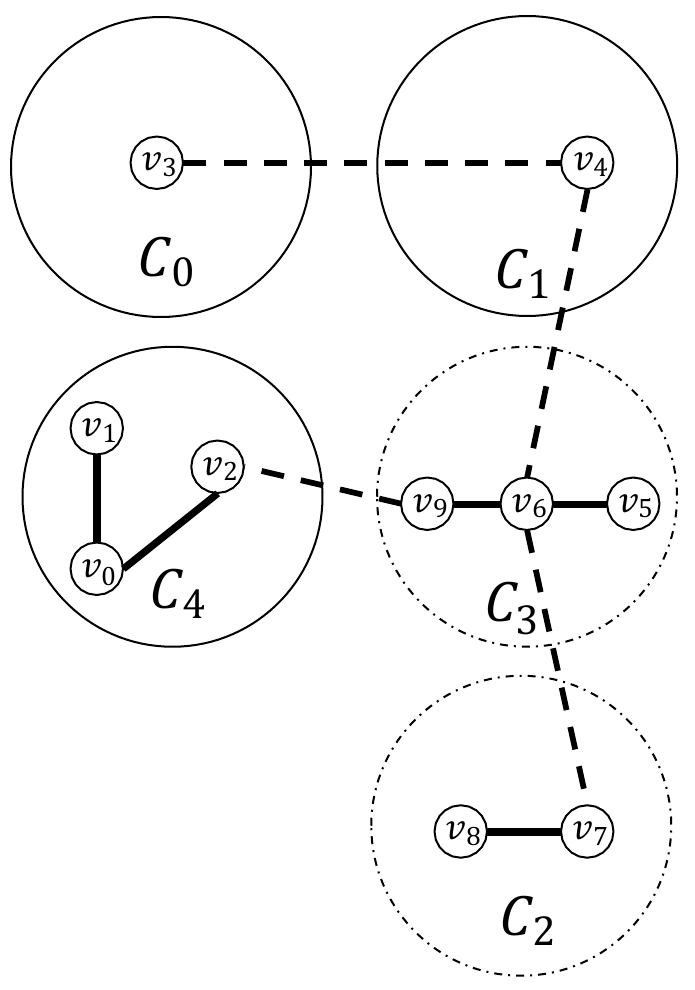}}	
	\hfill			
	\caption{The idea of the recursive formula  $f_{T}(10,4;(v_7)))  = f_{T_{1}}(9,3;(v_7))+ f_{T_{1}}(9,4;(v_7))  + f_{T_{ 1}}(9,4;(v_7,v_6) )    $.}	\label{fig:rec}				
\end{figure*}

\begin{lemma}\label{lemma:38}
	If $v_x$ is not in $\bfv_\ell$ then,	
	\begin{align*}
	f_{T}(n,t;\bfv_\ell)  & =  f_{T_{1}}(n-1,t;\bfv_\ell)  \\
	& + f_{T_{ 1}}(n-1,t;\bfv_{\ell+1}(v_y) )   + f_{T_{1}}(n-1,t-1;\bfv_\ell).
	\end{align*}
	If $v_x$ is in $\bfv_\ell$, and without loss of generality $v_x =v_{i_{\ell-1}} $, then
	\begin{align*}
	& f_{T}(n,t;\bfv_\ell)  =  f_{T_{1}}(n-1,t;\bfv_\ell(v_y))  +  f_{T_{1}}(n-1,t-1;\bfv_{\ell-1}).
	\end{align*}
\end{lemma}

The proof of Lemma~\ref{lemma:38} can be found in Appendix~\ref{app:5}. An example that illustrates this recursive formula is now presented.

\begin{example}\label{ex:rec}
	For $n=10$, we illustrate in Fig.~\ref{fig:rec}$(a)$ a tree $T\in \mathbf{T}(10)$. 
	In this example, $t=4$ and $\ell=1$, $v_x = v_5,v_y = v_6$ and $\bfv_\ell = (v_7)$. 
	Let $T_1 \in \mathbf{T}(9)$ be a tree which is derived from $T$ by removing the node $v_5$. After an erasure of $4$ edges, the  multiplication of the five connected components is counted in $f(n,t)$. Fig.~\ref{fig:rec}$(b),(c)$ and $(d)$ represent the idea of the formula $f_{T}(n,t;\bfv_\ell)   =  f_{T_{1}}(n-1,t;\bfv_\ell)   + f_{T_{ 1}}(n-1,t;\bfv_{\ell+1}(v_y) )   + f_{T_{1}}(n-1,t-1;\bfv_\ell)$. The dashed edges in Fig.~\ref{fig:rec}$(b),(c)$ and $(d)$ represent the erased edges from $T$, yielding a forest with five connected components $C_0,C_1,C_2,C_3,$ and $C_4$.
	An example of possible erasure including the edge $\langle v_5,v_6 \rangle$ is shown in Fig.~\ref{fig:rec}$(b)$. This example emphasizes the case which corresponds to the multiplication $|C_0|\cdot|C_1|\cdot|C_3|\cdot|C_4|$ that is also counted in $ f_{T_{1}}(n-1,t;(v_7))$ since $|C_0|=1$. 
	Fig.~\ref{fig:rec}$(c)$ and $(d)$ similarly emphasize the case in which an erasure of $4$ edges does not include the edge  $\langle v_5,v_6 \rangle$. While Fig.~\ref{fig:rec}$(c)$ emphasizes the multiplication $|C_0|\cdot|C_1| \cdot\Big(|C_3|-1\Big)\cdot|C_4|$, which is counted in $ f_{T_{1}}(n-1,t-1;(v_7))$ (since $v_5$ is not in $T_1$), Fig.~\ref{fig:rec}$(d)$ emphasizes  the multiplication $|C_0|\cdot|C_1| \cdot|C_4|$, which is counted in $f_{T_{ 1}}(n-1,t;(v_7,v_6) )$. Hence, $|C_0|\cdot|C_1|\cdot|C_3|\cdot|C_4|$ is also counted  in the case that the edge $\langle v_5,v_6 \rangle$ is not erased.
\end{example}

Finally, the upper bound for $f_{T}(n,t;\bfv_\ell) $ is presented, while the proof is shown in Appendix~\ref{app:6}. 

\begin{lemma}\label{lemm:39}
	For any tree $T \in \mathbf{T}(n),n\geq 1$ and a vector of $0\leq \ell \leq t+1\leq n$ nodes $\bfv_\ell = (v_{i_0},v_{i_1},\dots,v_{i_{\ell - 1}})$,
	\begin{align*}
	f_{T}(n,t;\bfv_\ell)  \leq \binom{n+t-\ell}{2t+1-\ell}.
	\end{align*}
\end{lemma}

From Lemma~\ref{lemm:39} it is immediately deduced that for all $T \in \mathbf{T}(n)$,
\begin{align}\label{eq:21}
\sum_{(i_0,i_1,\dots,i_{t}) \in P_T(n,t)} i_0i_1\cdots i_t = f_{T}(n,t)  \leq \binom{n+t}{2t+1}.
\end{align}

Using~\eqref{eq:21} the tighter upper bound for the recursive formula in Theorem~\ref{theo35} is shown in the following theorem.
\begin{theorem}\label{theo45}
	For any $T \in \mathbf{T}(n)$ it holds that
	\begin{align*}
	& \sum^{t}_{i=0} 	\binom{n-2-t+i}{i} V_T(n,t-i)   \leq n^{t-1}\binom{n+t}{2t+1}.
	\end{align*}
\end{theorem}

From Theorem~\ref{theo45} and Theorem~\ref{theo:35} we immediately deduce the following corollary.
\begin{corollary}\label{theo44}
	For any $T \in \mathbf{T}(n)$ it holds that
	\begin{align*}
	\sum^{t}_{i=0} \binom{n-2-t+i}{i} \Big(	 V_T(n,t-i) - V^{\bline}(n,t-i)  \Big) \leq 0.
	\end{align*}
\end{corollary}
Even though by Corollary~\ref{theo43}, 
\begin{align*}
\sum^{t}_{i=0} \binom{n-2-t+i}{i} \Big(	 V_T(n,t-i) - V^{\star}(n,t-i)  \Big) \geq 0,
\end{align*}
and by Corollary~\ref{theo44},
\begin{align*}
\sum^{t}_{i=0} \binom{n-2-t+i}{i} \Big(	 V_T(n,t-i) - V^{\bline}(n,t-i)  \Big) \leq 0,
\end{align*}
it does not imply that for all $n$ and $t$, $V^{\star}(n,t) \leq  V_T(n,t) \leq V^{\bline}(n,t) $. For example, if $t=n-2$, $V^{\star}(n,t) = n^{n-2}$ while $V^{\bline}(n,t) < n^{n-2}$, since one can check that there are two path trees $T_1,T_2 \in \mathbf{T}(n)$ such that $d_\cT(T_1,T_2) = n-1$. However, we conjecture that for fixed  $t$ and large enough $n$, it holds that $V^{\star}(n,t) \leq  V_T(n,t) \leq V^{\bline}(n,t) $.

\section{Constructions of Codes over Trees}\label{tree:const}

In this section we show several constructions of codes over trees. The first is the construction of $\ut(n,\lfloor n/2 \rfloor,n-1)$ codes, and the second is the construction of  $\ut(n,n,n-2)$  codes. The third and our main result in this section is the construction of $\ut(n,M,d)$  codes for fixed $d$ where $M  = \Omega ( n^{n-2d})$.  For positive integers $a$ and $n$ we will use the notation $\langle a \rangle_n$ to denote the value of $(a \mod n) $. 

\subsection{A Construction of $\ut(n,\lfloor n/2 \rfloor,n-1)$ Codes }\label{sub:n-2}
A path tree $T=(V_n,E)$ with the edge set
$$E = \{ (v_{i_j},v_{i_{j+1}}) ~|~ j\in[n-1], i_j\in[n]  \},$$
will be denoted by $T = (v_{i_0},v_{i_1},\dots,v_{i_{n-1}})$, i.e., the nodes $v_{i_0}$ and $v_{i_{n-1}}$ are leaves and the rest of the nodes have degree $2$. 
Note that the number of path trees over $n$ nodes is $n!/2$, so every path tree has two representations in this form and we will use either one of them in the sequel. For $s \in [\lfloor n/2 \rfloor ] $, denote by $T_s = (V_n,E)$ the path tree 
\begin{align*}
T_s = \begin{cases}
\Big(v_{\langle s \rangle_n},v_{\langle s-1 \rangle_n},v_{\langle s+1 \rangle_n},\dots ,v_{\langle s+ \frac{n-1}{2} \rangle_n}\Big):\textmd{if $n$ is odd}, \\
\Big(v_{\langle s \rangle_n},v_{\langle s-1 \rangle_n},v_{\langle s+1 \rangle_n}, \dots ,v_{\langle s- \frac{n}{2} \rangle_n}\Big):\textmd{if $n$ is even}.
\end{cases}
\end{align*}

\begin{example}\label{ex:hamilton}
	For $n=10$ we show an example of the path tree $T_0$. By looking at the lower half of the circle in this figure, i.e. nodes $v_0,v_9,v_8,v_7,v_6,v_5$, there is a single edge connecting two vertices on this half circle. The path tree $T_1$ is received by rotating anticlockwise the nodes on this circle by one step. Note that all the edges in $T_0$ and $T_1$ are disjoint and this property holds also for the other path trees $T_2,T_3,T_4$.
	\begin{figure}[h!]\label{fig:hamilton}
		\hfill
		\subfigure[The $T_0$ tree.]{\includegraphics[width=43.2mm]{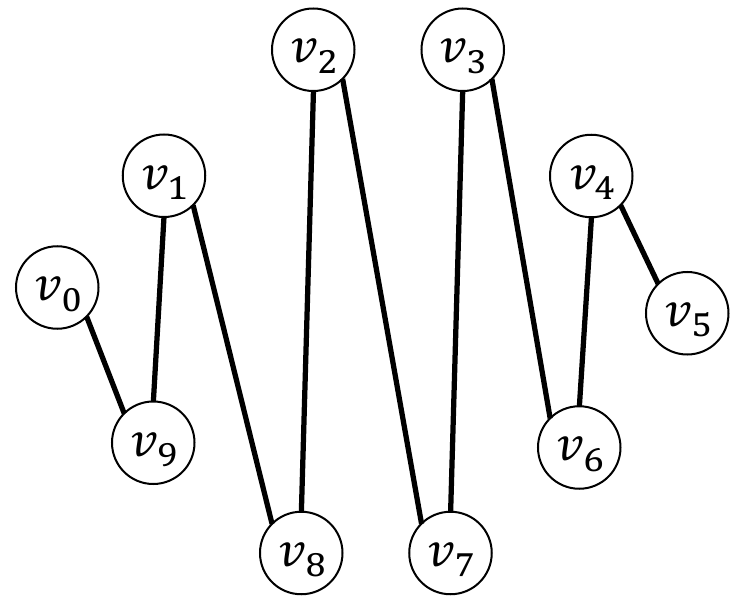}}\label{fig:graph example3.1}
		\hfill
		\subfigure[The $T_1$ tree.]{\includegraphics[width=43.2mm]{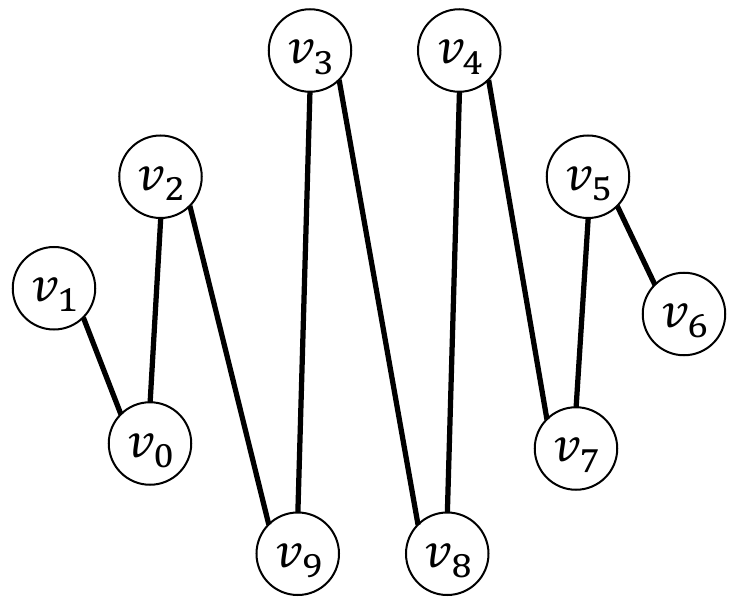}}\label{fig:graph example3.2} 
		\hfill
		\caption{This code contains $5$ trees, $T_0,T_1,T_2,T_3,$ and $T_4$.}	
	\end{figure}
	
\end{example}

The construction of a $\ut(n,\lfloor n/2 \rfloor,n-1)$ code is given as follows. This construction is motivated by the factorization of the complete graph into mutually disjoint Hamiltonian paths; see~\cite{Hartsfield,Lucas}. Even though this result is well known, for completeness we present it here along with its proof. 
\begin{Construction}\label{const:tree0}
	For all $n\geq 3$ let $\cC_{\cT_1}$ be the following code over trees
	\begin{equation*}
	\cC_{\cT_1} = \{T_s = (V_n,E)| s \in [\lfloor n/2 \rfloor ] \}.
	\end{equation*}
\end{Construction}
\begin{theorem}\label{th:trees0}
	The code $\mathcal{C}_{\cT_1}$ is a $\ut(n,\lfloor n/2 \rfloor,n-1)$ code.
\end{theorem}
\begin{IEEEproof}
	Clearly, since for all distinct $s_1,s_2\in[\lfloor n/2 \rfloor]$ it holds that $s_1\neq s_2+\lfloor n/2 \rfloor $, it is deduced that $|\mathcal{C}_{\cT_1}| = \lfloor n/2 \rfloor$. Next we prove that this code can correct $\rho = n-2$ edge-erasures, by showing that $d_\cT(\cC_{\cT_{1}}) > n-2$.
	
	Assume on the contrary that $d_\cT(\cC_{\cT_{1}}) \leq n-2$. Therefore, there are two distinct numbers $s_1,s_2 \in [\lfloor n/2 \rfloor]$  such that the trees $T_{s_1} = (V_n,E_1),T_{s_2} = (V_n,E_2) \in \mathcal{C}_{\cT_1}$ hold $|E_1\cap E_2| \geq 1$. Therefore, there exist two integers  $t_1,t_2\in [\lfloor n/2 \rfloor]$ such that one of the following cases hold:
	\begin{enumerate}
		\item $\langle v_{\langle s_1 + t_1 \rangle_n},v_{\langle s_1 - (t_1+1) \rangle_n} \rangle = \langle v_{\langle s_2 + t_2 \rangle_n},v_{\langle s_2 - (t_2+1) \rangle_n} \rangle  $,
		\item $\langle v_{\langle s_1 + t_1 \rangle_n},v_{\langle s_1 - (t_1+1) \rangle_n} \rangle = \langle v_{\langle s_2 - t_2 \rangle_n},v_{\langle s_2 + t_2 \rangle_n} \rangle  $,
		\item $\langle v_{\langle s_1 - t_1 \rangle_n},v_{\langle s_1 +t_1 \rangle_n} \rangle = \langle v_{\langle s_2 - t_2 \rangle_n},v_{\langle s_2 + t_2 \rangle_n} \rangle  $.
	\end{enumerate}
	We will eliminate all those options as follows. 
	\begin{enumerate}
		\item If $\langle s_1 + t_1 \rangle_n = \langle s_2 + t_2 \rangle_n$ and  $\langle s_1 - (t_1+1) \rangle_n = \langle s_2 - (t_2+1) \rangle_n$ then by summing those equations we deduce that $\langle 2s_1 -1 \rangle_n = \langle 2s_2 -1 \rangle_n$. Therefore, we deduce that $s_1 = s_2$ which is a contradiction. Similar proof shows that it is impossible to have $\langle s_1 + t_1 \rangle_n = \langle s_2 - (t_2+1) \rangle_n$ and $\langle s_2 + t_2 \rangle_n = \langle s_1 - (t_1+1) \rangle_n$.
		\item If $\langle s_1 + t_1 \rangle_n = \langle s_2 - t_2 \rangle_n$ and  $\langle s_1 - (t_1+1) \rangle_n = \langle s_2 +t_2 \rangle_n$ then by summing those equations we deduce that $\langle 2s_1 -1 \rangle_n = \langle 2s_2 \rangle_n$. Since $s_1,s_2 \in [\lfloor n/2 \rfloor]$, if $s_1\neq 0 $ then {$ 2s_1 -1 <n-1$ and  $ 2s_2 <n-1$. Clearly, $\langle 2s_1 -1 \rangle_n$ is odd and $\langle s_2 \rangle_n$ is even, (since both of them smaller than $n$) so it is deduced that they are distinct}. If $s_1 = 0$ then $\langle 2s_1 -1 \rangle_n = n-1$ but since $s_2 \in [\lfloor n/2 \rfloor]$ it holds that $2s_2 < n-1$ and therefore we get again that  $\langle 2s_1 -1 \rangle_n \neq  \langle 2s_2 \rangle_n$, which is a contradiction. Similar proof shows that it is impossible to have $\langle s_1 + t_1 \rangle_n  = \langle s_2 +t_2 \rangle_n $ and $\langle s_1 - (t_1+1) \rangle_n= \langle s_2-t_2 \rangle_n$.
		\item If $\langle s_1 - t_1 \rangle_n = \langle s_2 - t_2 \rangle_n$ and  $\langle s_1 + t_1 \rangle_n = \langle s_2 +t_2 \rangle_n$ then by summing those equations we deduce that $\langle 2s_1 \rangle_n = \langle 2s_2 \rangle_n$. Therefore, we deduce that $s_1 = s_2$ which is a contradiction. Similar proof shows that it is impossible to have $\langle s_1 - t_1 \rangle_n = \langle s_2 + t_2 \rangle_n$ and $\langle s_1 + t_1 \rangle_n = \langle s_2 - t_2 \rangle_n$.
	\end{enumerate}
\end{IEEEproof}
In this construction the result $A(n,n-1) \geq  \lfloor n/2 \rfloor$ is shown, and since by~\eqref{boundC2_0}, $A(n,n-1) \leq   n/2 $ it is deduced that $A(n,n-1) =  \lfloor n/2 \rfloor$.  

\subsection{A Construction of $\ut(n,n,n-2)$ Codes }
For convenience, a star $T$ with a node $v_{i}$ of degree $n-1$ will be denoted by $T_{v_{i}}$.
The construction of a $\ut(n,n,n-2)$ code will be as follows.
\begin{Construction}\label{const:tree1}
	For all $n\geq 4$ let $\cC_{\cT_2}$ be the following code 
	\begin{equation*}
	\cC_{\cT_2} = \{T_{v_i} = (V_n,E)| i\in [n] \}.
	\end{equation*}
\end{Construction}
Clearly, the code $\cC_{\cT_2}$ is a set of all stars over $n$ nodes. Next we prove that this code is a $\ut(n,n,n-2)$ code.
\begin{theorem}\label{th:trees1}
	The code $\mathcal{C}_{\cT_2}$ is a $\ut(n,n,n-2)$ code.
\end{theorem}

\begin{IEEEproof}
	Let $T_{v_i}=(V_n,E), i\in [n]$, be a codeword-tree of $\cC_{\cT_2}$ with a node $v_i$ of degree $n-1$. Since $T_{v_i}$ is a star, after the erasure of $n-3$ edges from $T_{v_i}$, the node $v_i$ will have degree $2$ and all the nodes $v_j\in T_{v_i}$, $j\neq i$ will have degree of at most $1$. Therefore the node $v_i$ can be easily recognized and the codeword-tree $T_{v_i}$ can be corrected. 
\end{IEEEproof}
In this trivial construction we showed that $A(n,n-2) \geq  n$ and since by Theorem~\ref{th:trees1_1}, $A(n,n-2) \leq  n$  it is deduced that $A(n,n-2) =  n$.  

\subsection{A Construction of $\ut(n, \Omega( n^{n-2d}),d)$ Codes}\label{subsec:const1}
In this section we show a construction of $\ut(n, \Omega( n^{n-2d}),d)$ codes for any positive integer $d \leq n/2$. Note that according to Corollary~\ref{cor1}, for fixed $d$, $ A(n,d) = \cO( n^{n-1-d}) $ and by Corollary~\ref{cor:2} it will be deduced that $A(n,d) = \Omega( n^{n-2d}) $.

For a vector $\bfu \in \F^m_2$ denote by $w_H(\bfu)$ its Hamming weight, and for two vectors $\bfu,\bfw  \in \F^m_2$,  $d_H(\bfu,\bfw )$ is their Hamming distance. A binary code $\cC$ of length $m$ and size $K$ over $\F_2$ will be denoted by $(m,K)$ or $(m,K,d)$, where $d$ denotes its minimum Hamming distance. If $\cC$ is also linear and $k$ is its dimension, we denote the code by $[m,k]$ or $[m,k,d]$.

Let $E_n$ be the set of all $\binom{n}{2}$ edges as defined in~\eqref{eq:En}, with a fixed order. For any set $E\subseteq E_n$, let $\bfv_E$ be its {characteristic vector} of length $\binom{n}{2}$ which is indexed by the edge set $E_n$ and every entry has value one if and only if the corresponding edge belongs to $E$. That is, 
\begin{align*}
(\bfv_E)_e = \begin{cases}
1, & e \in E\\
0, & otherwise
\end{cases}.
\end{align*}

The construction of $\ut(n,M,d)$ code over trees will be as follows.

\begin{Construction}\label{const:tree3_1}
	For all $n\geq 1$ let $\cC$ be a binary code $(\binom{n}{2},K,2d-1)$. Then, the code $\cC_{\cT_3}$ is defined by
	\begin{equation*}
	\cC_{\cT_3} = \{T\in \mathbf{T}(n) ~|~ \bfv_E \in \cC \}.
	\end{equation*}
\end{Construction}

\begin{theorem}\label{th:trees2_2}
	The code $\mathcal{C}_{\cT_3}$ is a $\ut(n,M,d)$ code over trees.
\end{theorem}
\begin{IEEEproof}
	By Theorem~\ref{th:dist}, a code over trees  $\cC_\cT$ with parameters $\ut(n,M)$ has minimum distance $d$ if and only if $\cC_\cT$ can correct any $d-1$ edge erasures. Notice also that since $ \cC $ is a code with Hamming distance $2d-1$, it can correct at most any $d-1$ substitutions.
	
	Let $T = (V,E)$ be a codeword-tree of $\cC_{\cT_3}$ with its binary edge-vector $\bfv_E$. Suppose that $T$ experienced at most $d-1$ edge erasures, generating a new forest $F$ with the edge set $E'$. Since $E'\subseteq E$ and $|E'|\geq |E|-(d-1)$, it holds that $d_H(\bfv_{E'},\bfv_E) \leq d-1$ and the vector $\bfv_E$ can be corrected using a decoder of $\cC$. 
\end{IEEEproof}

The next corollary summarizes the result of this construction.
\begin{corollary}\label{cor:2}
	For positive integer $n$ and fixed $d$, $ A(n,d) =\Omega( n^{n-2d}) $ and the redundancy is $r(n,d) \leq  (d-1)\log(n) + \cO(1)$.
\end{corollary}

\begin{IEEEproof}
	Applying BCH codes  {(see Chapter 5.6 in~\cite{Roth})} in Construction~\ref{const:tree3_1} for all $n\geq 1$, linear codes $[\binom{n}{2},k,2d-1]$ are used with redundancy 
	$$r=(d-1)\log (\binom{n}{2})+ \cO(1) = 2(d-1)\log(n) + \cO(1)$$ redundancy bits. The $2^r$ cosets of the $\cC$ codes are also binary $(\binom{n}{2},2^k,2d-1)$ codes. Note that each tree $T$ from $\mathbf{T}(n)$ can be mapped by {Construction~\ref{const:tree3_1}} to exactly one of these cosets. Thus, by the pigeonhole principle, there exists a code $\cC_{\cT_3}$ of cardinality at least
	$$ \frac{n^{n-2}}{2^{2(d-1)\log(n) + \cO(1)}}= \frac{n^{n-2}}{\alpha n^{2d-2}} =  \frac{1}{\alpha}n^{n-2d},$$ 
	for some constant $\alpha$. Thus, we also deduce that $$r(n,d) \leq 2(d-1)\log(n) + \cO(1).$$
	
\end{IEEEproof}

\begin{remark}
	{We note that the use of BCH codes can be changed to any linear codes. In fact, it is possible to use in Construction~\ref{const:tree3_1} a code correcting $d-1$ asymmetric errors}. However, we chose to use symmetric error-correcting codes since the use of asymmetric error-correcting codes does not improve the asymptotic result and in order to derive the result in Corollary~\ref{cor:2} we needed linear codes. 
\end{remark}

In this section we showed a family of codes with $\Omega(n^{n-2d})$ codeword-trees where $d\leq n/2$. Next we show a  construction of  codes over trees with $\Omega(n^2)$ codeword-trees where $d$ is almost $3n/4$.

\subsection{A Construction of $\ut(n, \frac{n-1}{2}\cdot  \lfloor \frac{n-1}{m} \rfloor,\lfloor  \frac{3n}{4} \rfloor - \lceil\frac{3n}{2m}\rceil - 2  )$ Codes}\label{subsec:const2}
In this section, for a prime $n$, we show a construction of $\ut(n,\frac{n-1}{2}\cdot  \lfloor \frac{n-1}{m} \rfloor, \lfloor  \frac{3n}{4} \rfloor - \lceil\frac{3n}{2m}\rceil - 2 )$ codes, where $m$ is a positive integer such that $3 \leq  m \leq n-1$. By Corollary~\ref{cor:2},  $ A(n,d) = \Omega( n^{n-2d}) $ where $d\leq n/2$. Here we extend this result by showing that for $d$ approaching $\lfloor 3n/4 \rfloor$, there exists a code with $\Omega(n^2)$ codeword-trees. First, several definitions are presented.

\begin{figure}[h!]
	\centering
	\includegraphics[width=80mm]{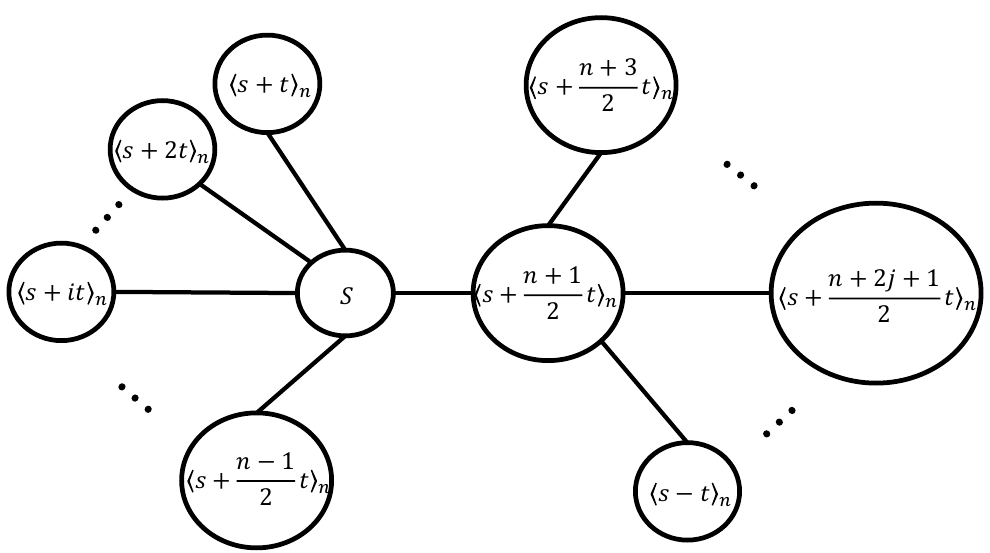}
	\caption{A two-star tree.  The nodes are marked by numbers $i\in[n]$ instead of nodes $v_i$. Note that 
		by the definition of $E^{(-)}_{s,t}$, the node marked by $\langle s - t \rangle_n$ is exactly the node marked by {$\langle s+\frac{n+2j+1}{2}t \rangle_n$}, where $j=\frac{n-3}{2}$. }\label{fig:dustar}
\end{figure}

A \textit{two-star tree} over $n$ nodes is a tree who has exactly $n-2$ leaves. 
For a prime $n$ and integers $s,t \in [n]$ where $t\neq0$, denote the following two edge sets
\begin{align*}
E^{(+)}_{s,t} & = \Big\{ \langle v_{s},v_{\langle s + it \rangle_n} \rangle | 1 \leq i\leq \frac{n+1}{2} \Big\}, \\
E^{(-)}_{s,t} & =  \Big\{ \langle v_{\langle s+\frac{n+1}{2}t \rangle_n},v_{\langle s+\frac{n+2j+1}{2}t \rangle_n}\rangle | 1 \leq j\leq  \frac{n-3}{2} \Big\}.
\end{align*}
Denote by $T_{s,t} = (V_n,E_{s,t})$ the two-star tree with the edge set
\begin{align*}
E_{s,t} =  E^{(+)}_{s,t} \cup E^{(-)}_{s,t} .
\end{align*}
It is possible to verify that indeed according to this definition $T_{s,t}$ is well defined and is a two-star tree. Furthermore, It will be shown in Theorem~\ref{th:tree4} that each pair $(s,t)$ defines a unique tree $T_{s,t}$. The nodes $ v_{s}$ and $ v_{\langle s+\frac{n+1}{2}t \rangle_n} $ are called \textit{the central nodes} of $T_{s,t}$. Also note that 
\begin{align}\label{degs}
\deg( v_{s}) = \frac{n+1}{2}, ~~& \deg( v_{\langle s+\frac{n+1}{2}t \rangle_n} ) = \frac {n-1}{2}.
\end{align}
In Fig.~\ref{fig:dustar} we illustrate a two-star tree $T_{s,t}$.

For a prime $n$ and an integer $ 1 \leq t  \leq \lfloor \frac{n-1}{m}\rfloor$, where $3 \leq m \leq n-1$  and $\alpha  \in \{ \frac{n+1}{2},\frac{n-1}{2}\}$, denote by $W(n,t,\alpha)$ the set
\begin{align}\label{Wt}
W(n,t,\alpha) = \{\langle t \rangle_n,\langle 2t \rangle_n,\langle 3t \rangle_n,\dots,\langle \alpha t \rangle_n \}.
\end{align}

First we state the following claim.
\begin{claim}\label{claim:25}
	For any two positive real numbers $a,b$ such that $a<b$, the number of integers $j$ such that $a < j \leq b$ is at most $\ceil{b-a}$.
\end{claim}

The following lemma is now presented.

\begin{lemma}\label{claim:10}
	Let $n$ be a prime number, $\alpha = \frac{n+1}{2}$, and $t_1,t_2$ be two distinct integers $1\leq t_1,t_2 \leq  \lfloor \frac{n-1}{m}\rfloor $. Then 
	$$ |W(n,t_1,\alpha)\cap W(n,t_2,\alpha)| <  \Big\lceil \frac{n}{4}\Big\rceil + \Big\lceil\frac{3n}{2m} \Big\rceil + 1. $$
\end{lemma}
\begin{IEEEproof}
	It is sufficient to prove this claim for $t_1=1$, since all the other cases are proved by relabeling $t_1$ to $1$ and $t_2$ to $t_2-t_1+1$. 
	In this case, 
	$$W(n,1,\alpha) = \Big\{1,2,\dots,\frac{n+1}{2}\Big\} =  \Big[\frac{n+3}{2}\Big]\setminus \{0\}.$$ 
	Thus, denote $t = t_2$ and since $0 \notin W(n,t,\alpha)$, it is sufficient to prove that for all $2 \leq t \leq \lfloor \frac{n-1}{m}\rfloor$, 
	$$\Big| \Big[\frac{n+3}{2}\Big]\cap W(n,t,\alpha) \Big| <  \Big\lceil \frac{n}{4}\Big\rceil + \Big\lceil\frac{3n}{2m} \Big\rceil  + 1.$$
	For an integer $k$ such that $1 \leq k \leq \frac{n+1}{2n}t$, let $A_k$ be the set	
	$$ A_k = \{\langle jt \rangle_n   ~|~  (k-1)\frac{n}{t} < j \leq k\frac{n}{t} \}.$$
	Note that  $ jt = (k-1)n + \langle jt \rangle_n$ and also
	$$   W(n,t,\alpha) = \bigcup^{\lfloor \frac{n+1}{2n}t \rfloor}_{k=1} A_k,$$
	where all $A_k$'s are mutually disjoint. Moreover, for all $1 \leq k \leq \frac{n+1}{2n}t$, 
	$$|A_k| \leq  \Big\lceil k\frac{n}{t} -  (k-1)\frac{n}{t} \Big\rceil = \lceil n/t \rceil,$$
	which holds due to Claim~\ref{claim:25}. 
	Hence, 
	\begin{align*}
	\Big| A_k\cap  \Big[\frac{n+3}{2}\Big] & \Big|  \leq \Big| A_k\cap  \Big[\frac{n-1}{2}\Big] \Big|+2 \\
	& \stackrel{(a)}{\leq} |\{\langle jt \rangle_n   ~|~  (k-1)\frac{n}{t} < j \leq (k-0.5)\frac{n}{t} \}| +2 \\
	& \stackrel{(b)}{\leq} \Big\lceil(k-0.5)\frac{n}{t} - (k-1)\frac{n}{t} \Big\rceil=  \Big\lceil \frac{n}{2t} \Big\rceil + 2,
	\end{align*}
	where $(a)$ holds since for all $\langle jt \rangle_n \in A_k \cap [\frac{n-1}{2}]$ it holds that  $0< \langle jt \rangle_n < \frac{n-1}{2}$ and hence
	\begin{align*}
	(k-1)n & < jt = (k-1)n + \langle jt \rangle_n \\
	& < (k-1)n +\frac{n-1}{2} < (k-0.5)n.
	\end{align*}
	Equality $(b)$ holds by Claim~\ref{claim:25}.
	Since $t \leq \lfloor \frac{n-1}{m} \rfloor$, it is deduced that
	\begin{align*}
	\Big| W(n,t,\alpha)\cap  \Big[ \frac{n+3}{2}\Big] \Big| 
	& \leq 	\Big|\bigcup^{\lfloor\frac{n+1}{2n}t\rfloor}_{k=1} A_k\cap  \Big[\frac{n+3}{2}\Big] \Big| \\
	&\leq  \frac{n+1}{2n}t \Big( \Big\lceil \frac{n}{2t} \Big\rceil + 2 \Big)\\
	& \leq   \frac{n+1}{2n}t \Big(\frac{n}{2t}+3\Big)   \\
	& =  \frac{n+1}{4} + \frac{3}{2}\frac{n+1}{n}t\\
	&  \leq  \frac{n+1}{4} + \frac{3}{2}\frac{n+1}{n}\frac{n-1}{m}\\
	&=  \frac{n+1}{4} + \frac{3}{2}\frac{n^2-1}{nm}\\
	& < \Big\lceil \frac{n}{4}\Big\rceil + \Big\lceil\frac{3n}{2m} \Big\rceil +1.
	\end{align*}
	
\end{IEEEproof}	
Note that this lemma holds also for $ \alpha =  \frac{n-1}{2}$. We state the following corollary which is derived directly from Lemma~\ref{claim:10}.
\begin{corollary}\label{cor:46}
	Assume that $W$ is a subset of one of the sets $W(n,t,\alpha)$, where $1\leq t\leq \lfloor \frac{n-1}{m}\rfloor$ and $\alpha = \frac{n+1}{2}$. If $|W|\geq  \lceil \frac{n}{4} \rceil + \lceil \frac{3n}{2m}  \rceil + 1$, then the value of $t$ can be uniquely determined. 
\end{corollary}

This corollary holds also for  $ \alpha =  \frac{n-1}{2}$. We proceed by introducing several more definitions.
For all $1\leq t \leq \lfloor \frac{n-1}{m} \rfloor$  denote the following set
\begin{align*}
B_t = \left\{ \Big\langle \frac{n+1}{2}it  \Big\rangle_n \middle| 
\begin{array}{cc}
i \in [n], i ~\textrm{is odd}
\end{array}
\right\}
\end{align*}
and the set $A_{n,m}$ to be
\begin{align*}
A_{n,m} = \left\{  (s,t)  \middle| 
\begin{array}{cc}
s \in B_t,1\leq t \leq \lfloor \frac{n-1}{m} \rfloor
\end{array}
\right\}.
\end{align*}
Note that for every fixed $1\leq t \leq \lfloor \frac{n-1}{m} \rfloor $, it holds that $|B_t| = \frac{n-1}{2}$.   Thus,
\begin{align}\label{eq:34}
|A_{n,m}| = \frac{n-1}{2}\cdot  \Big\lfloor \frac{n-1}{m} \Big\rfloor.
\end{align}
Next, the following lemma is presented.
\begin{lemma}\label{cor:22}
	For any $a\in[n]$, it holds that $\Big(\langle a-\frac{n+1}{2}t \rangle_n,t \Big) \in A_{n,m}$ if and only if $ (a,t) \notin A_{n,m}$.
\end{lemma}

\begin{IEEEproof}
	If $\Big(\langle a-\frac{n+1}{2}t \rangle_n,t \Big) \in A_{n,m}$ then $\langle a-\frac{n+1}{2}t \rangle_n \in B_t$. Therefore, there is an odd $i\in[n]$ such that $$\Big\langle a-\frac{n+1}{2}t\Big \rangle_n  = \Big\langle\frac{n+1}{2}it \Big\rangle_n.$$
	Thus, $a = \langle\frac{n+1}{2}(i+1)t \rangle_n $ when $i+1$ is even. Therefore, $a \notin B_t$ which leads to $ (a,t) \notin A_{n,m}$. The opposite direction is proved similarly.
\end{IEEEproof}

The construction of a $\ut(n,\frac{n-1}{2}\cdot  \lfloor \frac{n-1}{m} \rfloor,\lfloor  \frac{3n}{4} \rfloor - \lceil\frac{3n}{2m}\rceil - 2  )$ code will be as follows.
\begin{Construction}\label{const:tree2}
	For a prime  $n\geq 3$ let $\cC_{\cT_4}$ be the following code over trees
	\begin{equation*}
	\cC_{\cT_4} = \{T_{s,t} = (V_n,E_{s,t})~| ~(s,t)\in  A_{n,m} \}.
	\end{equation*}
\end{Construction}

\begin{theorem}\label{th:tree4}
	The code $\mathcal{C}_{\cT_4}$ is a $\ut(n,\frac{n-1}{2}\cdot \lfloor \frac{n-1}{m} \rfloor ,\lfloor  \frac{3n}{4} \rfloor - \lceil\frac{3n}{2m}\rceil  - 2 )$ code over trees.
\end{theorem}

\begin{IEEEproof}
	First, it is deduced above in~\eqref{eq:34} that  $| A_{n,m}| = \frac{n-1}{2}\cdot \lfloor \frac{n-1}{m} \rfloor$. We now prove that {$$|\mathcal{C}_{\cT_4}| = |A_{n,m}| =  \frac{n-1}{2}\cdot \Big\lfloor \frac{n-1}{m} \Big\rfloor.$$}
	It is clear that $|\mathcal{C}_{\cT_4}| \leq  |A_{n,m}|$ and assume in the contrary that $|\mathcal{C}_{\cT_4}| <  |A_{n,m}|$. 
	Thus, there are two distinct pairs $(s,t),(s',t') \in A_{n,m}$ such that $T_{s,t} = T_{s',t'}$, which implies that the central nodes of $T_{s,t}$ and $T_{s',t'}$ are identical. Since $\deg(s) = \deg(s')$, the nodes $v_s$ and $v_{s'}$ represent the same center node, so it  is deduced that $s=s'$. From that, by the definition of the second central node, it is immediately implied that $t=t'$ which results with a contradiction.

	Next, we show that $d=\lfloor  \frac{3n}{4} \rfloor - \lceil\frac{3n}{2m}\rceil - 2  $ by showing that it is possible to correct $\rho = d-1$ edge erasures due to Theorem~\ref{th:dist}. Assume that  $\rho$ edges are erased in a tree $T_{s,t} \in\mathcal{C}_{\cT_4}$.   
	We separate the proof for two cases.\\
	\textbf{Case 1:} after the erasure, both central nodes have degree of at least two, and will be denoted by $v_a$ and $v_b$.
	If $a = s$ and $b = \langle s+\frac{n+1}{2}t \rangle_n$, then 
	$$\langle (a-b)\cdot2 \rangle_n = \Big\langle \Big(s-(s+\frac{n+1}{2}t)\Big)\cdot2\Big\rangle_n =  \langle -t\rangle_n.$$
	Similarly, if $a =  \langle s+\frac{n+1}{2}t \rangle_n$ and $b =s$, then
	$$\langle (a-b)\cdot2 \rangle_n = \Big\langle \Big((s+\frac{n+1}{2}t)-s\Big)\cdot2 \Big\rangle_n = t.$$
	Since $t\leq \lfloor \frac{n-1}{m} \rfloor$, it is deduced that $\lceil \frac{n-1}{m} \rceil < \langle -t\rangle_n \leq  n-1$, so only one of these options is valid and $t$ is easily determined. Moreover, it is now determined which one of the values $a$ or $b$ is equal to $s$, and thus, $T_{s,t}$ is corrected.\\
	\textbf{Case 2:} after the erasure, one of the central nodes has degree of at most one. Denote by $v_a$ the central node with degree of at least two. Let $\alpha$ be a number such that if $a=s$ then $\alpha = \frac{n+1}{2}$ and if $a =  \langle s+\frac{n+1}{2}t \rangle_n$ then $\alpha = \frac{n-1}{2}$.
	Note that since $\rho$ edges were erased, $v_a$ has degree of at least
	\begin{align*}
	(n-1) - \rho -1  & = (n-1) - (\Big\lfloor  \frac{3n}{4}\Big\rfloor - \Big\lceil \frac{3n}{2m}\Big\rceil  - 3) -1   \\
	&= \Big\lceil  \frac{n}{4} \Big\rceil+ \Big\lceil \frac{3n}{2m} \Big\rceil + 1 .
	\end{align*} Thus, there are integers $i_1,i_2,\dots, i_{(n-2) - \rho } \in [n]$ such that the edge set
	$$  E = \{ \langle v_a,v_{\langle a+i_jt \rangle_n} \rangle |  1\leq j \leq (n-2) - \rho \}  $$
	consists of all the edges connected to $v_a$ and were not erased.
	Let $W(n,t,\alpha)$ be the set defined in~\eqref{Wt}, and let $W$ be the set
	\begin{align*}
	W = \left\{ \langle i_jt \rangle_n \in W(n,t,\alpha)  \middle|
	\begin{array}{cc}
	1\leq j \leq(n-2) - \rho ,\\ 
	\langle v_a,v_{\langle a+i_jt \rangle_n} \rangle\in E
	\end{array}
	\right\}.
	\end{align*}	
	Since $|W| = (n-2) - \rho  =\lceil  \frac{n}{4} \rceil + \lceil  \frac{3n}{2m} \rceil + 1 $, by Corollary~\ref{cor:46}, the value of $t$ is uniquely determined. 	Therefore, the codeword-tree $T_{s,t}$ is either $T_{a,t}$ or $T_{\langle a-\frac{n+1}{2}t \rangle_n,t}$. By Lemma~\ref{cor:22}, it holds that $\Big(\langle a-\frac{n+1}{2}t \rangle_n,t \Big) \in A_{n,m}$ if and only if $ (a,t) \notin A_{n,m}$. Thus, $T_{a,t} \in \cC_{\cT_4}$ if and only if $T_{\langle a-\frac{n+1}{2}t \rangle_n,t} \notin \cC_{\cT_4}$, and by finding either $T_{a,t}$ or $T_{\langle a-\frac{n+1}{2}t \rangle_n,t}$ in $\cC_{\cT_4}$ we find the codeword-tree $T_{s,t}$.
\end{IEEEproof}
Note that according to Theorem~\ref{th:tree4}, it is possible to construct codes of cardinality $\Omega(n^2)$, while the minimum distance $d$ approaches $\lfloor 3n/4 \rfloor$ and $n$ is a prime number. In Theorem~\ref{th:trees2} we showed that $A(n,n-3) = \cO(n^2)$, while from Theorem~\ref{th:tree4}, $A(n,d) = \Omega(n^2)$, when $d$ approaches $\lfloor 3n/4 \rfloor$ and $n$ is prime. Thus, it is interesting to study the values of $d$ for {such that} $A(n,d) = \Theta(n^2)$. 


\section{Conclusion}\label{sec:conc}
In this paper, we initiated the study of codes over trees over the tree distance. Upper bounds on such codes were presented together with specific code construction for several parameters of the number of nodes and minimum tree distance. For the tree ball of trees, it was shown that the star tree reaches the smallest size, while the maximum is achieved for the path tree. This guarantees that for a fixed value of $t$, the size of every ball of a tree is lower, upper-bounded from below, above by $\Omega(n^{2t})$,  $\cO(n^{3t})$, respectively. Furthermore, it was also shown that the average size of the ball is $\Theta(n^{2.5t})$. We also showed that optimal codes over trees ranged between $\cO(n^{n-d-1})$ and $\Omega(n^{n-2d})$. 

While the results in the paper provide a significant contribution in the area of codes over trees, there are still several interesting problems
which are left open. Some of them are summarized as follows.
\begin{enumerate}
	\item Improve the lower and upper bounds on the size of codes over trees, that is, the value of $A(n,d)$. 
	\item Find an optimal construction for $d=n-3$. 
	\item Study codes over trees under different metrics such as the tree edit distance.
	\item Study the problem of reconstructing trees based upon several forests in the forest ball of trees; for more details see~\cite{Davies}.
\end{enumerate}

\section*{acknowledgement}
The authors would like to thank the associate editor Amin Gohari and the three anonymous reviewers for their very helpful comments.

\appendices

\section{}\label{app:0}
\begin{customlemma}{\ref{lem29}.} 
	For a positive integer $n$ it holds that 
	\begin{align*}
	\sum_{T\in \mathbf{T}(n) } V_T(n,1) = \sum_{F\in {\mathbf{F}(n,2)}  } (V_F(n,1))^2- 	(n-2)n^{n-2} .
	\end{align*}	
\end{customlemma}
\begin{IEEEproof}
	{
		The following sequence of equalities will be explained below,
		\begin{align*}
		&\sum_{T\in \mathbf{T}(n) } V_T(n,1) =\\
		\stackrel{(a)}{=}& \sum_{T\in \mathbf{T}(n) } \Big(\sum_{F\in \cP_T(n,1) } (V_F(n,1) -1) +1\Big) \\
		= & \sum_{T\in \mathbf{T}(n) } \sum_{F\in \cP_T(n,1) } V_F(n,1) -\left(\sum_{T\in\mathbf{T}(n)  } \sum_{F\in \cP_T(n,1) }1 \right) +n^{n-2} \\
		\stackrel{(b)}{=} & \sum_{T\in \mathbf{T}(n) } \sum_{F\in \cP_T(n,1) } V_F(n,1) - 	(n-1)n^{n-2}  +n^{n-2} \\
		\stackrel{(c)}{=} & \sum_{F\in {\mathbf{F}(n,2)} } \sum_{T\in \cB_F(n,1) }  V_F(n,1) - 	(n-1)n^{n-2}  +n^{n-2} \\
		= & \sum_{F\in {\mathbf{F}(n,2)}  } (V_F(n,1))^2- 	(n-2)n^{n-2}.
		\end{align*}
		In equality $(a)$ we explain why 
		\begin{align*}
		V_T(n,1) -1 = | \cB_T(n,1)\setminus \{T\}| =\sum_{F\in \cP_T(n,1) } (V_F(n,1) -1).
		\end{align*}
		Note that for all $T, T' \in \mathbf{T}(n)$ such that $d_{\cT}(T,T') = 1$, there exists exactly one forest $F\in \mathbf{F}(n,2)$ such that $F\in \cP_T(n,1)\cap \cP_{T'}(n,1)$.
		Thus, each $ T' \in\cB_T(n,1) \setminus \{T\}$ can be generated from  $T$  uniquely by removing and adding exactly one edge. Equivalently, each such a tree is counted by adding an edge to a forest $F \in  \cP_T(n,1)$. By doing so for all forests in $\cP_T(n,1)$, while subtracting 1 for the tree $T$, equality $(a)$ holds.
		In  equality $(b)$ it is deduced that $$ \sum_{T\in \mathbf{T}(n) } \sum_{F\in \cP_T(n,1) } 1 = (n-1)n^{n-2}.$$ Lastly, in equality $(c)$, by taking pairs of trees and forests, the order of summation is changed,
		\begin{align*}
		\sum_{T\in \mathbf{T}(n) } \sum_{F\in \cP_T(n,1) } 1& = \sum_{F\in {\mathbf{F}(n,2)} } \sum_{T\in \cB_F(n,1) } 1.
		\end{align*}}
\end{IEEEproof}

\section{}\label{app:1.1}
\begin{customtheorem}{\ref{theo35}.}
	For any $T \in \mathbf{T}(n)$ it holds that
	\begin{align*}
	& \sum^{t}_{i=0} 	\binom{n-2-t+i}{i} V_T(n,t-i)  =n^{t-1} \hspace{-3ex}\sum_{(i_0,i_1,\dots,i_{t}) \in P_T(n,t)} i_0i_1\cdots i_t.
	\end{align*}
\end{customtheorem}
\begin{IEEEproof}
	By definition, for $t\geq 1$, $S_T(n,t) = V_T(n,t) - V_T(n,t-1)$. Thus,
	\begin{align*}
	& \sum^{t}_{i=0} 	\binom{n-1-t+i}{i} S_T(n,t-i) = \\
	& \sum^{t-1}_{i=0} 	\binom{n-1-t+i}{i} \Big( V_T(n,t-i) - V_T(n,t-1-i) \Big)\\
	& + 	\binom{n-1}{t}V_T(n,0) 	 = 	V_T(n,t) \\
	& + \sum^{t}_{i=1} V_T(n,t-i)\Big(	\binom{n-1-t+i}{i}  - 	\binom{n-2-t+i}{i-1} \Big) \\
	& \stackrel{(a)}{=} 	V_T(n,t) + \sum^{t}_{i=1} V_T(n,t-i)\binom{n-2-t+i}{i} \\
	& = \sum^{t}_{i=0} 	\binom{n-2-t+i}{i} V_T(n,t-i),
	\end{align*}	
	where $(a)$ holds by the identity $\binom{n}{k}+\binom{n+1}{k} = \binom{n+1}{k+1}$. 
	Using the result of Corollary~\ref{cor:10}, we conclude the proof.
\end{IEEEproof}

\section{}\label{app:1.2}
\begin{customlemma}{\ref{lemma:30}.} 
	For any positive integer $\alpha$, if
	\begin{align*}
	& \sum^{t}_{i=0} 	\binom{n-2-t+i}{i} V_T(n,t-i)  = \Omega(n^{\alpha t}),
	\end{align*}	
	and $V_T(n,0) = 1$, then  $V_T(n,t) =  \Omega(n^{\alpha t})$.
\end{customlemma}

\begin{IEEEproof}
	This lemma is proved by induction on $t$.\\
	\textbf{Base:} for $t=0$, $V_T(n,0) = n^0 = 1$ which is true by the definition.\\
	\textbf{{Inductive} Step:} suppose that the lemma holds for all $0\leq t' \leq t-1$. 
	Thus,
	\begin{align*}
	\Omega(n^{\alpha t})& = \sum^{t}_{i=0} 	\binom{n-2-t+i}{i} V_T(n,t-i)  \\
	&  =  V_T(n,t) + \sum^{t}_{i=1} 	\binom{n-2-t+i}{i}  V_T(n,t-i)   \\
	& =   V_T(n,t) +\sum^{t}_{i=1} 	\binom{n-2-t+i}{i}  \Omega(n^{\alpha(t-i)})   \\
	& =   V_T(n,t) +\sum^{t}_{i=1} 	\Omega(n^{i})  \Omega(n^{\alpha(t-i)})   \\
	& =  V_T(n,t) + \Omega (n^{\alpha (t-1)+1}).
	\end{align*}	
	Therefore we deduce that 
	\begin{align*}
	V_T(n,t) & = \Omega(n^{\alpha t})  -  \Omega (n^{\alpha (t-1)+1}) =\Omega(n^{\alpha t}) .
	\end{align*}
\end{IEEEproof}

\section{}\label{app:2}
\begin{customclaim}{\ref{claim:4}.}
	For a positive integer $n$ and a fixed $t$ it holds that
	\begin{align*}
	& \sum^{n-1}_{i=1}\binom{n}{i} i^{i} (n-i)^{n-i} \Theta (i^{t/2})  =  \Theta(n^{t/2})\sum^{n-1}_{i=1}\binom{n}{i} i^{i} (n-i)^{n-i}.
	\end{align*}
\end{customclaim}

\begin{IEEEproof}
	The upper bound is derived immediately, 
	\begin{align*}
	& \sum^{n-1}_{i=1}\binom{n}{i} i^{i} (n-i)^{n-i} \Theta (i^{t/2})  =  \cO(n^{t/2})\sum^{n-1}_{i=1}\binom{n}{i} i^{i} (n-i)^{n-i}.
	\end{align*}
	Next, the lower bound is proved by,
	\begin{align*}
	&  \sum^{n-1}_{i=1}\binom{n}{i} i^{i} (n-i)^{n-i} \Omega (i^{t/2}) \\
	&  \geq  \sum^{n-1}_{i=\lfloor \frac{n-1}{2} \rfloor }\binom{n}{i} i^{i} (n-i)^{n-i} \Omega (i^{t/2})  \\
	&  =   \Omega(n^{t/2})\sum^{n-1}_{i=\lfloor \frac{n-1}{2} \rfloor }\binom{n}{i} i^{i} (n-i)^{n-i} \\
	&  =   \Omega(n^{t/2})\sum^{n-1}_{i=1  }\binom{n}{i} i^{i} (n-i)^{n-i}.	
	\end{align*}
\end{IEEEproof}

\section{}\label{app:1}
\begin{customtheorem}{\ref{theo:29}.}
	The size of the sphere for a star satisfies
	\begin{align*}
	S^{\star}(n,t)= \binom{n-1}{t}(n-1)^{t-1}(n-t-1),
	\end{align*}
	and the size of the tree ball of trees for a star satisfies
	\begin{align*}
	V^{\star}(n,t)= \sum^t_{j=0}\binom{n-1}{j}(n-1)^{j-1}(n-j-1).
	\end{align*}	
\end{customtheorem}

\begin{IEEEproof}
	Let  $T \in \mathbf{T}(n)$ be a star tree, and denote the function
	\begin{align*}
	H(n,t)= \binom{n-1}{t}(n-1)^{t-1}(n-t-1).
	\end{align*}
	We say that $\frac{d}{dn}(f(n))$ is the derivative of $f(n)$ with respect to $n$. Thus,
	\begin{align*}
	& \sum^{t}_{i=0} 	\binom{n-1-t+i}{i} H(n,t-i) \\%
	& = \sum^{t}_{i=0} 	\binom{n-1-t+i}{i}  \binom{n-1}{t-i}(n-1)^{t-1-i}(n-t-1+i)  \\
	& \stackrel{(a)}{=} \sum^{t}_{i=0} 	\binom{n-1}{t}  \binom{t}{i}(n-1)^{t-1-i}(n-t-1+i)   \\
	& = \binom{n-1}{t} \sum^{t}_{i=0} 	  \binom{t}{i}(n-1)^{t-1-i}(n-t-1+i)   \\
	& = \binom{n-1}{t} \Big(\sum^{t}_{i=0} 	  \binom{t}{i}(n-1)^{t-1-i}(n-1) \\
	&~~~~~~~~~~~~~~~~-   \sum^{t}_{i=0}  \binom{t}{i}(n-1)^{t-1-i}(t-i)\Big)  \\
	& = \binom{n-1}{t} \Big(\sum^{t}_{i=0} 	  \binom{t}{i}(n-1)^{t-i}-   \frac{d}{dn}\Big(\sum^{t}_{i=0}  \binom{t}{i}(n-1)^{t-i}\Big)\Big) \\	
	& \stackrel{(b)}{=}  \binom{n-1}{t}\Big( n^{t}-tn^{t-1} \Big) = \binom{n-1}{t} n^{t-1}  (n-t) \\
	& \stackrel{(c)}{=}  n^{t-1} \sum_{(i_0,i_1,\dots,i_{t}) \in P_T(n,t)} i_0i_1\cdots i_t,
	\end{align*}
	where $(a)$ holds by known formula 
	$$\binom{a-(b-c)}{c}  \binom{a}{b-c} = \binom{a}{b}  \binom{b}{c}$$
	(i.e. $a=(n-1),b=t$, and $c=i$), and $(b)$ holds by the binomial theorem, which is, $$\sum^{t}_{i=0} \binom{t}{i}(n-1)^{t-i} = (n-1+1)^t = n^t.$$
	Equality $(c)$ holds due to~\eqref{eq:32}.
	Thus, by Corollary~\ref{cor:10}, it is deduced that $S^{\star}(n,t) = H(n,t)$.
	Next, 
	\begin{align*}
	V^{\star}(n,t)= \sum^t_{j=0}\binom{n-1}{j}(n-1)^{j-1}(n-j-1),
	\end{align*}
	which is derived by the fact that for every $T\in \mathbf{T}(n)$, $$ V_{T}(n,t)= \sum^t_{i=0} S_T(n,i).$$
\end{IEEEproof}

\section{}\label{app:4}
\begin{customclaim}{\ref{claim:5}.}
	The following properties hold
	\begin{enumerate}
		\item It holds that	
		\begin{align*}
		& \sum_{ (c_0,\dots,c_{t})  \in P_T(n,t;\bfv_\ell) } c_{\ell}\cdots c_{t} \\
		& = \sum_{ (c_0,\dots,c_{t})  \in 	A_T(n,t;\bfv_\ell,v_y) }\hspace{-3ex} c_{\ell}\cdots c_{t}  +\hspace{-3ex} \sum_{ (c_0,\dots,c_{t})  \in P_T(n,t;\bfv_{\ell+1}(v_y)) } \hspace{-3ex}c_{\ell}\cdots c_{t}.
		\end{align*}
		\item  If $v_x$ is not in $\bfv_\ell$ then,	
		\begin{align*}
		&    \sum_{ (c_0,\dots,c_t)  \in 	A_{T_1}(n-1,t;\bfv_\ell,v_y)  } \hspace{-3ex}c_{\ell}\cdots c_t =\hspace{-3ex}   \sum_{ (c'_0,\dots,c'_t)  \in 	A_T(n,t;\bfv_\ell,v_x)  }\hspace{-3ex} c'_{\ell}\cdots c'_{t}.
		\end{align*}
		\item If $v_x$ is not in $\bfv_\ell$ then,
		\begin{align*}
		&  \sum_{ (c_0,\dots,c_{t})  \in P_T(n,t;\bfv_{\ell+1}(v_x)) } c_{\ell}\cdots c_{t} \\	
		& = \sum_{ (c'_0,\dots,c'_{t})  \in P_{T_1}(n-1,t;\bfv_{\ell+1}(v_y)) } (c'_{\ell}+1)c'_{\ell+1}\cdots c'_{t}  \\
		& +  \sum_{ (c_0,\dots,c_{\ell-1},1,c_{\ell+1},\dots, c_t) \in P_T(n,t;\bfv_{\ell+1}(v_x)) } 1\cdot c_{\ell+1}\cdots c_{t}.
		\end{align*}
		\item If $v_x=v_{i_{\ell-1}}$ then 
		\begin{align*}
		& \sum_{ (c_0,\dots,c_t)  \in P_T(n,t;\bfv_\ell(v_x))} c_\ell\cdots c_t \\
		& =  \sum_{ (c'_0,\dots,c'_t)  \in P_{T_1}(n-1,t;\bfv_\ell(v_y)) } c'_\ell  \cdots c'_t   \\
		& +  \sum_{ (c_0,\dots,c_{\ell-2},1,c_{\ell},\dots, c_t) \in P_T(n,t;\bfv_{\ell}(v_x)) } c_{\ell}\cdots c_{t}.
		\end{align*}
	\end{enumerate}
\end{customclaim}

\begin{IEEEproof}
	\begin{enumerate}
		\item By definition of the set $A_T(n,t;\bfv_\ell,v_y)$ it holds  that $A_T(n,t;\bfv_\ell,v_y) \subseteq P_{T}(n,t;\bfv_\ell)$. Moreover, 
		\begin{align*}
		&\sum_{ (c_0,c_1,\dots,c_{t})  \in  P_{T}(n,t;\bfv_\ell) \setminus 	A_T(n,t;\bfv_\ell,v_y) } c_{\ell}\cdots c_{t}\\
		& = \sum_{ (c_0,c_1,\dots,c_{t})  \in P_{T}(n,t;\bfv_{\ell+1}(v_y)) } c_{\ell}\cdots c_{t},
		\end{align*}
		and the proof is concluded.
		\item  Again, $	A_T(n,t;\bfv_\ell,v_x) \subseteq P_{T}(n,t;\bfv_\ell)$. Since $v_x$ is not in $\bfv_\ell$ it holds that $v_x$ and $v_y$ are always in the same connected component with respect to $	A_T(n,t;\bfv_\ell,v_x)$, and thus, $|	A_T(n,t;\bfv_\ell,v_x)| = |	A_{T_1}(n-1,t;\bfv_\ell,v_y)|$. Moreover, since there is an index $j\in[\ell]$ such that $v_x,v_y \in C_j$ in $T$, it holds that $(c_0,\dots ,c_j,\dots ,c_t) \in A_T(n,t;\bfv_\ell,v_x)$ if and only if $(c_0,\dots (c_j-1),\dots ,c_t)  \in 	A_{T_1}(n-1,t;\bfv_\ell,v_y)$. Hence, this difference does not affect the equality, which concludes this proof. 
		\item 
		Assume that $v_x$ and $v_y$ are in the same connected component $C_\ell$ with respect to $T$.
		In this case since $v_x,v_y \in C_\ell$, it holds that $(c_0,\dots,c_\ell,\dots ,c_{t}) \in P_{T}(n,t;\bfv_{\ell+1}(v_x))$ if and only if $(c_0,\dots,c_\ell-1,\dots,c_{t})  \in P_{T_1}(n-1,t;\bfv_{\ell+1}(v_y)) $. 
		Thus, the following expression 
		$$\sum_{ (c'_0,c'_1,\dots,c'_{t})  \in P_{T_1}(n-1,t;\bfv_{\ell+1}(v_y)) } (c'_{\ell}+1)c'_{\ell+1}\cdots c'_{t}$$
		corresponds to all cases where the edge $\langle v_x,v_y\rangle$ was not removed, and 
		$$\sum_{ (c_0,\dots,c_{\ell-1},1,c_{\ell+1},\dots, c_t) \in P_{T}(n,t;\bfv_{\ell+1}(v_x)) } 1\cdot c_{\ell+1}\cdots c_{t},$$
		corresponds to all cases where the edge $\langle v_x,v_y\rangle$ was removed. Hence, the sum of the two expressions equals to
		$$\sum_{ (c_0,c_1,\dots,c_{t})  \in P_{T}(n,t;\bfv_{\ell+1}(v_x)) } c_{\ell}\cdots c_{t}.$$
		\item 
		Assume that $v_x$ and $v_y$ are in the same connected component $C_{\ell-1}$ with respect to $T$. Since $v_x,v_y \in C_{\ell-1}$, it holds that $(c_0,\dots,c_{\ell-1},\dots ,c_{t}) \in P_{T}(n,t;\bfv_{\ell}(v_x))$  if and only if $(c_0,\dots,c_{\ell-1}-1,\dots,c_{t})  \in P_{T_1}(n-1,t;\bfv_{\ell}(v_y)) $.
		Thus, the following expression 
		$$\sum_{ (c'_0,c'_1,\dots,c'_t)  \in P_{T_1}(n-1,t;\bfv_\ell(v_y)) } c'_\ell  \cdots c'_t,$$
		corresponds to all cases where the edge $\langle v_x,v_y\rangle$ was not removed, and 
		$$\sum_{ (c_0,\dots,c_{\ell-2},1,c_{\ell},\dots, c_t) \in P_{T}(n,t;\bfv_{\ell}(v_x)) } c_{\ell}\cdots c_{t}.$$
		corresponds to all cases where the edge $\langle v_x,v_y\rangle$ was removed. Again we get that the sum of the two expressions equals to 
		$$  \sum_{ (c_0,\dots,c_t)  \in P_T(n,t;\bfv_\ell(v_x))} c_\ell\cdots c_t.$$
	\end{enumerate}
\end{IEEEproof}

\section{}\label{app:5}
\begin{customlemma}{\ref{lemma:38}.}
	If $v_x$ is not in $\bfv_\ell$ then,	
	\begin{align*}
	f_{T}(n,t;\bfv_\ell)  & =  f_{T_{1}}(n-1,t;\bfv_\ell)  \\
	& + f_{T_{ 1}}(n-1,t;\bfv_{\ell+1}(v_y) )   + f_{T_{1}}(n-1,t-1;\bfv_\ell).
	\end{align*}
	If $v_x$ is in $\bfv_\ell$, and without loss of generality $v_x =v_{i_{\ell-1}} $, then
	\begin{align*}
	& f_{T}(n,t;\bfv_\ell)  =  f_{T_{1}}(n-1,t;\bfv_\ell(v_y))  +  f_{T_{1}}(n-1,t-1;\bfv_{\ell-1}).
	\end{align*}
\end{customlemma}

\begin{IEEEproof}	
	In this proof, it is assumed that  $v_y$ is not in $\bfv_\ell$, although the proof is valid also for this case, where by the definition $f_{T_{ 1}}(n-1,t;\bfv_{\ell+1}(v_y) )  = 0$. First we prove the case where $v_x$ is not in $\bfv_\ell$.
	In this case, we have that
	\begin{align*}
	& f_{T_{1}}(n-1,t;\bfv_\ell)   + f_{T_{ 1}}(n-1,t;\bfv_{\ell+1}(v_y))      \\
	& \stackrel{(a)}{=} \sum_{ (c'_0,c'_1,\dots,c'_{t})  \in P_{T_1}(n-1,t;\bfv_\ell) } c'_{\ell}\cdots c'_{t} \\
	& + \sum_{ (c'_0,c'_1,\dots,c'_{t})  \in P_{T_1}(n-1,t;\bfv_{\ell+1}(v_y) ) } c'_{\ell+1}\cdots c'_{t} \\
	& \stackrel{(b)}{=} \sum_{ (c'_0,c'_1,\dots,c'_{t})  \in 	A_{T_1}(n-1,t;\bfv_\ell,v_y) } c'_{\ell}\cdots c'_{t}  \\
	& + \sum_{ (c'_0,c'_1,\dots,c'_{t})  \in P_{T_1}(n-1,t;\bfv_{\ell+1}(v_y)) } c'_{\ell}\cdots c'_{t} \\
	& + \sum_{ (c'_0,c'_1,\dots,c'_{t})  \in P_{T_1}(n-1,t;\bfv_{\ell+1}(v_y)) } c'_{\ell+1}\cdots c'_{t} \\
	& = \sum_{ (c'_0,c'_1,\dots,c'_{t})  \in 	A_{T_1}(n-1,t;\bfv_\ell,v_y) } c'_{\ell}\cdots c'_{t}  \\
	& + \sum_{ (c'_0,c'_1,\dots,c'_{t})  \in P_{T_1}(n-1,t;\bfv_{\ell+1}(v_y)) } (c'_{\ell}+1)\cdots c'_{t} \\
	&  \stackrel{(c)}{=} \sum_{ (c_0,c_1,\dots,c_{t})  \in 	A_T(n,t;\bfv_\ell,v_x) } c_{\ell}\cdots c_{t}  \\
	& +  \sum_{ (c_0,c_1,\dots,c_{t})  \in P_{T}(n,t;\bfv_{\ell+1}(v_x))} c_{\ell}\cdots c_{t} \\
	& -  \sum_{ (c_0,\dots,c_{\ell-1},1,c_{\ell+1},\dots, c_t) \in P_{T}(n,t;\bfv_{\ell+1}(v_x)) } 1\cdot c_{\ell+1}\cdots c_{t}\\	
	& \stackrel{(d)}{=} \sum_{ (c_0,c_1,\dots,c_{t})  \in P_{T}(n,t;\bfv_\ell) } c_{\ell}\cdots c_{t} \\
	& -  \sum_{ (c_0,\dots,c_{\ell-1},1,c_{\ell+1},\dots, c_t) \in P_{T}(n,t;\bfv_{\ell+1}(v_x)) } 1\cdot c_{\ell+1}\cdots c_{t}.	
	\end{align*}
	Equality $(a)$ holds by definition of the function $f_{T_1}$.
	Equality $(b)$ holds due to Claim~\ref{claim:5}$(a)$.
	Equality $(c)$ holds by Claim~\ref{claim:5}$(b)$ and $(c)$.
	Equality $(d)$ holds again due to Claim~\ref{claim:5}$(a)$. Next we show that 
	\begin{align*}
	& f_{T_{1}}(n-1,t-1;\bfv_\ell) \\
	& = \sum_{ (c'_0,\dots c'_{\ell-1},c'_{\ell+1},\dots,c'_{t})  \in P_{T_1}(n-1,t-1;\bfv_\ell) } c'_{\ell+1}\cdots c'_{t} \\
	&  \stackrel{(a)}{=} \sum_{ (c_0,\dots,c_{\ell-1}, 1 , c_{\ell+1} ,\dots,c_{t})  \in P_{T}(n,t;\bfv_{\ell+1}(v_x)) } c_{\ell+1}\cdots c_{t},
	\end{align*}
	where $(a)$ holds since $(c_0,\dots c_{\ell-1},c_{\ell+1},\dots,c_{t})  \in P_{T_1}(n-1,t-1;\bfv_\ell)$ if and only if $(c_0,\dots c_{\ell-1},1,c_{\ell+1},\dots,c_{t})  \in P_{T}(n,t;\bfv_\ell)$.
	Thus,
	\begin{align*}
	& f_{T_{1}}(n-1,t;\bfv_\ell)   + f_{T_{ 1}}(n-1,t;\bfv_{\ell+1}(v_y) )    \\
	&  + f_{T_{1}}(n-1,t-1;\bfv_\ell)  \\
	&= \sum_{ (c_0,c_1,\dots,c_{t})  \in P_{T}(n,t;\bfv_\ell) } c_{\ell}\cdots c_{t} \\
	& -  \sum_{ (c_0,\dots,c_{\ell-1},1,c_{\ell+1},\dots, c_t) \in P_{T}(n,t;\bfv_{\ell+1}(v_x)) } 1\cdot c_{\ell+1}\cdots c_{t} \\
	& + \sum_{ (c_0,\dots,c_{\ell-1}, 1 , c_{\ell+1} ,\dots,c_{t})  \in P_{T}(n,t;\bfv_{\ell+1}(v_x)) } 1\cdot  c_{\ell+1}\cdots c_{t} \\
	& =  \sum_{ (c_0,c_1,\dots,c_{t})  \in P_{T}(n,t;\bfv_\ell) } c_{\ell}\cdots c_{t} = f_{T}(n,t;\bfv_\ell).
	\end{align*}
	Similarly, if $v_x$ is in $\bfv_\ell$ and $v_x = v_{i_{\ell-1}}$,
	\begin{align*}
	&  f_{T_{1}}(n-1,t;\bfv_\ell(v_y))  +  f_{T_{1}}(n-1,t-1;\bfv_{\ell-1}) \\
	&  \stackrel{(a)}{=} \sum_{ (c'_0,c'_1,\dots,c'_t)  \in P_{T_1}(n-1,t;\bfv_\ell(v_y)) } c'_\ell \cdots c'_t  \\
	& + \sum_{ (c'_0,\dots,c'_{\ell - 2} ,c'_{\ell},\dots,c'_t)  \in P_{T_1}(n-1,t-1;\bfv_{\ell-1}) } c'_\ell \cdots c'_t  \\
	& \stackrel{(b)}{=} \sum_{ (c_0,c_1,\dots,c_t)  \in P_{T}(n,t;\bfv_\ell(v_x))} c_\ell\cdots c_t \\
	& -  \sum_{ (c_0,\dots,c_{\ell-2},1,c_{\ell},\dots, c_t) \in P_{T}(n,t;\bfv_{\ell}(v_x)) }  c_{\ell}\cdots c_{t}	\\
	& + \sum_{ (c_0,\dots,c_{\ell-2},1,c_{\ell},\dots,c_t)  \in P_{T}(n,t;\bfv_\ell(v_x))} c_\ell \cdots c_t \\
	& = \sum_{ (c_0,c_1,\dots,c_t)  \in P_{T}(n,t;\bfv_\ell(v_x)) } c_\ell \cdots c_t \\
	& =  f_{T}(n,t;\bfv_\ell(v_x))=  f_{T}(n,t;\bfv_\ell) ,
	\end{align*}
	where equality $(a)$ holds by the definition of $f_{T_1}$,
	equality $(b)$ holds due to Claim~\ref{claim:5}$(d)$, and since  $(c_0,\dots,c_{\ell - 2} ,c_{\ell},\dots,c_t)  \in P_{T_1}(n-1,t-1;\bfv_{\ell-1}) $ if and only if $(c_0,\dots,c_{\ell-2},1,c_{\ell},\dots,c_t)  \in P_{T}(n,t;\bfv_\ell(v_x))$. 
\end{IEEEproof}

\section{}\label{app:6}
\begin{customlemma}{\ref{lemm:39}.}
	For any tree $T \in \mathbf{T}(n),n\geq 1$ and a vector of $0\leq \ell \leq t+1\leq n$ nodes $\bfv_\ell = (v_{i_0},v_{i_1},\dots,v_{i_{\ell - 1}})$,
	\begin{align*}
	f_{T}(n,t;\bfv_\ell)  \leq \binom{n+t-\ell}{2t+1-\ell}.
	\end{align*}
\end{customlemma}
\begin{IEEEproof}
	Note that if $\ell = t+1$ by the definition of $f_{T}$
	\begin{align*}
	f_{T}(n,t;\bfv_{t+1})  \leq \binom{n-1}{t} = \binom{n+t-(t+1)}{2t+1-(t+1)}.
	\end{align*}
	As showed in~\eqref{eq:30}, if $n = t+1$, then 
	\begin{align}\label{eq:31}
	f_{T}(n,n-1;\bfv_\ell) = 1  = \binom{n+(n-1)-\ell}{2(n-1)+1-\ell},
	\end{align}
	so the lemma is correct for this two cases. 
	Thus it is left to prove the cases where $0 \leq \ell <t+1 <n$, and it will be shown by the induction on $n\geq 1$. \\
	\textbf{Base:} immediately derived from~\eqref{eq:31}. \\
	\textbf{{Inductive Step:}}	
	assume that for any tree $T \in \mathbf{T}(n-1),n\geq 1$ and a vector of $1\leq \ell \leq t+1\leq n-1$ nodes $\bfv_\ell = (v_{i_0},v_{i_1},\dots,v_{i_{\ell - 1}})$,
	\begin{align*} 
	& 	f_{T}(n-1,t;\bfv_\ell)  \leq \binom{n-1+t-\ell}{2t+1-\ell}.
	\end{align*}
	Let $T \in \mathbf{T}(n)$ and let $v_x$ be a leaf connected to a node denoted by $v_y$. Assume that $T_1 \in \mathbf{T}(n-1)$ is the tree generated by removing $v_x$ from $T$.
	For two integers $t$ and $\ell$ such that $0 \leq \ell <  t+1 < n$, let   $\bfv_\ell = (v_{i_0},v_{i_1},\dots,v_{i_{\ell - 1}})$ be a vector of $\ell$ nodes in $T$.	If $v_x$ is not in $\bfv_\ell$, using Lemma~\ref{lemma:38} and the induction assumption,  we deduce that
	\begin{align*}
	& f_{T}(n,t;\bfv_\ell)   =  f_{T_{1}}(n-1,t;\bfv_\ell)  \\
	& + f_{T_{ 1}}(n-1,t;\bfv_{\ell+1}(v_y))  + f_{T_{1}}(n-1,t-1;\bfv_\ell) \\
	& \leq  \binom{n-1+t-\ell}{2t+1-\ell} +  \binom{n-1+t-\ell-1}{2t+1-\ell-1} \\
	& +  \binom{n-1+t-1-\ell}{2t-1-\ell} \\
	& =   \binom{n-1+t-\ell}{2t+1-\ell} +  \binom{n-1+t-\ell}{2t-\ell} \\
	& =  \binom{n+t-\ell}{2t+1-\ell},
	\end{align*}
	where each equality holds by the identity $\binom{n}{k}+\binom{n+1}{k} = \binom{n+1}{k+1}$.
	Similarly, if $v_x\in \bfv_\ell$, and without loss of generality $v_x = v_{i_{\ell-1}}$, then
	\begin{align*}
	& f_{T}(n,t;\bfv_\ell)  =  f_{T_{1}}(n-1,t;\bfv_\ell(v_y))  +  f_{T_{1}}(n-1,t-1;\bfv_{\ell-1}) \\
	& \leq \binom{n-1+t-\ell}{2t+1-\ell} +  \binom{n-1+t-1-\ell+1}{2t-1-\ell+1} \\
	& =  \binom{n+t-\ell}{2t+1-\ell}. 
	\end{align*}	
\end{IEEEproof}

\end{document}